\documentclass[preprint2,onecolappendix,numberedappendix,appendixfloats]{aastex6}
\usepackage{graphicx}
\usepackage{physics}
\graphicspath{{figures/}}
\begin{document}

\title{Evidence for Different Disk Mass Distributions Between Early and Late-Type Be Stars in the
BeSOS Survey}

\author{C. Arcos\altaffilmark{1}, C. E. Jones\altaffilmark{2}, T. A. A. Sigut\altaffilmark{2,3}, S. Kanaan\altaffilmark{1} and M. Cur\'e\altaffilmark{1}}
\affil{\altaffilmark{1} Instituto de F\'isica y Astronom\'ia, Facultad de Ciencias, Universidad de Valpara\'iso. 
 Av. Gran Bretana 1111, Valpara\'iso, Chile. \\
 \altaffilmark{2} Department of Physics and Astronomy, The University of Western Ontario.
 London, Ontario, N6A 3K7, Canada. \\
  \altaffilmark{3} Centre for Planetary Science and Exploration, The University of Western Ontario.
 London, Ontario, N6A 3K7, Canada}

%%\altaffiltext{1}{catalina.arcos@uv.cl}

\begin{abstract}
 
The circumstellar disk density distributions for a sample of 63 Be southern stars from the BeSOS survey were found by modelling their H$\alpha$ emission line profiles. These disk densities were used to compute disk masses and disk angular momenta for the sample. Average values for the disk mass are 3.4$\times$10$^{-9}$ and 9.5$\times$10$^{-10}$ $M_{\star}$ for early (B0-B3) and late (B4-B9) spectral types, respectively. We also find that the range of disk angular momentum relative to the star are between 150-200 and 100-150 $J_{\star}/M_{\star}$, again for early and late-type Be stars respectively. The distributions of the disk mass and disk angular momentum are different between early and late-type Be stars at a 1\% level of significance. Finally, we construct the disk mass distribution for the BeSOS sample as a function of spectral type and compare it to the predictions of stellar evolutionary models with rapid rotation. The observed disk masses are typically larger than the theoretical predictions, although the observed spread in disk masses is typically large.

\end{abstract}

\keywords{stars: emission lines, Be  --- surveys ---  circumstellar matter}

\section{Introduction} \label{sec:intro}

A Be star is defined by \cite{Collins1987} as ``A non-supergiant B star whose spectrum has, or had at some time, one or more Balmer lines in emission". 
The accepted explanation for the emission lines is the presence of a circumstellar envelope (CE) of gas surrounding the central star analogous to the first model of a Be star proposed by \citet{Struve1931}. The material is expelled from the central star and placed in a thin equatorial disk with Keplerian rotation \citep{Meilland2007}. Different mechanisms such as rapid rotation \citep{Porter1996,Townsend2004,Domiciano2003,Fremat2005}, mass loss from the stellar wind \citep{Stee1994,Bjorkman1993,Cure2004,Silaj2014b}, binarity \citep{Okazaki2002,Romero2007,Oudmaijer2010}, magnetic fields \citep{Donati2001,Cassinelli2002,Neiner2003}, and stellar pulsations \citep{Rivinius2003} have been proposed to explain how the star loses enough mass to form the CE and how this material is placed in orbit, but it seems that more than one mechanism is required to reproduce the observations. Such mechanisms must continually supply enough angular momentum from the star to form and to maintain the disk. Given some mechanism to deposit material into the inner edge of the disk, the evolution of the gas seems well described by the viscous disk decretion model presented by \cite{Lee1991}, with angular momentum transported throughout the disk by viscosity \citep{Rivinius2013b}.

Be stars are variable on a range of different time scales associated with a variety of phenomenon occurring in the disk.
For example, short-term variations ($ \sim $ hours-days) in the emission lines are associated with non-radial pulsations, probably due to the high rotation rate of the central star \citep[e.g.][]{Rivinius2003,Rivinius2013b}; intermediate-term variations ($ \sim $ months-years) are seen in the cyclical variation between the violet and red peaks in doubled-peaked emission lines. Such variations are well represented by the global disk oscillation model \citep{okazaki1997,carciofi2009}.
Longer term variability, in some cases the emission lines disappear and/or are formed again on timescales of years to decades, is associated with the formation and dissipation of the disk (see section 5.3.1 of \citealp{Rivinius2013} for several examples). 

Spectroscopy of the emission lines can be used to get information about the geometry, kinematics and physical properties of the disk. A very convenient model, in agreement with observations, is to assume that the density in the disk's equatorial plane falls with a power law with exponent, $n$, and follows a Gaussian model in the vertical direction (see details provided in section~\ref{sec:theory:a}). 

We use the density distribution described above, the radiative transfer code \texttt{BEDISK} and the auxiliary complementary code \texttt{BERAY} to solve the transfer equation along many rays ($\sim10^{5}$) through the star/disk configuration. A grid of calculated H$\alpha$ line profiles from models with different disk density distributions and stellar parameters are used to match the observed H$\alpha$ line profiles and provide constraints on the disk parameters. We apply this method to a sample of 63 stars from the BeSOS catalogue. We selected a fraction of the best fitting models and we obtained the distribution of the disk density parameters, mass and total angular momentum content in the disk, with results provided for both early- and late-type Be stars. 

This paper is organized as follows: Our program stars and reduction steps are given in Section~\ref{sec:sample}. Section~\ref{sec:theory} describes our theoretical models including the main assumptions of \texttt{BEDISK} and \texttt{BERAY} codes in Section~\ref{sec:theory:a}. Input parameters to create the grid of models are provided in Section~\ref{sec:theory:b}. Section~\ref{sec:results} describes our results from selecting best-fit disk density parameters from all our sample stars in two ways: visual inspection (Subsection~\ref{sec:results:b}) and a percentage of the best models (Subsection~\ref{sec:results:c}). Subsection~\ref{sec:results:d} gives the mass and angular momentum distributions of the disks. A discussion and conclusions of our main results are presented in Sections~\ref{sec:discussion} and ~\ref{sec:conclusion}, respectively. The Appendix displays H$\alpha$ spectra from our best-fit models for our program stars compared to observations.

\section{Sample and data reduction} \label{sec:sample}

\begin{figure}[ht]
%\figurenum{1}
\centering
\includegraphics[width=1\columnwidth]{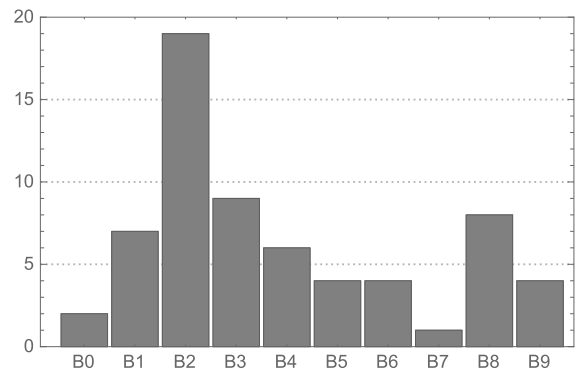}
\caption{Histogram of the sample of Be stars by spectral type. The distribution peaks at B2, which corresponds to \hbox{$\sim \\ 30 \%$} of the sample.}
\label{Histogram}
\end{figure}

We selected Be stars with B spectral type near or on the main sequence from the Be Stars Observation Survey (BeSOS\footnote{\url{http://besos.ifa.uv.cl}}) catalogue for our study. All Be targets in BeSOS website are confirmed as a Be star in the BeSS\footnote{\url{http://basebe.obspm.fr/basebe/}} catalogue or have an IR excess in the spectral energy distribution. This gives us a total of 63 Be stars. The sample distribution of spectral type is shown in the Figure~\ref{Histogram}. Approximately 30\% of our sample corresponds to the B2V spectral type. The same distribution was found previously by other authors \citep{Porter1996,Slettebak1982}, with B2V being the most frequently observed spectral type in Be stars. 
\\
BeSOS spectra were obtained using the Pontifica Universidad Catolica (PUC) High Echelle Resolution Optical Spectrograph (PUCHEROS) developed at the Center of Astro-Engineering of PUC \citep{Infante2010}. The instrument is mounted at the ESO 50 cm telescope of the PUC Observatory in Santiago, Chile, and has a spectral range of 390-730 nm with a spectral resolution of $\lambda/\Delta\lambda$ $ \sim $ 18000. Details about the instrument are provided in \cite{Vanzi2012}.  Observations were acquired between 2012 November and 2015 October. The exposure time was chosen to reach a S/N in the range 100-200 (as a consequence, the BeSOS catalogue has a limiting magnitude of V$<$6 in the sample selection criteria). 
For the wavelength calibration, exposures of ThAr lamps were used. The data reduction was performed using IRAF \citep{tody1993} following standard reduction procedures described in ``A User's Guide to Reducing Echelle Spectra with IRAF''\footnote{\url{http://www.astro.uni.wroc.pl/ludzie/molenda/echelle_iraf.pdf}}. 
The basic steps included removing bias and dark contributions,  flat fielding, order detection and extraction, fitting the dispersion relation, normalization, wavelength calibration, and heliocentric velocity corrections.

\section{Theoretical models} \label{sec:theory}

\subsection{Disk density and temperature structure} 
\label{sec:theory:a}

We calculated theoretical H$\alpha$ line profiles using two codes: \texttt{BEDISK}, a non-local thermodynamic equilibrium (non-LTE) code developed by \citet{Sigut2007}, and \texttt{BERAY} \citep{Sigut2011}, an auxiliary code that uses \texttt{BEDISK}'s output to solve the transfer equation along a series of rays ($ \sim $10$^{5}$) to produce model spectra.

There are two significant components that must be specified to model the physics of a star+disk system:  the density distribution of the gas in the disk and the input energy provided by the photo-ionizing radiation field of the central star. Assuming both, \texttt{BEDISK} code solves the statistical equilibrium equations for the ionization states and level populations using a solar chemical composition. Then, the code calculates the temperature distribution in the disk by enforcing radiative equilibrium. All calculations are made under the assumption that the vertical density distribution is fixed in approximate hydrostatic equilibrium, and the geometry of the disk is axisymmetric about the stars's rotation axis and symmetric on the midplane of the disk. 

The assumed density distribution has the form:
\begin{equation}
\rho(R,Z) = \rho_{0}\left( \frac{R}{R_{\star}}\right)^{-n} \exp\left({-(Z/H)^{2}}\right),
\label{EQ1}
\end{equation}
where $Z$ is the height above the equatorial plane, $R$ is the radial distance from the stars' rotation axis, $ \rho_{0} $ is the initial density in the equatorial plane, $n$ is the index of the radial power law, and $H$ is the height scale in the $Z$-direction and is given by
\begin{equation}
H = H_{0}\left(  \frac{R}{R_{\star}}\right) ^{3/2},
\label{EQ2}
\end{equation}
with the parameter $ H_{0}$ defined by,
\begin{equation}
H_{0} = \left( \frac{2 R_{\star}^{3} k T_{0}}{G M_{\star} \mu_{0} m_{H}}\right)^{1/2} ,
\label{EQ3}
\end{equation}
where $M_{\star}$ and $ R_{\star} $ are the stellar parameters, mass and radius, respectively; $G$ is the gravitational constant, $m_{H}$ is the mass of a hydrogen atom, $k$ is the Boltzmann constant, $ \mu_{0} $ is the  mean molecular weight of the gas and $T_{0}$ is  an isothermal temperature used only  to fix the vertical structure of the disk initially. This parameter was fixed at $T_{0} = 0.6  T_{eff}$ \citep{Sigut2009}.
Since Be stars are fast rotators, the rotational velocity of the star was assumed to be 0.8$v_{crit}$ for all spectral types, where $v_{crit}$ is given by

\begin{equation}
v_{crit} = \sqrt{\frac{2GM_{\star}}{3R_{\star}}}.
\label{EQ4}
\end{equation}

Finally, the rotation of the disk is assumed to be in pure Keplerian rotation \citep{Meilland2007}. For more details the reader is referred to \citet{Sigut2007}. 

\subsection{Input parameters and grid of models} \label{sec:theory:b}
We computed a grid of models using \texttt{BEDISK/BERAY} for a range of spectral classes from B0 to B9 in integer steps in spectral subtype in the main sequence stage. For early spectral types, we also computed models for B0.5 and B1.5  due to the large number of  B2V stars in our program stars (see Figure~\ref{Histogram}). We also included turbulent velocity ($v_{tur}$ = 2.0 km s$^{-1}$) into the disk for a more realistic model, since thin disks are likely to be turbulent \citep{Frank1992} which increases the Doppler width in line profiles. The stellar parameters were interpolated from \citet{cox2000} and are displayed in the Table~\ref{table1}. Each disk model was computed using 65 radial ($R$) and 40 vertical ($Z$) points. The spacing of the points in the grid is non-uniform, with smaller spacing near the star and in the equatorial plane where density is the greatest.
\citet{Jones2008} studied the disk density of classical Be stars by matching the observed interferometric H$\alpha$ visibilities with Fourier transforms of synthetic images produced by the \texttt{BEDISK} code. In their study, they suggest that the base density $\rho_{0}$ is typically between $10^{-12}$ to $10^{-10}$ g cm$^{-3}$ and the index power-law, $n$, normally ranges from 2 to 4 \citep{Waters1987}. 
 The outer radius of the H$\alpha$ emitting region has been estimated by several authors considering samples of Be stars as well as studies for individual stars (see Discussion Section~\ref{sec:discussion:b}). \cite{Hanuschik1986} found that a typical outer radius of the envelope region producing the secondary H$\alpha$ component is 20$R_{\star}$ and a similar value was found by \cite{Slettebak1992} of 18.9$R_{\star}$ for strong lines and 7.3$R_{\star}$ for weak lines. Measurements obtained using interferometric techniques determine the H$\alpha$ emitting region to be between $\sim$ 5.0 - 30.0 $R_{\star}$ \citep[e.g.][]{Tycner2005,Grundstrom2006}. 
 Given this, we computed models for a disk truncation radius, $R_{T}$, of: 6.0, 12.5, 25.0 and 50.0 $R_{\star}$, with base densities of : (0.1, 0.25, 0.5, 0.75, 1.0, 2.5, 5.0, 7.5, 10.0, 25.0) $\times$10$^{-11}$ g cm$^{-3}$ and $n$ from 2.0 to 4.0 in increments of 0.5, to adequately cover the full range of parameters space reported in the literature. Finally, the inclination angle $i$ was varied from 10$^{\circ}$ to 90$^{\circ}$, in steps of 10$^{\circ}$, with 90$^{\circ}$ replaced by 89$^{\circ}$ to avoid an infinity value. Thus with 9 $\rho_{0}$ values, 5 $n$ values, 9 $i$ values and 4 $R_{T}$ values, each spectral type is represented by a library of 1620 individual H$\alpha$ model line profiles.
To properly compare the synthetic profiles with our observations, every model was convolved with a Gaussian to match the resolving power of 18000 of our spectra. 

%________________________________________________________________
\begin{table}[h]
\caption{Adopted Stellar Parameters}             % title of Table 
 \label{table1}
\centering                          % used for centering table
\begin{tabular}{l c c c c}        % centered columns (4 columns)
\hline\hline                 % inserts double horizontal lines
ST & $T_{eff}$ & $\log$ g & $R_{\star}$ & $M_{\star}$ \\    % table heading 
   &    (K)    &          &    ($R_{\odot}$) & ($M_{\odot}$) \\
\hline                        % inserts single horizontal line
   B0V & 30000 & 4.0 &   7.40 &  17.50 \\      % inserting body of the table
   B0.5V &27800&4.0 &   6.93 &  15.43 \\      % inserting body of the table
   B1V & 25400 & 3.9 &   6.42 & 13.21 \\
   B1.5V & 23000 &4.0&  5.87& 11.04 \\
   B2V & 20900 & 3.9 &   5.33 & 9.11 \\
   B3V & 18800 & 4.0 &   4.80 & 7.60 \\
   B4V &  16800 & 4.0&   4.32&  6.62 \\
   B5V & 15200 & 4.0 &   3.90 & 5.90 \\
   B6V & 13800 & 4.0 &   3.56 & 5.17 \\
   B7V & 12400 & 4.1 &   3.28 & 4.45 \\
   B8V & 11400 & 4.1 &   3.00 & 3.80 \\
   B9V & 10600 & 4.1 &   2.70 & 3.29 \\
\hline                                   %inserts single line
\end{tabular}
\end{table}

\subsection{Behaviour of the H$\alpha$ emission line} \label{sec:results:aa}

\begin{figure*}[ht]
%\figurenum{2}
\centering
\includegraphics[scale=0.5]{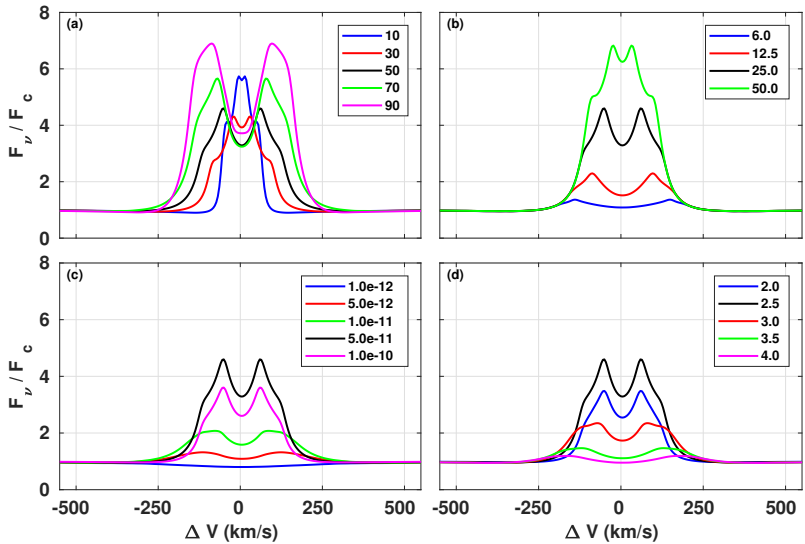} 
  \caption{Example of the variation of the emission H$\alpha$ line profiles by varying disk parameters. The reference model is shown in black in each panel for ease of comparison and corresponds to the disk parameters of $n$ = 2.5, $\rho_{0}$ = 5.0$\times10^{-11}$ g cm$^{-3}$, $R_{T} = $ 25.0 $R_{\star}$ and $i = 50^{\circ}$.  The fluxes are normalized to the continuum star+disk flux outside of the line. \textit{Top left:} inclination variation. \textit{Top right:} disk truncation radius variation. \textit{Bottom left:} base density variation. \textit{Bottom right:} power-law exponent variation.} 
   \label{Ha-variation}
\end{figure*}

\begin{figure*}[ht]
%\figurenum{3}
\centering
\includegraphics[scale=0.5]{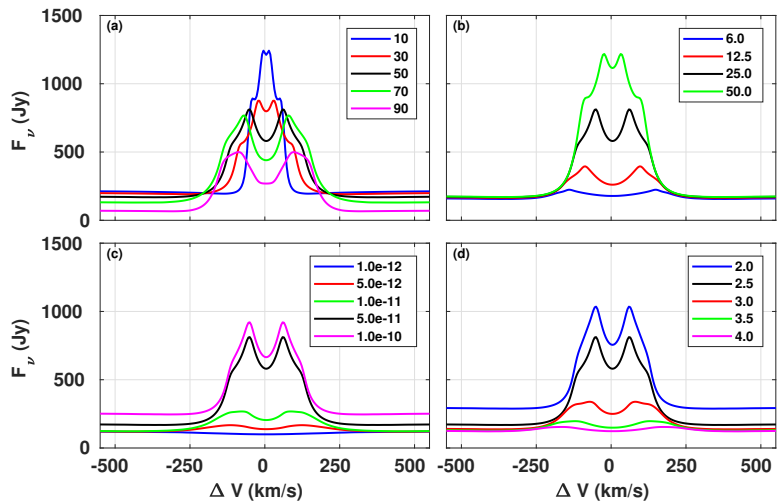} 
  \caption{Same as Figure~\ref{Ha-variation} but with the H$\alpha$ lines plotted as absolute fluxes in Janskys.} 
   \label{Ha-variation2}
\end{figure*}

Prior to beginning our statistical analysis, we illustrate the behavior of the predicted H$\alpha$ emission line profile as each of the four model parameters, $\rho_{0}$, $n$, $R_{T}$ and $i$, are varied. Figure~\ref{Ha-variation} shows the results, with the line profiles convolved down to a nominal resolution of $\lambda/\Delta\,\lambda=20000$. The fluxes are normalized by the continuum star+disk flux outside of the line. The reference model, shown in black in each panel, was chosen to be a disk with parameters $n=2.5$, $\rho_{0}$ = 5.0$\times 10^{-11}\,\rm g cm^{-3}$, $i = 50^{\circ}$ and $R_{T}=25.0\,R_{\star}$ surrounding a central B2V star. Panel~(a) shows the predicted lines obtained by varying the inclination from $10^{\circ}$ to $90^{\circ}$ in steps of $20^{\circ}$. The profile goes from a singly-peaked,``wine bottle" profile at $10^{\circ}$, to a doubly-peaked profile for higher inclinations. While the profile at line centre does not drop below the continuum at $i=90^{\circ}$, it does strongly satisfy the shell-star definition of \cite{Hanuschik1996} in which the peak to line centre flux ratio exceeds 1.5. Absorption below the continuum would result for less massive disks. Panel~(b) shows the result of varying the disk truncation radius; the flux increases strongly with the disk size and the emission peak separation becomes smaller for larger disks, as expected by the \cite{Huang1972} relation. Panel~(c) shows the effect of increasing the base density of the disk, $\rho_{0}$. The emission line strength increases with increasing $\rho_{0}$ up to the reference value of $5.0\times 10^{-11}\,\rm g cm^{-3}$, but then decreases for higher densities. This occurs because the line profile is the ratio of the total flux, line-plus-continuum, to the continuum flux alone. The line flux saturates with density first, causing the ratio to then decrease with increasing $\rho_0$ as the unsaturated continuum flux then increases faster. Finally, panel~(d) shows the effect of varying the power-law index of equatorial plane drop-off. The behaviour reflects both the effect of increased density seen in panel~(c) combined with a reduction in the emission peak separation since the disk density is concentrated closer to the star for larger $n$.

As noted in the previous paragraph, the H$\alpha$ line profiles shown as relative fluxes, {\it i.e.} divided by the predicted star+disk continuum, can show a more complex behaviour than might be expected because the line and continuum fluxes often have a different dependence on, say, the disk density. To clarify this point, Figure~\ref{Ha-variation2} shows the same line profiles as Figure~\ref{Ha-variation} but plotted as absolute fluxes in Janskys without continuum normalization. In panel~(a) of Figure~\ref{Ha-variation2}, the $i=90^{\circ}$ profile is now the weakest and the $i=0^{\circ}$ profile, the strongest. The disk contribution to the normalizing continuum decreases in proportion to the disk's projected area, i.e.\ $\cos(i)$, while for large inclinations, $i\sim 90^{\circ}$, the stellar continuum can be significantly obscured by the circumstellar disk. In panel~(b), there is a strong dependence of the line flux on $R_T$, whereas the continuum flux is essentially independent of $R_T$. This is because the continuum forms very close to the central star (inside of the $6\,R_*$, the smallest disk considered) whereas the optically thick H$\alpha$ line emission forms over a much larger portion of the disk. In panel~(c), the fluxes are now seen to scale in order with increasing $\rho_o$, and the saturation of the line flux as compared to the continued increase in the continuum flux is clear. Finally in panel~(d), the line fluxes are ordered with increasing flux with decreasing $n$, and the dependence of the continuum flux with the density-drop off in the disk is as expected.

\begin{figure*}[ht]
%\figurenum{4}
\centering
\includegraphics[scale=0.5]{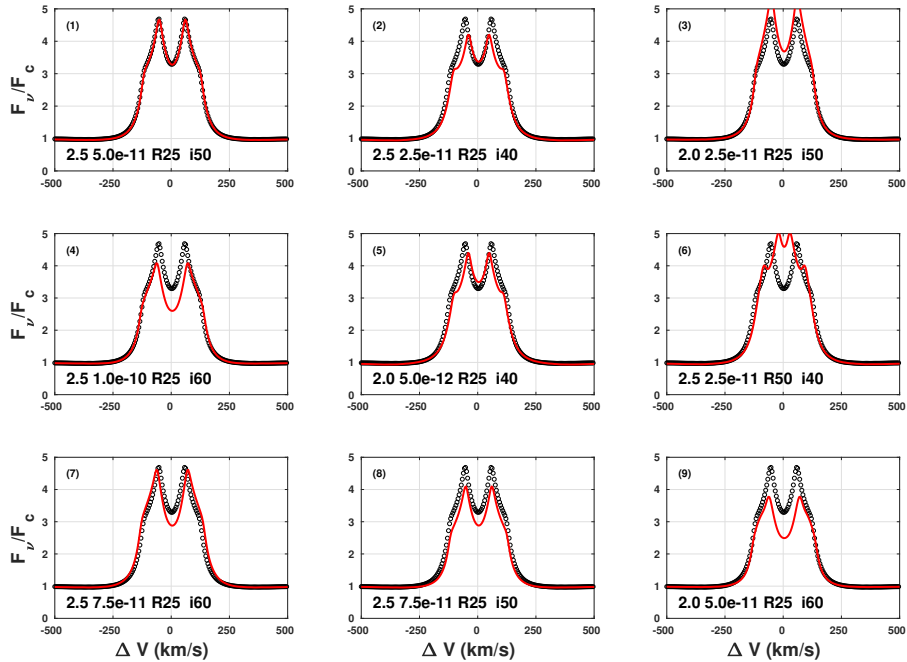} 
  \caption{The top nine most similar profiles in the B2V H$\alpha$ line library to the reference profile of Figure~\protect{\ref{Ha-variation}}. The first panel is an identical match, whereas panels (2) through (9) represent increasing differences as measured by the average percentage difference between the two profiles. The model parameters $(n,\rho_o,R_T,i)$ are as indicated at the bottom of each panel, and the reference parameters are those given in panel~1.}
   \label{similar-profiles}
\end{figure*}

Figure~\ref{Ha-variation} suggests that there is some degeneracy among the calculated H$\alpha$ line profiles, {\it i.e.} very similar relative flux line profiles can result from different combinations of the model parameters $(n,\rho_o,R_T,i)$. To explore this further, we have used the reference profile of Figure~\ref{Ha-variation} corresponding to $(n=2.5,\rho_0=5\times 10^{-11}\,\rm g cm^{-3},R_T=25\,R_*, i=50^\circ)$ as a simulated observed profile and searched the B2V profile library for the top nine closest model profiles as defined by the smallest average percentage difference between the model and ``observed" profile across the line: this figure-of-merit for the closeness of two line profiles is further discussed in the next section. Figure~\ref{similar-profiles} shows the results. While all nine profiles share the same $R_T$, there are small differences among the returned parameters, with $n$ ranging between $2.0$ and $2.5$, $\rho_o$, between $5.0\times 10^{-12}$ and $7.5\times 10^{-11}\;\rm g\,cm^{-3}$, and $i$ between $40^\circ$ and $60^{\circ}$. The variations in the parameters are correlated: typically, smaller $\rho_o$ values are associated with larger $n$ values. In the next section, we describe how we deal with this degeneracy in assigning model parameters to each star.

\section{Results} \label{sec:results}

\subsection{Selection of the best disk models} \label{sec:results:a}

\begin{figure*}[ht]
%\figurenum{5}
\centering
    \includegraphics[width=1\columnwidth]{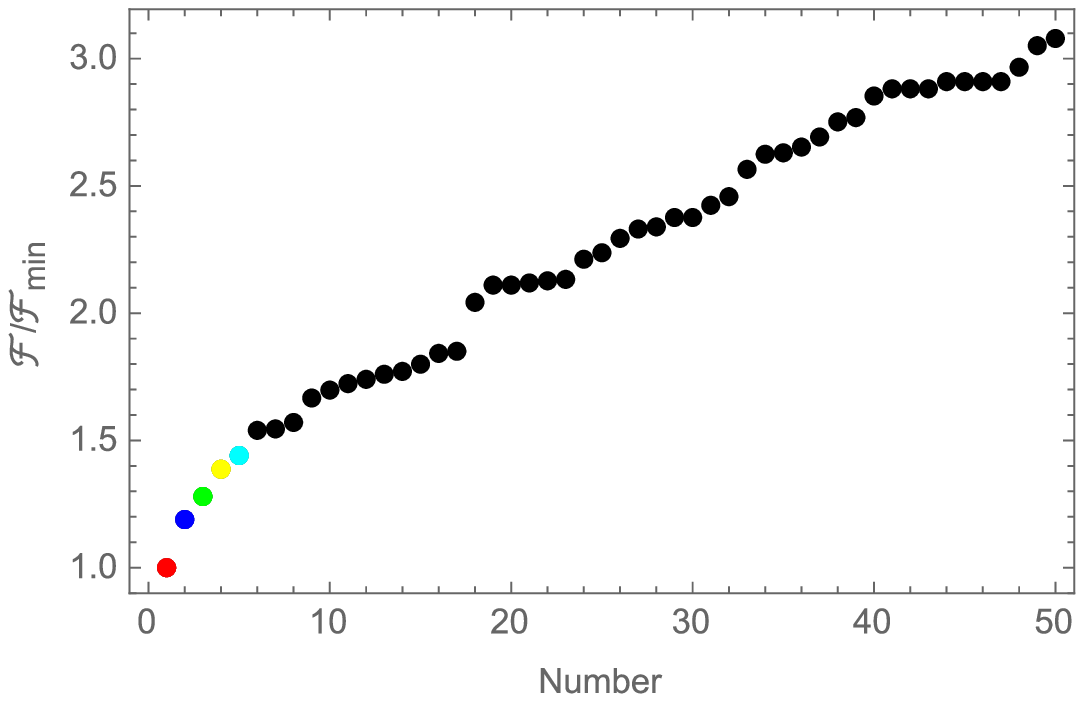}  
    \includegraphics[width=1\columnwidth]{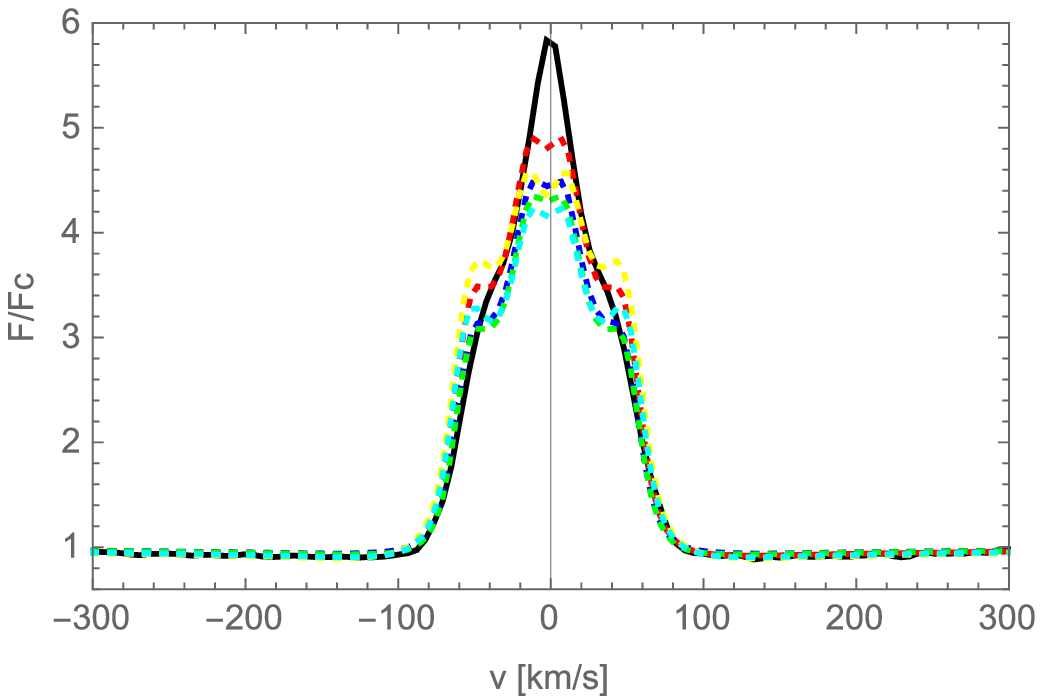} \\
    \vspace{0.5cm}
    \includegraphics[width=1\columnwidth]{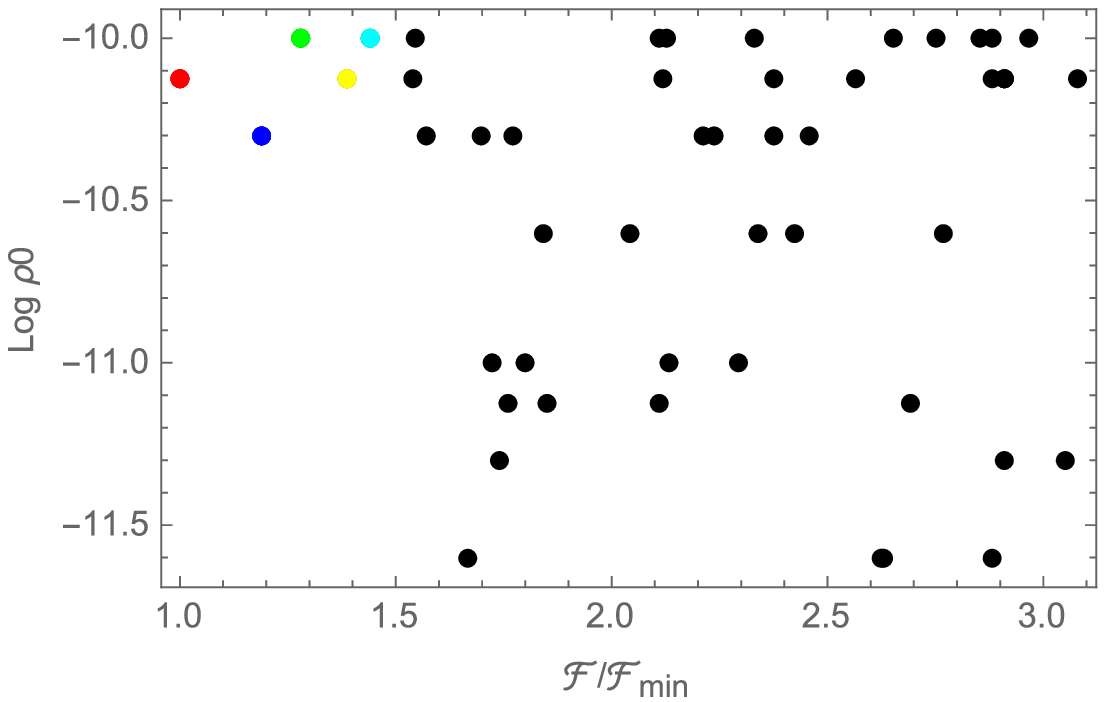} 
    \includegraphics[width=1\columnwidth]{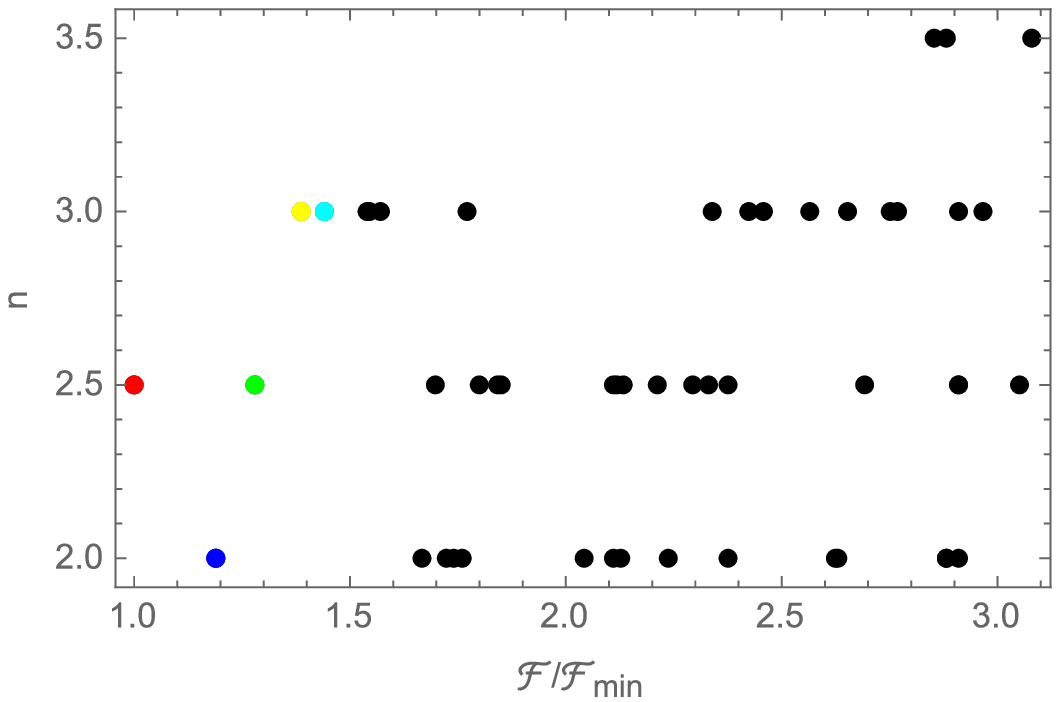} 
  \caption{Example of the selection method. The results correspond to the Be star HD58343 with an  inclination angle of $i = 10^{\circ}$ and $R_{T} = $ 25 $R_{\star}$. The first 5 best models are indicated with the $\mathcal{F}/\mathcal{F}_{min}$ value starting at 1.00 (red), 1.20 (blue), 1.30 (green), 1.40 (yellow) and 1.45 (cyan) in all panels. \textit{Top left:} $\mathcal{F}/\mathcal{F}_{min}$ of the 50 best models. \textit{Top right:}  H$\alpha$ line profiles models compared with the observation (black solid line). \textit{Bottom left:} $\log \rho_{0}$ values for the best 50 models. \textit{Bottom right:} $n$ values for the best 50 models.  } 
   \label{FOM-example}
\end{figure*}

The H$\alpha$ spectrum of each star in our sample was compared to the theoretical library for that spectral type using a script that systematically finds the best match to the observed profile. For each comparison, the percentage flux difference between the model and observation was averaged over the line to assign each comparison a figure-of-merit value (hereafter called $\cal F$), defined as
\begin{equation}
{\cal F} \equiv \sum_{i=1}^{i=N} w_i\,\frac{|F_i^{\rm obs}-F_i^{\rm mod}|}{F_i^{\rm mod}} \,
\end{equation}
where $F_i^{obs}$ is the observed relative line flux, $F_i^{\rm mod}$ is the model relative line flux, $w_i$ is a weight, discussed below, and the sum is over all wavelengths spanning the line. Several different weights were examined: uniform weighting $w_i=1$, line-center weighting $w_i=|F_i^{\rm mod}/F_c^{\rm mod}-1|$, and uniform weighting but using the sum of the square of flux differences divided by flux. For each spectrum, we tested the second option first, but also calculated the quality of the fits for other options as well, and by visual inspection we selected the best $\mathcal{F}$ method to adopt for each spectrum (which may be different for each star) to use in our results. 

Initially the best 50 matches out of the 1620 profiles using the smallest $\mathcal{F}/\mathcal{F}_{min}$ values were identified, where ${\cal F}_{min}$ is the minimum figure-of-merit of the best-fitting library profile. We show an example for a B2 spectral type in Figure~\ref{FOM-example} for the Be star HD58343. The upper left panel shows the best 50 models sorted by $\mathcal{F}/\mathcal{F}_{min}$ (black dots) with the best 5 models in red, blue, green, yellow, and cyan colors corresponding to $\mathcal{F}/\mathcal{F}_{min}$ of 1.00, 1.20, 1.30, 1.40 and 1.45, respectively. The best 5 models are different in the disk density parameters, but they have the same inclination angle, \hbox{$i =$ 10$^{\circ}$}, and the same disk truncation radius of \hbox{$R_{T} =$ 25.0$R_{\star}$} for this star. The upper right panel shows models of H$\alpha$ line profiles corresponding to each respective color as well as the observed profile shown in black. The main difference between these models appears in the flanks of the emission line. \cite{Hanuschik1986} classified typical emission profiles seen in Be stars at different inclination angles, where this particular ``wine bottle shape'' is usually seen at low inclinations. Moreover,  \cite{Hummel1994} reproduced emission line profiles using a Keplerian disk model for an optically thick disk ($\sim 10^{-10}$ g cm$^{-3}$) and he found for inclinations between $5^{\circ} \lesssim i \lesssim 30^{\circ}$, emission line profiles show inflection flanks. For high inclination angles, $i \gtrsim 75^{\circ}$, he noticed that a central depression plus a double peak profile is generated due to the velocity field present in the disk.
The lower left panel shows the behavior of $\log \rho_{0}$ vs $\mathcal{F}/\mathcal{F}_{min}$ where, in this particular case, we can see that higher values of $\rho_{0}$ dominate. The lower right panel is the same as the lower left panel except for $n$. 
In Figure~\ref{FOM-example}, the best model (red color) is well constrained by $\mathcal{F}/\mathcal{F}_{min}$ $=$ 1.00, however we notice that similar values of $\rho_{0}$ combined with different values of $n$ give us similar profiles of the emission line (for the same inclination angle and same disk truncation radius). For this reason we consider a range of models within a percentage of $\mathcal{F}/\mathcal{F}_{min}$ as described in Subsection~\ref{sec:results:c}.

\subsection{Best fit models by visual inspection} \label{sec:results:b}

 We chose the best model by visual inspection of the comparison plots between the models and the observations; such plots are shown in Appendix~\ref{ApA}, and the model parameters corresponding to this best fit are displayed in Table~\ref{table2} in the columns 4 to 8. Targets with a superscript $a$ indicate an $H\alpha$ absorption line in that star's spectrum. In some cases, the script was not able to suitably reproduce the core and wings of the emission line profile (see discussion section~\ref{sec:discussion:e} for possible explanations). However, we chose the fit that best represents the wings of the line (instead of the core) and classified them as poor fits. These cases are indicated with the superscript $pf$ in the Table~\ref{table2} and they are not considered in our analysis. 

Targets are sorted by HD number indicating the date of the observation and the $\mathcal{F}/\mathcal{F}_{min}$ value of the chosen model. Table~\ref{table2} also lists the H$\alpha$ equivalent width, EW, and the emission double-peak separation, $\Delta V_p$, measured from the observations. Some of the targets are represented by more than one observation due to variability, and they show changes in the line profile (peak height, violet-to-red peak ratio, etc). There are 22 such variable cases indicated by an asterisk symbol beside the star name below the plot (14 of these are in emission and 8 in absorption), and they were treated by keeping the inclination angle constant for the system, and each time fit with different models. In our program stars there are 15 Be stars with H$\alpha$ in absorption. We notice that in our sample all targets are confirmed as Be stars, so absorption profiles presented here are Be stars in disk-less phase or currently without a disk. We did not include absorption profiles in our analysis, nevertheless, our spectral library contains profiles with pure photospheric H$\alpha$ profiles. We display the values for systems with absorption in Table~\ref{table2} and in the plots in Appendix~\ref{ApB}.  

We provide our results separately for the emission profiles, absorption lines, and for the targets with poor fits. Overall, we have 42 Be stars with H$\alpha$ emission, 15 with absorption profiles, and 6 with poor fits. The systems with poor fits are displayed in Appendix~\ref{ApC}.

\subsection{Distribution of the disk density parameters: representative models} \label{sec:results:c}

In the previous section, we determined the best-fit disk density models for each of our program stars with H$\alpha$ in emission. In this section, we wish to look at the distribution of disk density parameters in this sample. From now on, every spectrum in emission for each target (if there is more than one) is considered by a separate, unique model. This give us a total of 61 emission models. As we explained in the previous section, there is a range of models for each star that fit the observed profile nearly as well as the best-fit model selected by visual inspection. Thus for any given star, we can systematically define a ``set" of best fit parameters by selecting all models with $\mathcal{F} \leq 1.25 \mathcal{F}_{min}$ resulting in $N$ models being selected. We note that by selecting a slightly larger range of $\mathcal{F}$, as Figure~\ref{FOM-example} demonstrates, the base density and the exponent of the disk surface density span a wide range of values especially for $\mathcal{F} \geq 1.50$. 
To define representative disk density parameters for each star, we choose a weighted-average over the $N$ selected models. For the disk parameter $X$, which could be $\rho_0$ or $n$, etc., we define
\begin{equation}
<X> \equiv \frac{1}{W}\,\sum_{i=1}^{N} \, w_i\, X_i \; ,
\end{equation}
where $W\equiv \sum_{i=1}^{N} \,w_i$ and the weights are chosen as
\begin{equation}
w_{i} \equiv \left(\mathcal{F}/\mathcal{F}_{min}\right)^{m}.
\end{equation}
The index $m$ was chosen to be equal to -10 so that significantly different weights are given to models ranging from 1 to 1.25, i.e.\ the weight assigned to $\mathcal{F}=1.25$ is $1.25^{-10}\approx$ 0.1. This procedure was applied to all the physical quantities obtained from the emission profiles which are presented below. In order to study the conditions under which the disk exists and its link with the spectral type, we distinguish in our study between early (B0-B3) and late (B4-B9) type Be stars.

The representative values (weighted-average) of the parameters governing the disk density ($n$ and $\rho_{0}$ in Eq.~\ref{EQ1}) of emission profiles are displayed in Figure~\ref{FOM-example}. The most frequent pairs are concentrated between $<n> \ \ \simeq$ 2.0 - 2.5 and $<\rho_{0}> \ \ \simeq (4.00-6.30) \times 10^{-11}\,\rm g\,cm^{-3}$ or $<\log \rho_{0}> \ \ \simeq$ -10.4 to -10.2.

\begin{figure}
%\figurenum{6}
\centering
    \includegraphics[width=1\columnwidth]{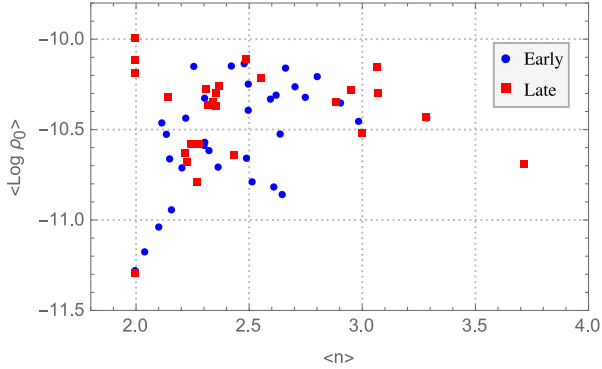}
	\caption{Distribution of the representative $<n>$ and \hbox{$<\log \rho_{0}>$} model values for systems with emission profiles.}
	\label{representative}
\end{figure}

We note that we detect emission profiles in the upper left triangular region of Figure~\ref{representative}. With increasing values of the density exponent and decreasing base density, corresponding to the lower right in Figure~\ref{representative}, it would be increasing difficult to detect emission due to reduced disk density. The lack of disk material for these stars made it impossible to constrain our models as mentioned above so we did not analyze any features for them. Moreover, some absorption profiles seemed to be pure photospheric lines, and some showed evidence of a possible formation/dissipation disk phase (see HD33328's spectrum, for example, in Appendix~\ref{ApB}).

\subsection{Distribution of disk mass and angular momentum} 
\label{sec:results:d}

From each star's fitted disk density parameters, we can estimate the mass of the disk by integrating the disk density law, Eq.~\ref{EQ1}, over the volume of the disk. For the radial extent of the disk, we chose the radius that encloses 90\% of the total flux of the H$\alpha$ line in an $i = 0^{\circ}$ (i.e.\ face-on disk) image computed with \texttt{BERAY}. This measure of the H$\alpha$ disk size was used in favor of the fitted $R_{T}$ as the latter was computed on a very coarse grid of only four values. To compute each disk mass, $<M_{d}>$, the representative values of the disk parameters were used which included the models with $\mathcal{F} < 1.25\mathcal{F}_{min}$. In addition to disk mass, the representative value of the total angular momentum content, $<J_{d}>$, of each disk was also computed, using the same disk density parameters and assuming pure Keplerian rotation for the disk. Representative values of the disk mass and angular momentum in stellar units are displayed in Table~\ref{table2} in columns 12 and 13, respectively.

\begin{figure}[t]
\centering
%\figurenum{7}
   \includegraphics[width=1\columnwidth]{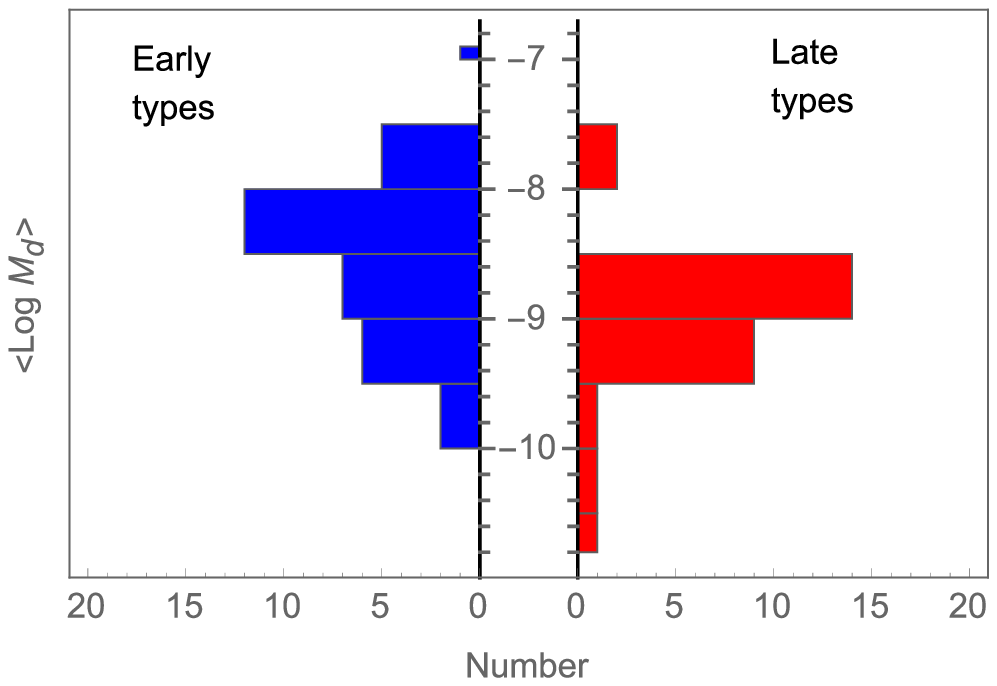} \\
    \includegraphics[width=1\columnwidth]{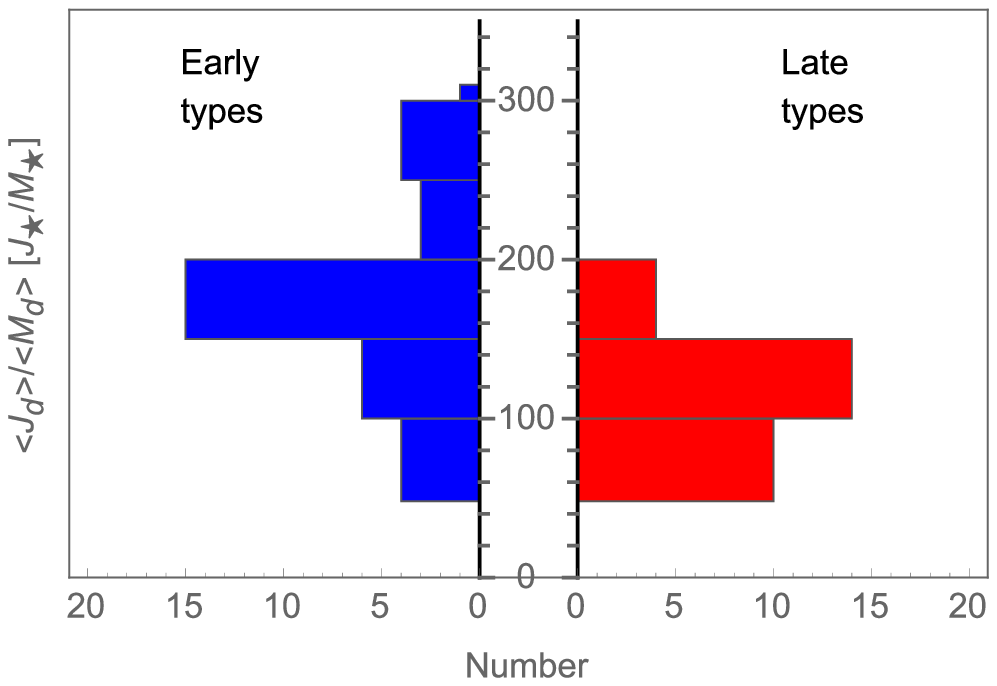}
	\caption{Distribution of the representative values of the disk mass (top panel) and angular momentum of the disk (bottom panel) compared to the central star.}
	\label{diskmass}
\end{figure}

Figure~\ref{diskmass} shows the distribution of both representative values, disk mass and disk specific angular momentum, $<J_{d}>/<M_{d}>$, for early and late stellar types. To normalize by the stellar angular momentum, the central star was assumed to rotate as a solid body at $0.8\,v_{\rm crit}$ with the critical velocity computed using Eq.~\ref{EQ4}. (See also Section~\ref{sec:discussion:c} for a discussion about the effect of the stellar rotation on $J_{d}$). 

The distribution of the disk mass in early types ranges from $1.0 \times 10^{-7}$ to $3.0 \times 10^{-10} M_{\star}$ (see top panel in Figure~\ref{diskmass}). For late types, values range from $1.7 \times 10^{-8}$ to $1.7 \times 10^{-11} M_{\star}$. The mean disk mass for the early-types is $3.4 \times 10^{-9} M_{\star}$, while for the late-types, the mean disk mass is $9.5 \times 10^{-10} M_{\star}$. 

The bottom panel in Figure~\ref{diskmass} shows the distribution of the specific angular momentum $<J_{d}>/<M_{d}>$ of the disk in units of stellar specific angular momentum. For early types, the most frequent range is \hbox{$<J_{d}>/<M_{d}> \\ \simeq$ 150 - 200} and corresponds to a \hbox{$<J_{d}> \ \ \sim (1.2 - 3.0)\times10^{-6} J_{\star}$} and a total mass of \hbox{$<M_{d}> \ \ \sim (3.2 - 9.1)\times10^{-9} M_{\star}$}. For late types the most frequent values ranges from 100 to 150 corresponding to a range value of \hbox{$<J_{d}> \\ \sim (1.0 - 5.0)\times10^{-7} J_{\star}$} and \hbox{$<M_{d}> \\ \sim (1.0 - 2.9)\times10^{-9} M_{\star}$}. 
In general, late types have lower values of $<M_{d}>$ and $<J_{d}>$ in comparison with early types. It should be kept in mind that while the model disk masses vary over a large range (with $M_{d}/M_*$ spanning $1.7\times10^{-11}$ to $1.0\times10^{-7}$), the range of model specific angular momentum is much less owing to the assumption of Keplerian rotation. The minimum and maximum values of $<J_{d}>/<M_{d}>$ in units of $J_*/M_*$ are 49 and 306, for a total variation of just over a factor of 6. \\

\subsection{Relation between H$\alpha$ equivalent width and disk mass} \label{sec:results:e}

\begin{figure}
%\figurenum{8}
\centering
    \includegraphics[width=1\columnwidth]{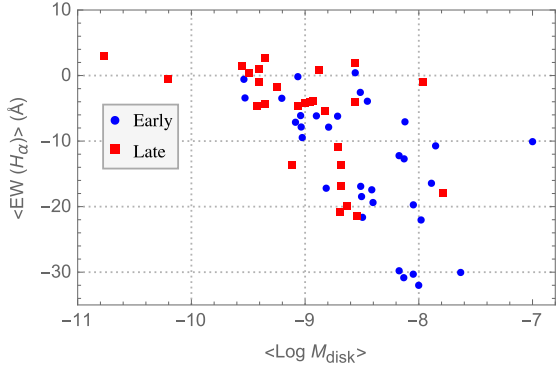} 
	\caption{Equivalent widths of the H$\alpha$ emission line profiles as a function of mass.}
	\label{EW}
\end{figure}

The relation between H$\alpha$ EW, and disk mass, \hbox{$<\log M_{d}>$}, separated by early and late-type Be stars, is shown in Figure~\ref{EW}. Negative values indicate that the net flux of the emission line is above the continuum level. While there is an overall trend for the most massive disks to have the largest H$\alpha$ EW, there is an extremely large dispersion. This is not unexpected; for any given power law index $n$ in Eq.~\ref{EQ1}, the H$\alpha$ EW will first increase with $\rho_0$, reach a maximum, and then decline \cite[see, for example,][]{Sigut2015}. This decline occurs because once the density becomes large enough, the continuum flux from the disk at the wavelength of H$\alpha$ rises more quickly than the line emission, so the equivalent width actually decreases with $\rho_0$ and so does the corresponding disk mass. The exact value of $\rho_0$ at which the H$\alpha$ equivalent width peaks is dependent on $n$; therefore, in a mix of models with differing $(\rho_0,n)$, there will not be a direct relationship between disk mass and H$\alpha$ EW. Finally, we note that the most massive disks and largest H$\alpha$ equivalent widths (absolute value) are found most frequently among the early-type Be stars.

\section{Discussion} \label{sec:discussion}

\subsection{Disk density} \label{sec:discussion:a}

We found a distribution of the representative values of the disk density parameters for early and late spectral types, which are displayed in Figure~\ref{representative}. Early stellar types cover values of $<n>$ between 2.0 and 3.0, while late stellar types reach values near 3.7. It appears that higher values of the power-law exponent are found for stars with lower effective temperature. This could explain the small emission disks seen in late type stars since with increasing $n$, the disk density falls faster with distance from the star. However, the average value of the representative values of the power-law exponent are essentially the same for early and late spectral types: $<\bar n_{early}> \ \ = 2.5 \pm 0.3$ and $<\bar n_{late}> \ \ = 2.5 \pm 0.4$.

Previous work in the literature using \texttt{BEDISK} was completed by \cite{Silaj2010}. They created a grid of disk models for B0, B2, B5 and B8 stellar types at three inclinations angles $i=20^{\circ}$, $45^{\circ}$ and $70^{\circ}$ for different disk densities. They modeled H$\alpha$ line profiles of a set of 56 Be stars (excluding Be-shell stars) and studied the effects of the density and temperature in the disk. Their results show a higher percentage of models ranging $\rho_{0}$ between $10^{-11}$ and $10^{-10}\;\rm g\,cm^{-3}$ and a significant peak of $n \sim 3.5$, which is slightly larger than the values of $n$ found in this study. We attribute this difference to the different methods used to compute the H$\alpha$ line profile. \cite{Silaj2010} used \texttt{BEDISK} to compute the line intensity escaping perpendicular to the equatorial plane in each disk annulus (i.e.\ rays for which $i=0^{\circ}$). They then assumed that this ray was representative of other angles considered, $i=20$, $45^{\circ}$ and $70^{\circ}$, and combined the $i=0^{\circ}$ rays with the appropriate Doppler-shifts and projected areas. Clearly this computation method becomes limited with larger viewing angles. In contrast, \texttt{BERAY}, used here, does not make any of these approximations, and it has been successfully used to model the H$\alpha$ lines of Be shell stars for which the inclination angle is large \citep{Silaj2014}. Eight Be-shell spectra were analyzed and values for $\rho_{0}$ between 10$^{-12}$ and 10$^{-10}$ g cm$^{-3}$ and $n$ between 2.5 and 3.5 were found. 

\cite{Touhami2011} used the assumption of an isothermal disk and the same density prescription as Equation~\ref{EQ1} to reproduce the color excess in the NIR of a sample of 130 Be stars. For the central star, they assumed an early-type star and adopted $n =$ 3.0 for all the models. They varied $\rho_{0}$ between 10$^{-12}$ and 2.0$\times$10$^{-10}$ g cm$^{-3}$, which is very similar to our range of $\rho_{0}$ variation. They set the inclination angle at $i = $45$^{\circ}$ and 80$^{\circ}$ used an outer disk radius of $\sim$ 14.6 and 21.4$R_{\star}$. Other studies also use the same scenario for the density distribution, where the base density of the disk is found to be between 10$^{-12}$ and 10$^{-10}$ g cm$^{-3}$ and the power-law exponent $n$ is usually in the range 2 - 4 (for a review of recent results the reader is referred to the section 5.1.3 of \citealp{Rivinius2013}). 

Recently \cite{Vieira2017} determined the disk density parameters $\rho_{0}$ and $n$ for 80 Be stars observed in different epochs, corresponding to 169 specific disk structures. They used the viscous decretion disk model to fit the infrared continuum emission of their observations, using infrared wavelengths. They found that the exponent $n$ is in the range between 1.5 and 3.5, where our most frequent values are between 2.0 and 2.5 for both early and late spectral types. They also found $\rho_{0}$ to range between 10$^{-12}$ and 10$^{-10}$ g cm$^{-3}$, which compares favorably with our average values of between (4.00-6.30) $\times$ 10$^{-11}$ g cm$^{-3}$, again for both early and late spectral types. \cite{Vieira2017} also established that the disks around early-type stars are denser than in late-type stars, consistent with our finding of more massive disks for the earlier spectral types.

Finally, we also notice that our models sometimes do not reproduce the wings of our H$\alpha$ observations. This may reflect our assumption of a single radial power law for the equatorial density variation in this disk. Alternatively, for earlier spectral types, this may reflect neglect of non-coherent electron scattering in the formation of H$\alpha$ \citep{Poeckert1979}. For example, \cite{Delaa2011} performed an interferometric study of two Be stars using a kinematic disk model neglecting the expansion in the equatorial disk. They were able to fit the wings and the core of the H$\alpha$ emission line by introducing a factor to estimate the incoherent scattering to their kinematic model. 

\subsection{Size of the emission region}  \label{sec:discussion:b}

The outer extent of the disk considered in the modelling of this work was assumed to be one of four values, 6.0, 12.5, 25.0 and 50.0$R_{\star}$.
From these values, the best fitting models have a disk truncation radius of 25.0$R_{\star}$ followed by 50.0$R_{\star}$. Nevertheless, as noted previously, a better estimate of the size of the H$\alpha$ emitting region is the equatorial radius that contains 90\% of the integrated H$\alpha$ flux in an $i=0^{\circ}$ image computed with \texttt{BERAY}, a quantity we denotes as $R_{90}$. We provide $R_{90}$ values in the column 11 on Table~\ref{table2}. These values, based on the integrated flux from our models, could used by other studies to conveniently compare with our results. 

As an additional check, we compare our $R_{90}$ disk sizes with a measure based on the observed separation of the H$\alpha$ emission peaks, as first suggested by \citet{Huang1972}, and tailored to our model assumptions. The basic idea of this method is that the double-peak separation is set by the disk velocity at it's outer edge, which we will denote $R_{H}$. If the observed peak separation is $\Delta V_p\;\rm km\,s^{-1}$, we have
\begin{equation}
\frac{1}{2} \left(\frac{\Delta V_p}{\sin i}\right) = \sqrt{\frac{GM}{R_H}} \; ,
\end{equation}
assuming Keplerian rotation for the disk and correcting the observed peak separation for the viewing inclination $i$. Hence,
\begin{equation}
\Delta V_p^2 = 4\, \left(\frac{GM}{R_H}\right) \sin^2 i \;.
\end{equation}
In this work, we assumed that all Be stars rotate at 80\% of their critical velocity; therefore, each star's equatorial velocity is
\begin{equation}
V_{eq} = 0.8\,\sqrt{\frac{GM}{(3/2)R_{\star}}} \;,
\end{equation}
where $R_{\star}$ is the stellar (polar) radius. Using this to eliminate $(GM)$ from the previous equation and solving for the disk size we have
\begin{equation}
\frac{R_H}{R_{\star}} = 9.375 \, \left(\frac{V_{eq}\sin i}{\Delta V_p}\right)^2
\end{equation}
As $V_{eq}\sin i$ is the star's $v\sin i$ value, we have approximately
\begin{equation}
\label{eq:HL}
\frac{R_H}{R_{\star}} \approx \left(\frac{3\,v\sin i}{\Delta V_p}\right)^2 \,.
\end{equation}
This equation is very similar to the form used by many authors to derive approximate disk sizes from observed spectra \citep[e.g.][]{Hanuschik1986,Hummel1994}. We note that the way we use Eqn.~\ref{eq:HL} is slightly non-standard: we do not measure $v\sin i$ directly from our spectra; instead, we adopt the $v\sin i$ of the best fit-model. As the H$\alpha$ profiles are essentially insensitive to $v\sin i$, we are using the observed peak separation $\Delta V_p$ and the best-fit value of $i$ for the viewing inclination.

\begin{figure}
%\figurenum{9}
\centering
    \includegraphics[width=1\columnwidth]{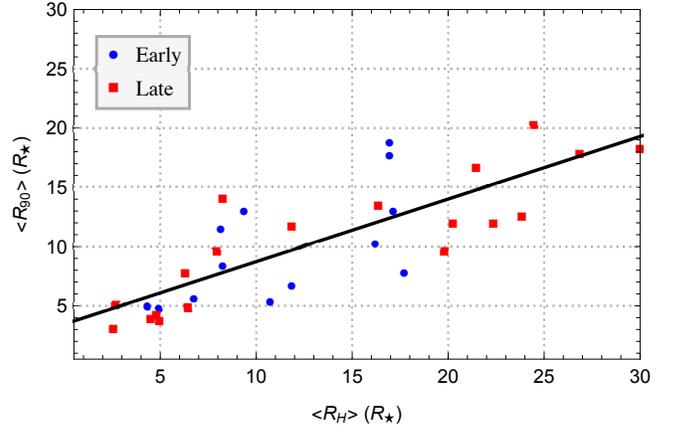} 
	\caption{Relation between the emitting size containing 90$\%$ of the integrated H$\alpha$ emission,  $R_{90}$, and the emitting size obtained from Huang's law, $R_{H}$.  A linear fit is represented by the solid line.}
	\label{Rdisk}
\end{figure}

The correlation between $R_{H}$ and $R_{90}$ is displayed in Figure~\ref{Rdisk}. For a few of our targets, we do not obtain a $R_{H}$ value because of a small $\Delta V_p$ or small inclination where Huang's law is not valid. The solid line indicates the linear fit over both early (blue circles) and late (red squares) stellar types considering values not larger than 50.0$R_{\star}$ and greater than 1$R_{\star}$ to be consistent with the input values used in the \texttt{BERAY} model. The relation between the representative values of the mentioned sizes is given by the linear equation \hbox{$<R_{90}>$ = (0.53 $\pm$ 0.07) $<R_{H}>$ + (3.45 $\pm$ 0.80)} in units of stellar radius, with a correlation of \hbox{$r_{corr}$ = 0.611} with confidence intervals calculated using a bootstrapping method. We notice that the most frequent disk sizes values calculated by Huang's relation for early and late spectral types are concentrated less than 5$R_{\star}$ and the values containing 90\% of the H$\alpha$ flux for early and late spectral types are concentrated between 10 and 15$R_{\star}$.

Many other measurements of the Be star disk sizes have been reported in the literature. \cite{Hanuschik1986} measured the $\Delta V_p$ and the FWHM in the H$\alpha$ emission line of 24 southern Be stars and using Huang's law he estimated an outer emitting size of $\sim$ 10$R_{\star}$. Similar values were found by \cite{Slettebak1992} for 41 Be stars, they obtained an outer emitting size in the range $\sim$ 7 - 19$R_{\star}$ for the H$\alpha$ emission line.
Using interferometric techniques \cite{Tycner2005} studied the relation between the total flux emission of H$\alpha$ line and the physical size of the emission region in 7 Be stars, finding for the first time a clear correlation between these both quantities. For early stellar types they found an extended emitting size of $ \sim$ 18.0 to 21.0$R_{\star}$ while for stars with lower effective  temperatures they found smaller values of $\sim$ 6.0$R_{\star}$ to 14.0$R_{\star}$ (with an exception for $\psi$ Per of $\sim$ 32.0$R_{\star}$). An alternative way to estimate the emitting region based on the H$\alpha$ half-width at half-maximum was proposed by \cite{Grundstrom2006}. They compared their results with the interferometric measures of the  H$\alpha$ emitting size in the literature and they obtained lower values between $\sim$ 5.0 to 10.0$R_{\star}$.

Our very low values of $R_{H}$ from observed emission profiles (less than \hbox{1$R_{\star}$} and not considered in the analysis) come from very large $\Delta V_p$ values. If the star is rotating near its critical rotation, the gas could accumulate near the star and consequently the emission region of the H$\alpha$ line could be of the order of a few stellar radii. 

Overall, our results for $R_{T}$, either from the representative models or from Huang's law, show general agreement with previous works in the literature, giving higher values for early stellar types and lower values for late-Be types.  

\subsection{Mass and angular momentum of the disk}  \label{sec:discussion:c}

In Section~\ref{sec:results:d} we provided the range of the total disk mass and the total disk angular momentum for early and late stellar types. Our results gave us higher values of $<J_{d}>$ and $<M_{d}>$ for early types in comparison with late types. This was expected considering that late stellar types have, in general, smaller disks. Considering the whole sample without distinction between early and late stellar types, we estimate that the total angular momentum content in the disk is approximately $10^{-7}$ times the angular momentum of the central star and the mass of the disk is approximately $10^{-9}$ times the mass of the central star. 

\cite{Sigut2015} studied the disk properties of the late Be shell star Omicron Aquarii (o~Aqr, B7IVe) combining contemporaneous interferometric and spectroscopy H$\alpha$ observations with near-infrared (NIR) spectral energy distributions. They compared the values obtained by each technique for different disk parameters. From H$\alpha$ spectroscopy, values of $R_{T}$, $M_{d}$ and $J_{d}$ are higher than those obtained from the NIR, while $\rho_{0}$ and $n$ are lower than NIR. From their results, the comparison between values obtained from spectroscopy, interferometry and NIR spectral distributions, give similar or consistent values for $M_{d}$ and $J_{d}$, but the disk density parameters $(\rho_0,n)$, showed in a range of values. As a result, for o~Aqr, \cite{Sigut2015} found values of $J_{d} \sim$ 1.6$\times$10$ ^{-8} J_{\star}$ and a total mass of $M_{d} \sim$ 1.8$\times$10$ ^{-10} M_{\star}$. These values are consistent with our results in Figure~\ref{representative}, but are at lower end of the distribution for late stellar types.

\begin{figure}
%\figurenum{10}
\centering
   \includegraphics[width=1\columnwidth]{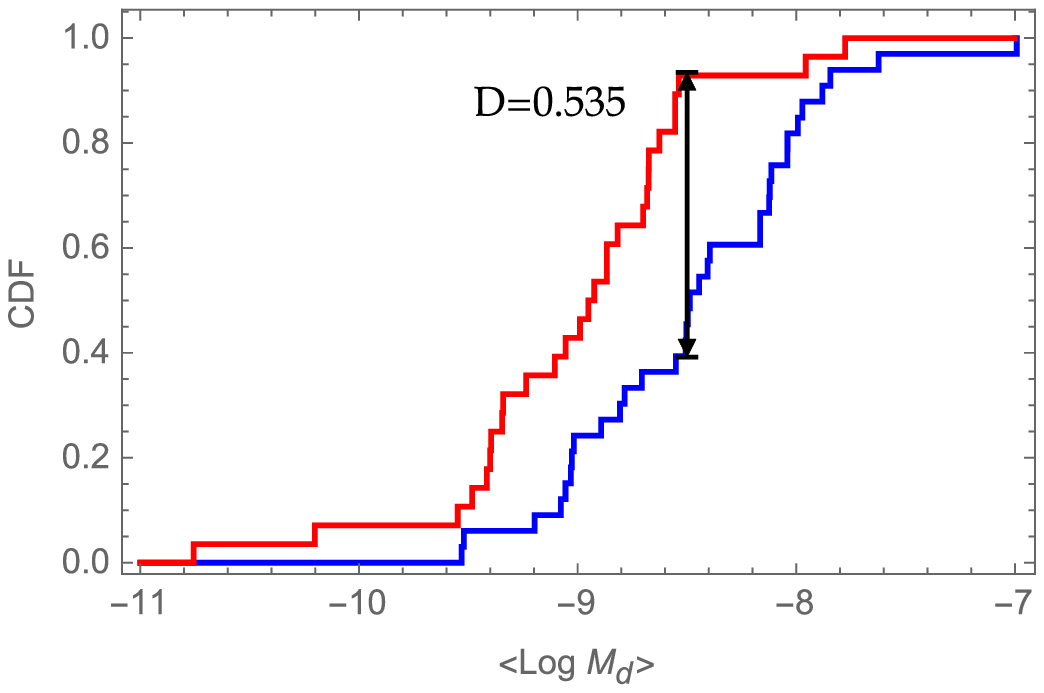} \\
    \includegraphics[width=1\columnwidth]{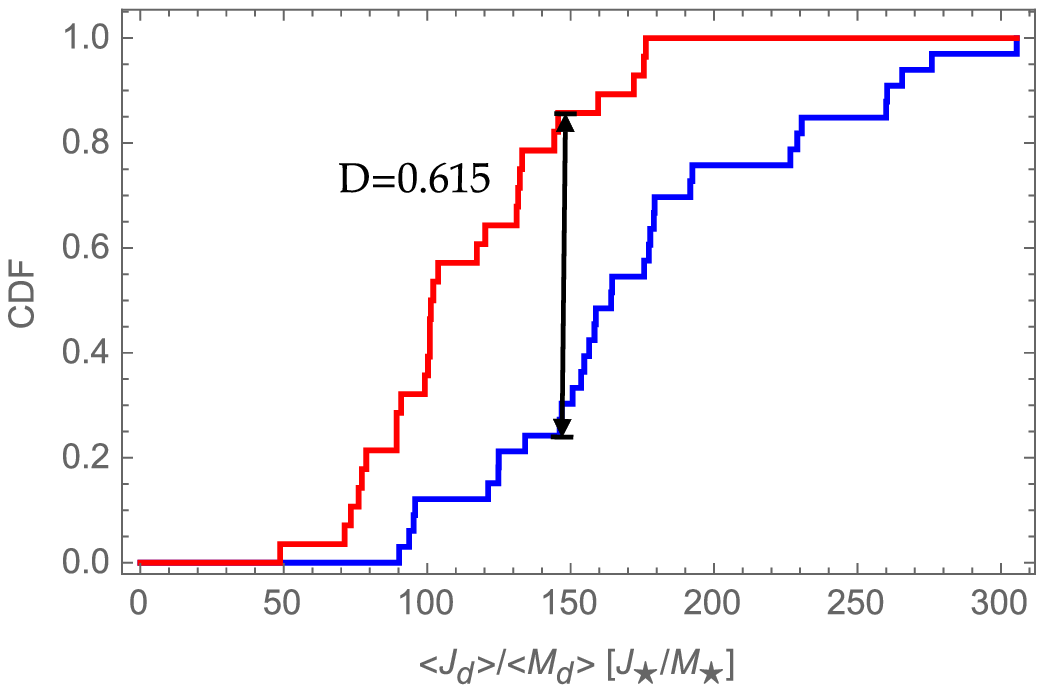}
  \caption{Cumulative distribution functions for the mass and total angular momentum of the disk. Blue (lower curve) and red (upper curve) colors represent early and late spectral types, respectively. Maximum distance between CDFs is indicated in each plot. The upper panel shows the CDFs comparison of both samples for the $<\log M_{d}>$ and the lower panel shows the same, but for the $<J_{d}>/<M_{d}>$ distribution. A KS test demonstrated for both, disk mass and disk angular momentum, that early and late samples come from different distributions at 1\% level.}
    \label{CDF} 
\end{figure}

As we mentioned earlier in Section~\ref{sec:results:c}, we distinguish our results between early (B0-B3) and late (B4-B9) type Be stars. Recall that the parameters associated with these stars are listed in Table~\ref{table1}. In order to study the effects of the central star on the distributions of disk mass and angular momentum for early and late spectral types, we performed a two-tailed Kolmogorov-Smirnov (KS) test with the null hypothesis that both samples come from the same distribution. Figure~\ref{CDF} shows the cumulative distribution functions (CDFs) for disk mass (upper panel) and total disk angular momentum per disk mass (bottom panel). For disk mass, the maximum distance, $D_{m}$, between CDFs for early and late types gives $D_{m} =  0.535$ and considering a significance level at 0.01, the critical value, $D_{c}$, is 0.50 for the 61 emission models. Hence we conclude that early and late samples of disk mass come from different distributions. The largest value for the maximum distance between CDFs for $<J_{d}>/<M_{d}>$, gives $D_{m} =  0.615$, again rejecting the null hypothesis that the distributions are the same at 1\% level. Therefore, our samples show that early-type Be stars are more likely to have massive disks with higher values of total angular momentum than late-type Be stars. 

We note that our results could be influenced by the choice of stellar rotation rate of 0.8$v_{crit}$ for all the luminosity classes in our models. Various studies have attempted to determine these rates more precisely, with a consensus that they are rapid rotators, but it is still not clear how close to critical these rates are. \cite{Porter1996} compared the observational distribution of a sample of $v \sin i$ values of Be-shell stars ($\sin i \sim 1$) with a theoretical distribution. He determined that these Be-shell stars rotate at 70\%-80\% of their critical rotation. \cite{Huang2010} studied the effect of the stellar rotation on the disk formation in ``normal'' B stars as a function of stellar mass, by comparing with Be stars in the literature. They found that the rotational velocity needed to create a Be star varies strongly with the stellar mass. For low-mass B stars (less than 4$M_{\odot}$ or later than B6 V) the upper-rotational limit is very close to the break-up velocity $\sim$ 0.96, while for high-mass B stars (more than 8.6$M_{\odot}$ or earlier than B2 V) the upper-rotational limit is near to 0.63$v_{crit}$. To test the significance of our choice of 0.8$v_{crit}$ on our angular momentum distribution of our sample, we adopted both limiting values of the break-up velocity, 0.63 and 0.96 for early and late stellar types, respectively. For early types, the disk angular momentum is under-estimated ($J_{\star} \sim 0.80/0.63 \simeq 1.3$) by $J_{disk}/J_{\star} \sim$ 0.8 times, while for late types are overestimated ($0.80/0.96 \simeq$ 0.8) by 1.2 times. Multiplying by these factors for the early- and late-type distributions of $<J_{d}>/<M_{d}>$, respectively, we found a total range distribution between $\sim$ 64 and 245, and a maximum distance value of $D_{m} = 0.879$, which also rejects the null hypothesis that both samples come from the same distribution within a 1\% level of significance. \\

\begin{figure}
%\figurenum{11}
\centering
    \includegraphics[width=1\columnwidth]{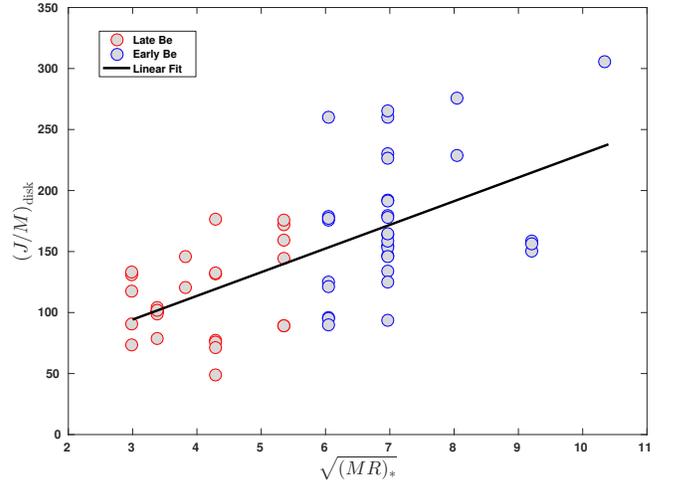} 
  \caption{The disk specific angular momentum of the BeSOS sample stars versus the square root of the stellar mass times the stellar radius. All qunatities are in solar units. The linear fit to the data has a correlation coefficient of $r=+0.63$.}
    \label{NewFigBetween10and11} 
\end{figure}

A key ingredient in the specific angular momentum distribution for Be star disks is the underlying Keplerian rotation law, well established for Be stars \citep{Rivinius2013b}. As the overall scale of the disk's Keplerian rotation is set by the parameters of the central star ($M_*, R_*$), a portion of the variation in disk specific angular momentum must simply reflect the change of stellar mass and radius with spectral type. To quantify this,\footnote{We are thankful to the anonymous referee for suggesting this line of reasoning.} we note that the disk specific angular must scale as $J/M\sim r\,v_{\rm K}(r)$ where $r$ is a characteristic radius for the disk, and $v_{\rm K}$ is the Keplerian velocity at this point. We may write this as $J/M \sim \sqrt{GM_*\,R_*\,(r/R_*)}$ by introducing the stellar radius $R_*$. If the characteristic disk size $(r/R_*)$ is constant with spectral type, we have $J/M\sim \sqrt{M_* R_*}$. Figure~\ref{NewFigBetween10and11} plots the disk specific angular momentum found for our sample versus the quantity $\sqrt{M_* R_*}$ from Table~\ref{table1}. While there is a wide dispersion, the linear trend is very clear, with a correlation coefficient of $r=+0.63$. Therefore, as expected, a significant portion of the variation in the disk specific angular momentum is due to the variation of the central star parameters via the overall scale of the disk's Keplerian rotation. The large scatter about this linear trend, typically a factor of $2-3$, must then reflect the different disk sizes and the distribution of the disk mass with radius, controlled mainly by the parameter $n$.

\subsection{Cumulative distribution of the inclination angles}  \label{sec:discussion:d}

\begin{figure}
%\figurenum{11}
\centering
    \includegraphics[width=1\columnwidth]{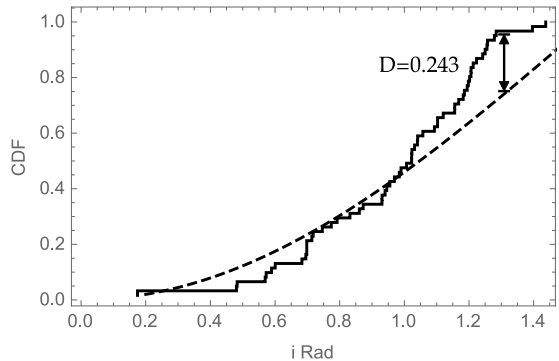} 
  \caption{Cumulative distribution of the representative inclination angles (solid line) versus the expected distribution $1-\cos(i)$ (dashed line). A KS test showed that the sample is not drawn from the expected distribution with a significance level of $\alpha = $0.01.}
    \label{CDFi} 
\end{figure}

An interesting consequence of the H$\alpha$ modelling is that the inclination of the system can be determined. Figure~\ref{CDFi} shows the CDF of the derived representative values of the inclination angles versus the expected $1-\cos(i)$ distribution, assuming that the rotation axes are randomly distributed. Using a one-sample KS test, we find that our data do not follow the expected distribution. Defining the null hypothesis $H_{0}:$ ``the inclination data comes from the $1-\cos(i)$ distribution'' and at significance level $\alpha = $0.01, the maximum distance, $D_{m}$ is $0.243$, while the critical value for our sample of 61 emission models is $D_{c} =  0.209$, therefore since $D_{m} > D_{c}$, $H_{0}$ is rejected with a 1\% level. This rejection, that our inclination angles distribution is not random, is not surprising as the selection criteria for Be stars in surveys are often biased against shell stars seen at high inclinations \citep{Rivinius2006}. This indeed seems to be the case for our sample as the observed CDF of Figure~\ref{CDFi} does not contain the expected fraction of high-inclination objects; in particular, our sample has only 8 Be shell stars.

\subsection{Comparison with disk mass predictions of models of stellar evolution with rotation}

\begin{figure*}[t]
%\figurenum{12}
\centering
    \includegraphics[width=12cm]{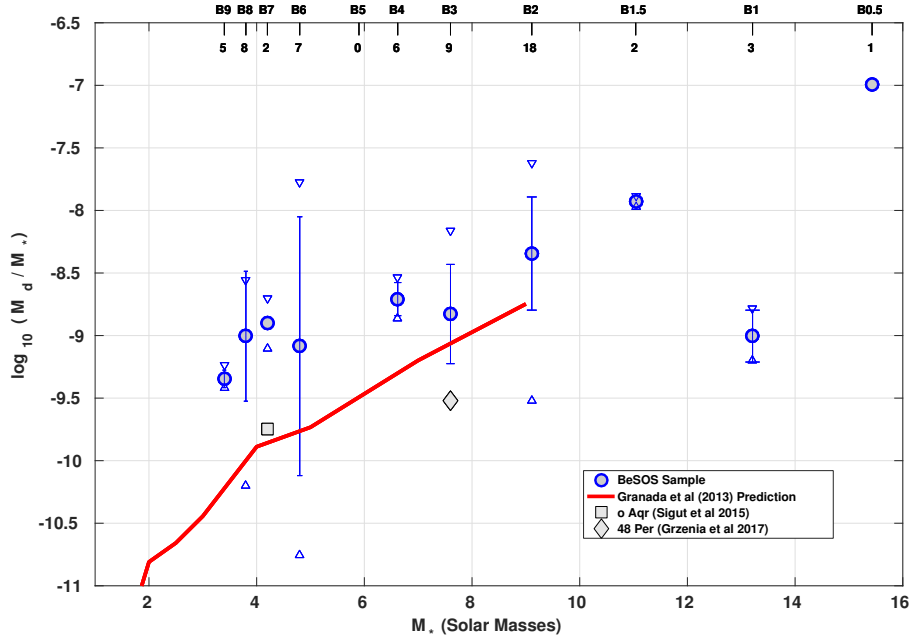} 
  \caption{Comparison of the average H$\alpha$ disk masses found in the current work (shaded blue circles) as a function of stellar mass (bottom axis) or spectral type (top axis). The average decretion disk masses of \cite{Granada2013} (Table~6), predicted from stellar evolutionary models rotating at $\Omega_{\rm crit}=0.95$, is given by the red line. The $1\sigma$ variation in the disk masses of the current work are shown as the error bars (shown only if the number of sample stars at that spectra type is 3 or more), and the associated triangles give the maximum and minimum disk masses found. The number directly below each spectra type is the number of stars in the BeSOS sample at that spectral type. The black square and black diamond are H$\alpha$ disk mass estimates for o~Aqr \citep{Sigut2015} and 48~Per \citep{Grzenia2017}, respectively.} 
    \label{granada} 
\end{figure*}

In this section, we compare the disk mass distribution derived for the BeSOS sample as a function of spectral type with the predictions of \citet{Granada2013}. While the hydrodynamical origin of the Be star disk ejection mechanism(s) is unknown, there is a broad consensus that rapid stellar rotation, likely reaching the critical value, is the ultimate driver for disk ejection in isolated Be stars \citep{Rivinius2013b}. Models of stellar evolution with rotation do predict episodes of critical rotation during main sequence evolution due to the internal transport of angular momentum. Under the assumption that disk ejection removes the excess surface angular momentum at critical rotation, and using the formalism of \citet{Krticka2011} for the ejected disk and its angular momentum transport, \citet{Granada2013} compute the main sequence evolution of B stars with masses from 2 to 9 $M_\odot$ and follow the required disk ejections over the main sequence. While these models make many assumptions (such as the details of the angular momentum transport and the initial ZAMS rotation rate and profile) which may not be realistic, they do predict average disk masses as a function of spectral type. In Figure~\ref{granada}, we compare the disk masses obtained from the BeSOS survey stars with the predictions of \citet{Granada2013}. Shown are the average disk mass, its $1 \sigma$ variation, and the minimum and maximum disk masses, all for each spectral type. In the observational sample, there is often a very wide range of disk masses at each spectral sub-type, typically at least an order of magnitude. The observed average disk mass is always above the \citet{Granada2013} prediction, although the theoretical prediction typically falls within the observed range of disk masses. The predicted curve shows an increasing trend with earlier spectral type (or increased stellar mass). This is reflected in the BeSOS sample, although the number of stars with spectral types earlier than B2 is small (6 out of 63 stars). Also shown in the figure are the disk mass estimates for o~Aqr \citep{Sigut2015} and 48~Per \citep{Grzenia2017}, based on modelling of the H$\alpha$ emission profile (as in the current work), coupled with simultaneous modeling of interferometric visibilities and near-IR spectral energy distributions. These two, higher-precision disk mass estimates fall closer the predicted trend, although again within the observed variation of the BeSOS sample. We note that the current disk mass estimates are really lower limits as we are sensitive only to the H$\alpha$ emitting gas. Given the uncertainties in the theoretical modeling, a more detailed a comparison may be unwarranted at this point. However, the distribution of Be star disk masses may develop into a powerful diagnostic constraint on rotating models of stellar evolution.

\subsection{Observed profiles with poor fits} \label{sec:discussion:e}

Appendix~\ref{ApC} contains all the fits that we consider poor and do not reproduce the features in the observed H$\alpha$ line profiles. All targets are in emission and are early-type stars (between B0 and B2), with the exception of HD83953, a B5V star. The shape of the emission profiles are very similar, showing wide profiles reaching velocities of the order of 600 - 700 km s$^{-1}$. Our methodology was not able to find a good agreement between the observations and the models because these profiles do not have a symmetric central emission and are very wide. For example, in the entire sample of emission profiles (Appendix~\ref{ApA}), only three stars are classified between B0 and B1.5 and these three are evolved: HD68980 (B1.5 III), HD143275 (B0.3 IV) and HD212571 (B1 III-IV), with velocities between $\sim$ 300 - 500 km s$^{-1}$, and with almost symmetric profiles. On the other hand, we note that HD35439 (B1 Vn), HD50013 (B1.5 V) and HD110432 (B0.5 IVpe) show variation in the intensity peak, where HD35439 shows a clear V/R variation. 
In the literature two of the six stars are binary stars classified as a $\phi$ Per-type. These types of systems consist of an early B-type main sequence star as the primary and a hot subdwarf star as the secondary, both surrounded by an envelope. It is believed that the secondary at some time was a more massive star that has lost a large percentage of its mass (by mass-transfer to the primary) leaving a hot helium core. The primary star is increasing its mass and angular momentum, due to the mass-transfer interaction, as result a large $v \sin i$ value is observed. HD41335 (HR2142) was recently highly studied by \cite{Peters2016}, who used a large set of ultraviolet and H$\alpha$ observations to measure radial velocities of the primary star to compute an orbit. For the system, Be + sdO, they find a mass ratio $M_{2}/M_{1} = 0.07 \pm 0.02$ and for the companion they found a projected rotational velocity $v \sin i <$  30 km s$^{-1}$, an effective temperature greater than 43 $\pm$ 5 kK, a mass estimation of 0.7$M_{\odot}$ and radius greater than 0.13$R_{\odot}$ with a luminosity of $\log L/L_{\odot} > 1.7$. To explain the variations of the shell line absorption they proposed a circumbinary disk model, where the companion intersects with the boundaries of a gap in the disk of the primary star causing a tidal wave. Thus the gas moving in these regions interacts with the dense gas producing shocks. \cite{Peters2016} state that this model could operate in other Be binaries only if the disk of the primary star is massive enough with considerable density near the companion, if it has a high orbital inclination ($i$ = 90$^{\circ}$) and if the companion has low mass to create a wide gap so the gas can move across it. For HD41335 we have four observations between 2012 November and 2015 February. The H$\alpha$ emission line does not show peak intensity variations in this period. From the HeI 6678 $\mathrm{\AA}$ line we cannot determine if variability is present. The second $\phi$ Per-type star proposed is HD63462 (Omicron Puppis), a bright B1 IV type. This star shows intensity variations in H$\alpha$ and from the V/R variation two quasi-period are obtained: 2.5 and 8 years. \cite{Koubsky2012} also found a particular variation in the HeI 6678 $\mathrm{\AA}$ line. They described this variation as: ``an emission component swaying from the red side of the profile to the blue one and back''. Their observations were obtained between 2011 November and 2012 April. We inspected our spectra, which are observed in 2013 February and 2015 October, and while there are no variations in the H$\alpha$ emission line, the HeI 6678 $\mathrm{\AA}$ line shows the same pattern described by \cite{Koubsky2012}. A red peak is seen at 6682 $\mathrm{\AA}$ in 2013 and a blue peak is seen at 6675 $\mathrm{\AA}$ in 2015. \cite{Koubsky2012} estimated the periodicity of the radial velocities obtained at H$\alpha$, HeI 6678 $\mathrm{\AA}$ and Paschen emission lines (P14, P13 + Ca II and P12) determining an orbital period of 28.9 days. They also found a relation between the velocity and the emission intensity of the HeI line, as the velocity increases the intensity is strongest and vice versa. They did not find any direct evidence of spectral lines from the hot subdwarf companion, and for this reason they suggest that Omicron Puppis is a Be + sdO type. $\phi$ Per-type systems could potentially test the hypothesis that Be stars could be formed by binary interactions, however these systems are difficult to detect due to the faint companion and for this reason observations in the ultraviolet range are required. The disk density parameters for the best-fitting models for all of these objects were not included in our analysis.

\section{Conclusions}  \label{sec:conclusion}

We modeled the observed H$\alpha$ line profiles of 63 Be stars from the BeSOS catalogue. Compared to synthetic libraries computed with the \texttt{BEDISK} and \texttt{BERAY} codes, good matches were found for 57 objects, 42 with H$\alpha$ in emission and 15, in absorption. The remaining 6 objects had poor fits that did not reproduce the features of the emission line. From the 41 H$\alpha$ emission line objects, we modeled each available observational epoch giving a total of 61 matched line profiles. Our results were used to constrain to the range of values for the base density and power-law exponent of the disk density model given in Eq.~\ref{EQ1} for all 61 observations. We determined the best fit model for each observation which are displayed in Table~\ref{table2} and in the corresponding plots shown in the Appendix section. Moreover, we obtained a distribution of the best representative models with $\mathcal{F} \leq 1.25 \mathcal{F}_{min}$ on which we base our average results. 

The most frequent values for the base density are between $<\log\rho_{0}> \ \ \sim$  -10.4 and -10.2 and for the power-law exponent are between $<n> \ \ \sim$ 2.0 and 2.5. Combined with an estimate for the size of the H$\alpha$ disk, the sample distribution for disk mass and disk angular momentum (assuming Keplerian rotation for the disk) were found, with typical values of $<M_{d}>/<M_{\star}>\sim 10^{-7}$ and $<J_{d}>/<J_{\star}>\sim 10^{-9}$. We find that disk mass and angular momentum distributions were different between early (B0 - B3) and late (B4 - B9) spectral type at 1\% level of significance. Finally, we compare our disk masses as a function of spectral type in Figure~\ref{granada} with the theoretical predictions of \citet{Granada2013} based on stellar evolution calculations incorporating rapid rotation. Our average H$\alpha$ disk masses (which are lower limits to the total disk masses) are always larger than the theoretical predictions, although the variation at each spectral type is quite large, typically more than an order of magnitude.

Our estimates for the H$\alpha$ disk radius ($R_{90}$, the radius that encloses 90$\%$ of the line emission) are compared to Huang's well-known law relating the disk size to the double-peak separation in the profile. A linear correlation is found with a correlation coefficient of \hbox{$r_{corr} = 0.652$}, but there is a large dispersion, which is attributed to the large disk sizes obtained due to the largest $\Delta V_p$ and/or smallest $v \sin i$ values from the models used in the Huang's relation. The concentration of such values is less than 5$R_{\star}$ for Huang's law and between 15 and 20$R_{\star}$ for $R_{90}$ and is dominated by early-type Be stars. Several studies about similarities and differences between early and late type Be stars have been carried out recently. \cite{Kogure2007} suggested that early-type Be stars have more extended envelopes compared with the late-type Be stars from their analysis of H$\alpha$ equivalent widths by spectral type consistent with the findings presented here.  

Finally we find that the derived inclination angles from the H$\alpha$ profile fitting do not follow the expected random distribution. This is attributed to the under-representation of Be shell stars in the BeSOS survey.

Numerous studies have found that the mean $v \sin i$ values increase for late-type main sequence Be stars \citep[e.g.][]{Zorec1997,Yudin2001,cranmer2005}. In our case, we fixed the rotation of the star to be 80$\%$ of the critical value, consistent with \cite{Chauville2001}. Clearly, the study of Be stars is still in continuous development. In future we plan to re-analyze the sample by including more lines in the visible range (i.e., H${\beta}$, H${\gamma}$), as well as investigating the spectral energy distributions and $v \sin i$ values. 

\acknowledgments
The authors would like to thank the anonymous referee for insightful questions and suggestions that helped improve this paper. This research was supported by the DFATD, Department of Foreign Affairs, Trade and Development Canada, International scholarship program Chile-Canada; 
C.A acknowledges Gemini-CONICYT project No.32120033, Fondo Institucional de Becas FIB-UV, Becas de Doctorado Nacional CONICYT 2016 and PUC-observatory for the telescope time used to obtain the spectra presented in this work. 
C.E.J and T.A.A.S acknowledge support from NSERC, National Sciences and Engineering Research Council of Canada. 
S.K thanks the support of Fondecyt iniciaci\'on grant N 11130702.
C.A, S.K and M.C acknowledges the support from Centro de Astrof\'isica de Valpara\'iso. 

\newpage

\begin{deluxetable*}{ccc|ccccc|cc|ccc}
\tabletypesize{\scriptsize}
\tablewidth{0pt} 
\tablenum{2}
\tablecaption{Summary of the best fit model by visual inspection and representative models ($\mathcal{F} /\mathcal{F}_{min} \leq $ 1.25) of each spectrum for each star. \\
The reader is refereed to section~\ref{sec:results:b} for the selection details. \label{table2}}
\tablehead{
\multicolumn{3}{c}{} & \multicolumn{5}{c|}{Best model} & \multicolumn{2}{c|}{Observation} &
\multicolumn{3}{|c}{Representative model}\\
\cline{4-13}
\colhead{HD} & \colhead{Sp.T} & \colhead{date} & \colhead{$\mathcal{F} /\mathcal{F}_{min}$} & \colhead{i} & \colhead{n} & \colhead{$\rho_{0}$}& \colhead{$R_{T}$} & \colhead{EW} & \colhead{$\Delta V_p$} & \colhead{$<R_{90}>$} & \colhead{$<M_{d}/M_{\star}>$} & \colhead{$<J_{d}/J_{\star}>$} \\
\colhead{} &  \colhead{} & \colhead{(yyyy-mm-dd) } & \colhead{} & \colhead{($\deg$)} & \colhead{} & \colhead{(g cm$^{-3}$) } & \colhead{ ($R_{\star}$)} & \colhead{ ($\mathrm{\AA}$)} & \colhead{(km s$ ^{-1} $) } & \colhead{($R_{\star}$)} & \colhead{} & \colhead{}
}
\colnumbers
\startdata
10144  & B6 Vpe   & 2012-11-13 & -   &  -   &   - &   -     & -   & -0.8  &  719.5 & - & - & - \\   
       &          & 2013-01-18 & 1.2 &  70  & 3.0 & 7.5e-12 & 6.0 & -0.9  &  485.1 & 10.4 & 1.8e-11 & 8.5e-10\\   
	   & 		  & 2013-07-24 & 1.0 &  70  & 3.5 & 2.5e-11 & 6.0 & -0.5  &  361.8 & 13.5 & 2.8e-10 & 2.2e-08\\  
	   & 		  & 2013-10-29 & 1.0 &  70  & 4.0 & 7.5e-11 & 6.0 & -1.2  &  353.6 & 14.6 & 4.0e-10 & 3.0e-08\\    
	   & 		  & 2014-01-29 & 1.0 &  70  & 2.0 & 5.0e-11 & 6.0 & -1.7  &  345.3 & 12.9 & 3.3e-10 & 2.3e-08\\    
\hline
33328$^{a}$  &  B2 IVne & 2012-11-13 & 1.0  & 60 & 4.0 & 7.5e-12 & 25.0 & 1.7 & 703.0 & - & - & - \\ 
             &          & 2013-01-18 & 1.0  & 60 & 4.0 & 2.5e-12 & 6.0  & 0.1 & 534.5 & - & - & - \\  
             &          & 2015-02-25 & 1.0  & 60 & 4.0 & 7.5e-12 & 25.0 & 1.9 & 657.8 & - & - & - \\  
\hline 
35165   & B2 Vnpe  & 2014/2015 blue & 1.0 &  80  & 2.0 & 5.0e-11 & 12.5 & -12.1  &  283.7 & 45.0 & 8.4e-10 & 1.0e-07\\           
		& 		   & 2014/2015 red  & 1.1 &  80  & 2.0 & 1.0e-11 & 6.0  & -12.8  &  312.4 & 45.0 & 8.4e-10 & 1.0e-07\\   
\hline              
35411$^{a}$ &  B1 V + B2 & 2012-11-13 & 1.0 & 80 & 4.0 & 7.5e-12 & 25.0 & 2.15 &  0 & - & - & - \\   
            &            & 2013-01-18 & 1.0 & 80 & 3.5 & 1.0e-12 & 50.0 & 3.1  &  0 & - & - & - \\ 
            &            & 2013-02-26 & 1.0 & 80 & 4.0 & 1.0e-12 & 6.0  & 3.0  &  0 & - & - & - \\ 
            &            & 2015-02-25 & 1.0 & 80 & 4.0 & 7.5e-12 & 25.0 & 2.4  &  0 & - & - & - \\ 
\hline 
35439$^{pf}$  &  B1 Vpe &   2012-11-13 & 1.0 & 50 & 2.5 & 2.5e-11 & 50.0  & -27.7  & 209.7 & - & - & - \\ 
              &         &   2013-01-18 & 1.0 & 50 & 2.5 & 2.5e-11 & 50.0  & -28.6  & 185.0 & - & - & - \\ 
              &         &   2013-02-26 & 1.0 & 50 & 2.5 & 2.5e-11 & 50.0  & -30.2  & 193.2 & - & - & - \\ 
              &         &   2015-02-25 & 1.0 & 70 & 2.0 & 5.0e-12 & 50.0  & -25.6  & 152.1 & - & - & - \\ 
\hline
37795       & B9 V  & 2012-11-13 & 1.0 &  40  & 3.0  & 2.5e-10 & 50.0 & -9.3  &  106.9 & 46.4 & 3.8e-10 & 4.5e-08\\  
            &       & 2013-01-18 & 1.0 &  40  & 3.0  & 2.5e-10 & 50.0 & -9.7  &  82.2 & 53.3 & 4.5e-10 & 5.9e-08\\  
            &       & 2015-02-25 & 1.0 &  40  & 3.0  & 2.5e-10 & 50.0 & -9.0  &  82.2 & 50.2 & 4.0e-10 & 5.4e-08\\  
\hline
41335$^{pf}$&  B2 Vne   &  2012-11-13 & 1.0  & 80 & 2.0  & 5.0e-12  & 25.0  & -25.9 &  152.1 & - & - & - \\ 
            &           &  2013-01-18 & 1.0  & 80 & 2.0  & 5.0e-12  & 25.0  & -27.1 &  111.0 & - & - & - \\ 
            &           &  2013-02-26 & 1.0  & 80 & 2.0  & 5.0e-12  & 25.0  & -26.7 &  115.1 & - & - & - \\ 
            &           &  2015-02-27 & 1.0  & 80 & 2.0  & 5.0e-12  & 25.0  & -26.9 &  115.1 & - & - & - \\ 
 \hline
 42167      & B9 IV     & 2014-01-30 & 1.0 &  70  & 2.0  & 2.5e-10 & 6.0 & -2.0  &   160.3 & 32.6 & 5.8e-10 & 5.3e-08 \\  
            &           & 2015-02-25 & 1.0 &  70  & 2.0  & 2.5e-10 & 6.0 & -1.7  &   209.7 & 32.6 & 5.8e-10 & 5.3e-08 \\  
\hline
 45725      & B4 Veshell & 2015-02-26 & 1.0 &  70  & 2.0 & 5.0e-12 & 25.0 & -30.2 &  164.4 & 87.4 & 2.1e-09 & 3.7e-07\\  
\hline  
48917    & B2 IIIe  & 2014-01-29 & 1.0 &  60  & 2.0 & 5.0e-12 & 25.0 & -24.6 & 86.3 & 103.7 & 3.3e-09 & 6.3e-07\\ 
         &          & 2015-10-23 & 1.0 &  60  & 2.0 & 5.0e-12 & 25.0 & -27.1 & 90.4 & 103.7 & 3.3e-09 & 6.3e-07 \\    
\hline  
50013$^{pf}$    & B1.5 Ve & 2012-11-13 & 1.0 & 50 & 2.5  & 2.5e-11  & 50.0 & -24.1  &  94.6 & - & - & - \\ 
                &      &    2013-02-26 & 1.0 & 60 & 2.0  & 5.0e-12  & 50.0 & -22.2  &  98.7 & - & - & - \\ 
                &      &    2014-03-21 & 1.0 & 50 & 2.5  & 2.5e-11  & 50.0 & -24.0  &  65.8 & - & - & - \\ 
                &      &    2015-02-25 & 1.0 & 60 & 2.0  & 5.0e-12  & 50.0 &  -25.2 &  65.8 & - & - & - \\
                &      &    2015-10-23 & 1.0 & 60 & 2.0  & 5.0e-12  & 50.0 & -28.9  &  74.0 & - & - & - \\
\hline  
52918$^{a}$     & B1 V & 2014-01-29 &   1.0  &  60    &  4.0     &   1.0e-11 &  25.0  & 1.37  & 678.4 & - & - & - \\
\hline
56014    & B3 IIIe  &  2014-01-29 red & 1.0 &  80  & 2.5 & 1.0e-11 & 6.0 &  -2.0   &  390.6 & 23.4 & 3.0e-10 & 2.7e-08\\ 
         &          & 2014-01-29 blue & 1.0 &  80  & 2.5 & 5.0e-12 & 12.5 & -2.0   &  390.6 & 23.4 & 3.0e-10 & 2.7e-08\\     
\hline    
56139   & B2 IV-Ve  & 2013-02-27 & 1.0 &  30  & 2.0 & 2.5e-11 & 25.0 & -20.7   & 0 & 105.5 & 9.1e-09 & 1.7e-06\\    
        &           & 2015-02-27 & 1.0 &  30  & 2.0 & 2.5e-11 & 25.0 & -16.7   & 0 & 105.5 & 9.1e-09 & 1.7e-06\\  
        &		    & 2015-11-14 & 1.0 &  30  & 2.0 & 5.0e-11 & 25.0 & -10.2   & 0 & 73.7 & 1.4e-08  & 2.3e-06\\   
\hline        
57150    & B2 Ve + B3 IVne  & 2014-01-29 & 1.0 &  60  & 2.0 & 5.0e-12 & 50.0 &  -30.2   & 0 & 189.2 & 6.8e-09 & 1.8e-06\\    
\hline
57219$^{a}$    &  B3 Vne & 2014-01-29 &  1.0 & 80 &  3.5  & 7.5e-12 & 25.0 &  2.3  & 0 & - & - & - \\
\hline 
58343   & B2 Vne   & 2013-02-27 & 1.0 & 10  & 2.5 & 7.5e-11 & 25.0 & -7.2 & 0 & 71.7 & 7.7e-09 & 1.2e-06  \\ 
\hline
58715   & B8 Ve  & 2013-02-27 & 1.0 &  50  & 3.5 & 2.5e-10 & 25.0 & -7.2 &  127.4 & 35.6 & 1.2e-09 & 1.2e-07\\  
        &        & 2015-02-25 & 1.0 &  50  & 3.5 & 2.5e-10 & 25.0 & -7.3 &  115.1 & 35.6 & 1.2e-09 & 1.2e-07\\  
\hline
60606    & B2 Vne  & 2012-11-13 & 1.0 &  70  & 3.0 & 1.0e-10 & 25.0 & -21.3 & 143.9 & 62.2 & 3.2e-09 & 5.1e-07\\   
         &       &   2013-01-19 & 1.0 &  70  & 3.0 & 1.0e-10 & 25.0 & -22.8 & 152.1 & 62.2 & 3.2e-09 & 5.1e-07\\   
         &       &   2013-02-26 & 1.0 &  70  & 3.0 & 1.0e-10 & 25.0 & -18.9 & 135.7 & 62.2 & 3.2e-09 & 5.1e-07\\   
\hline        
63462$^{pf}$   & B1 IVe & 2013-02-27 & 1.0 & 70 & 2.0 & 5.0e-12 &  12.5 & -10.9 &  94.6 & - & - & - \\
               &        & 2015-10-23 & 1.0 & 50 & 2.5 & 1.0e-11 &  50.0 & -11.6 &  94.6 & - & - & - \\ 
\hline
68423     & B6 Ve  & 2014-03-21 & 1.0 &  10  & 2.0 & 2.5e-10 & 50.0 & -6.2    &  49.3 & 49.1 & 1.1e-08 & 1.4e-06\\   
\hline
68980    & B1.5 III   & 2013-02-27 &  1.0  &  40 & 2.0 & 5.0e-12 & 50.0 & -23.2  & 41.1 & 214.2 & 1.1e-08 & 2.9e-06\\    
         & 	          & 2015-02-26 & 1.0   & 40  & 2.5 & 2.5e-11 & 50.0 & -19.6  & 45.2 & 130.2 & 1.3e-08 & 3.0e-06\\   
\hline  
71510$^{a}$  &  B2 Ve & 2014-01-29 &   1.0   &  70    &  3.0 & 2.5e-12 & 12.5 & 2.6  & 0 & - & - & - \\ 
             &      &   2014-03-19 &   1.0   &  70    &  4.0 & 7.5e-12 & 6.0 & 2.6   & 0 & - & - & - \\ 
             &      &   2015-02-26 &   1.0   &  70    &  2.0 & 1.0e-12 & 6.0 & 2.25  & 0 & - & - & - \\  
\hline
75311  & B3 Vne  & 2014-03-19 & 1.0  &  60  & 3.0 & 7.5e-11 & 50.0 & -0.6   &   287.8 & 26.0 & 2.8e-09 & 2.7e-07 \\   
\hline
78764   & B2 IVe   & 2014-01-30 & 1.0 & 40  & 2.5 & 7.5e-11 & 12.5 &  -4.8   & 131.6 & 42.1 & 3.6e-09 & 5.3e-07 \\    
        &          & 2014-03-19 & 1.0 & 40  & 2.5 & 7.5e-11 & 12.5 &  -4.2   & 139.8 & 42.1 & 3.6e-09 & 5.3e-07 \\    
\hline
83953$^{pf}$&  B5V & 2013-02-27 & 1.0 & 70 & 3.0 & 1.0e-10 & 50.0  & -20.6  & 160.3 & - & - & - \\
\hline
89080   & B8 IIIe  & 2013-02-27 & 1.1 &  70  & 2.0 & 2.5e-12 & 25.0 & -7.2   & 164.4 & 35.6 & 8.8e-10 & 8.8e-08  \\  
        &          & 2014-05-09 & 1.1 &  70  & 2.0 & 2.5e-12 & 25.0 & -7.0   & 143.9 & 35.6 & 8.8e-10 & 8.8e-08   \\  
\hline 
89890$^{a}$     &  B3 IIIe & 2014-01-30 &   1.0 &   70    &   3.0    &    5.0e-12 &  50.0  & 1.7  & 0 & - & - & - \\
                &       & 2014-03-19 &     1.0 &   80    &   3.5    &    7.5e-12 &  25.0   & 2.3  & 0 & - & - & - \\
                &       & 2015-02-27 &     1.0 &   80    &   3.0    &    5.0e-12 &  50.0   & 1.7  & 0 & - & - & - \\
                &       & 2015-05-06 &     1.0 &   70    &   3.5    &    7.5e-12 &  25.0   & 1.9  & 0 & - & - & - \\
\hline
91465   &  B4 Vne  & 2013-02-26 & 1.0 &  70  & 2.0 & 5.0e-12 & 25.0 & -28.4 &  131.6 & 82.4 & 2.4e-09 & 3.8e-07 \\   
		& 		   & 2014-05-09 & 1.0 &  70  & 2.0 & 1.0e-10 & 50.0 & -24.9 &  135.7 & 63.1 & 2.1e-09 & 3.1e-07\\   
		& 		   & 2015-02-27 & 1.0 &  70  & 2.0 & 1.0e-10 & 50.0 & -22.9 &  94.6 & 63.1 & 2.1e-09 & 3.1e-07\\   
        & 		   & 2015-05-06 & 1.1 &  70  & 2.0 & 5.0e-12 & 25.0 & -30.4 &  98.7 & 97.4 & 2.9e-09 & 5.0e-07\\   
\hline        
92938$^{a}$ &  B4 V & 2014-01-30 & 1.0 &  80 &  4.0     &  7.5e-12  & 12.5 & 2.4  & 0 & - & - & - \\ 
            &       & 2015-02-27 & 1.0 &  80 &  4.0     &  7.5e-12  & 12.5 & 2.6  & 0 & - & - & - \\  
            &       & 2015-05-06 & 1.0 &  80 &  4.0     &  7.5e-12  & 12.5 & 4.3  & 0 & - & - & - \\ 
\hline
93563     & B8.5 IIIe  & 2014-01-30 & 1.2 &  70  & 3.5 & 1.0e-10 & 50.0 & -8.1   & 296.0 & 22.5 & 6.3e-11 & 5.0e-09  \\    
          & B8.5 IIIe  & 2015-05-06 & 1.2 &  70  & 3.5 & 1.0e-10 & 50.0 & -9.7   & 135.7 &  22.5 & 6.3e-11 & 5.0e-09    \\    
\hline
102776   & B3 Vne   & 2014-01-30 & 1.0 &  60  & 3.0 & 5.0e-11 & 50.0 & -12.2 & 98.7  & 52.1 & 9.6e-10 & 1.2e-07 \\   
		 & 		    & 2014-03-19 & 1.0 &  60  & 2.5 & 1.0e-11 & 50.0 & -9.7  & 185.0 & 90.6 & 9.4e-10 & 1.7e-07 \\   
         & 		    & 2015-02-27 & 1.1 &  60  & 2.0 & 2.5e-12 & 25.0 & -7.1  & 185.0 & 85.5 & 1.3e-09 & 2.3e-07  \\   
		 & 		    & 2015-05-06 & 1.0 &  60  & 2.0 & 2.5e-12 & 50.0 & -7.4  & 119.2 & 91.4 & 2.0e-09 & 3.5e-07 \\   
\hline
103192  & B9 IIIsp  & 2014-03-19 & 1.2  &  60  & 3.0 & 7.5e-12 & 50.0 & -1.4 & 259.0 & 13.4 & 4.6e-10	& 3.4e-08 \\    
        &           & 2015-02-26 & 1.2  &  60  & 3.0 & 7.5e-12 & 50.0 & 1.2  & 263.1 & 13.4 & 4.6e-10	& 3.4e-08 \\    
        &           & 2015-05-07 & 1.2  &  60  & 3.0 & 7.5e-12 & 50.0 & 2.0  & 234.3 & 13.4 & 4.6e-10	& 3.4e-08\\    
\hline 
105382$^{a}$ &  B6 IIIe & 2014-01-30 &  1.0  &  80    &  3.5     &  5.0e-12 &  25.0 & 1.3  &  0 & - & - & - \\
&      &   2015-05-07 & 1.0 & 80  &  3.0 &  2.5e-12 &  25.0  & 2.5   & 0 & - & - & - \\
\hline
105435    & B2 Vne  & 2014-01-30 & 1.0 &  60  & 2.5 & 1.0e-10 & 50.0 & -37.0   & 0 & 157.9 & 1.0e-08 & 2.3e-06 \\   
		  & 		& 2015-02-25 & 1.0 &  60  & 2.0 & 5.0e-12 & 50.0 & -33.1  & 0 & 198.5 & 9.1e-09 & 2.4e-06 \\    
		  & 		& 2015-05-06 & 1.0 &  60  & 2.0 & 5.0e-12 & 50.0 & -31.0  & 0 & 198.5 & 9.1e-09 & 2.4e-06  \\    
\hline
107348  & B8 Ve  & 2014-01-30 & 1.0 &  50  & 3.0 & 2.5e-10 & 25.0 &  -10.2 &  82.2 & 30.1 & 1.5e-09 & 1.5e-07\\   
		& 		 & 2015-05-07 & 1.1 &  50  & 3.0 & 5.0e-11 & 25.0 &  -6.9  &  123.3 & 37.4 & 1.0e-09 & 1.0e-07\\  
\hline 
110335   & B6 IVe  & 2014-01-30 & 1.1 &  70  & 3.0 & 2.5e-10 & 25.0 &  -19.3    &   69.9 & 55.8 & 2.1e-09 & 2.8e-07\\    
         &         & 2015-05-07 & 1.1 &  70  & 3.0 & 2.5e-10 & 25.0 &  -18.3    &   90.4 & 55.8 & 2.1e-09 & 2.8e-07 \\    
\hline
110432$^{pf}$ & B0.5 IVpe & 2014-01-31 &  1.0 & 80 & 2.0 & 7.5e-12 & 25.0 & -30.2  & 197.3 & - & - & - \\ 
              &           & 2015-05-06 &  1.0 & 80 & 2.0 & 7.5e-12 & 25.0 & -28.6  & 102.8 & - & - & - \\  
\hline
112078$^{a}$  &  B3 Vne & 2014-01-31 &    1.0  & 30  & 2.5   &  1.0e-12 &   50.0  & 2.2   & 0 & - & - & - \\
\hline 
120324  & B2 Vnpe  & 2014-01-31 & 1.0 &  50  & 2.0 & 5.0e-11 & 25.0 & -14.8   & 74.0 & 72.8 & 7.6e-09 & 1.2e-06\\   
		& 		   & 2015-02-25 & 1.0 &  50  & 2.5 & 7.5e-11 & 25.0 & -18.6   & 66.8 & 78.4 & 3.9e-09 & 5.7e-07 \\   
		& 		   & 2015-05-06 & 1.1 &  50  & 2.5 & 5.0e-11 & 25.0 & -21.0   & 0 & 97.0 & 4.0e-09 & 7.2e-07\\   
\hline
124195$^{a}$  &  B5 V & 2014-03-21 &    1.0 &  70    & 4.0   & 7.5e-12   & 50.0  & 2.2  & 0 & - & - & - \\
\hline 
124367  & B4 Vne  & 2014-01-31 & 1.1  &  70  & 2.0 & 5.0e-12 & 50.0 & -38.9 & 98.7 & 21.7 & 	1.4e-09	& 1.2e-07 \\   
\hline
124771$^{a}$   &  B4 V & 2014-03-21 &  1.1   & 70    &   4.0    &   5.0e-12 &   6.0 &  2.1   & 0 & - & - & - \\
\hline
127972  & B2 Ve   &   2014-01-31 & 1.0 &  80  & 2.5 & 7.5e-12 & 12.5 & -5.3 & 259.0 & 26.9 &	3.0e-10 &	2.8e-08 \\   
        &         &   2015-02-25 & 1.0 &  80  & 2.5 & 7.5e-12 & 12.5 & -3.7 & 349.5 & 26.9 &	3.0e-10 &	2.8e-08  \\   
        &         &   2015-07-15 & 1.0 &  80  & 2.5 & 7.5e-12 & 12.5 & -2.9 & 365.9 & 26.9 &	3.0e-10 &	2.8e-08  \\ 
\hline
131492   & B4 Vnpe  & 2014-03-21 & 1.0  &  70  & 3.0 & 1.0e-11 & 6.0 & -0.9   &  489.2 & 21.7 &	1.4e-09 &	1.2e-07\\  
\hline
135734 & B8 Ve  & 2013-07-24 & 1.1 &  60  & 2.0 & 2.5e-12 & 25.0 & -7.0 & 168.6 & 40.2 &	1.1e-09 &	1.2e-07 \\   
		& 		& 2015-02-25 & 1.1 &  60  & 2.5 & 1.0e-11 & 25.0 & -8.3 & 135.7 & 40.2 &	1.1e-09 &	1.2e-07  \\   
		& 		& 2015-07-15 & 1.1 &  60  & 2.5 & 1.0e-11 & 25.0 & -8.2 & 152.1 & 40.2 &	1.1e-09 &	1.2e-07 \\   
\hline 
138769$^{a}$&  B3 IVp & 2013-07-24 &  1.0 & 80 & 2.5 & 1.0e-12 & 12.5  & 4.2   & 0 & - & - & - \\ 
            &         & 2015-07-15 &  1.0 & 80 & 3.5 & 5.0e-12 & 50.0  & 3.1   & 0 & - & - & - \\  
\hline 
142184$^{a}$&  B2 V & 2013-07-24 & 1.0 & 60 & 4.0 & 5.0e-12 & 12.5 & 2.0  & 698.9 & - & - & - \\ 
            &       & 2014-03-21 & 1.0 & 80 & 4.0 & 2.5e-12 & 6.0  & 3.5  & 698.9 & - & - & - \\ 
\hline    
143275 & B0.3 IV & 2014-03-19 & 1.1 & 20 & 3.0 & 7.5e-11 & 50.0 & -11.3   & 0 & 143.6 &	1.0e-07 &	3.1e-05\\
\hline
148184    & B2 Ve   & 2013-07-24 & 1.0 &  30  & 2.0 & 1.0e-11 & 25.0 & -35.9   & 0 & 152.8 & 	2.4e-08	& 5.5e-06 \\    
&         & 2015-02-25 & 1.0 &  30  & 2.0 & 1.0e-11 & 25.0 & -34.9   & 0 & 152.8 & 	2.4e-08	& 5.5e-06  \\    
&         & 2015-05-06 & 1.0 &  30  & 2.0 & 1.0e-11 & 25.0 & -39.9   & 0 & 152.8 & 	2.4e-08	& 5.5e-06  \\ 
\hline   
157042  & B2 IIIne  & 2013-07-24 &  1.1  &  70  & 2.5 & 2.5e-11 & 12.5 & -20.2 &  160.3 & 55.0 &	1.6e-09	& 2.1e-07 \\ 
        &           & 2015-05-06 &  1.1  &  70  & 2.5 & 2.5e-11 & 12.5 & -22.9 &  213.8 & 55.0 &	1.6e-09	& 2.1e-07 \\  
\hline
158427   & B2 Ve  & 2015-05-06 & 1.0 &  70  & 2.0 & 5.0e-12 & 50.0 & -36.1  &  32.9 & 188.1 &	7.5e-09	& 2.0e-06\\   
\hline
167128   & B3 IIIpe  & 2013-07-24 & 1.0 &  40  & 3.5 & 7.5e-11 & 50.0   & -3.8   &   164.4 & 32.6 &	3.1e-09	& 3.9e-07\\  
\hline
205637    & B3 V  & 2012-11-14 & 1.1 &  89  & 2.0 & 1.0e-11 & 6.0 & -1.9   & 337.1 & 27.3 &	8.8e-10	& 8.4e-08   \\   
\hline
209014  & B8 Ve  & 2013-07-24  &  1.0  & 89  & 2.0   &   2.5e-10   &   12.5 & -8.0  & 242.6 & 29.3 & 1.1e-09 & 1.1e-07\\ 
        &        & 2015-10-23 &   1.0  & 89  & 2.0   &   2.5e-10   &   12.5 & -8.5   &  209.7 & 29.3 & 1.1e-09 & 1.1e-07\\
 \hline 
209409   & B7 IVe  & 2012-11-13 & 1.0 &  80  & 2.0 & 5.0e-12 & 25.0 & -18.9 &  143.9 & 58.2 & 7.9e-10 & 1.1e-07\\  
		 & 		   & 2015-10-24 & 1.2 &  80  & 2.0 & 5.0e-12 & 50.0 & -20.0 &  152.1 & 55.6 & 2.0e-09 & 2.4e-07\\  
\hline
212076   & B2 IV-Ve  & 2012-11-13 & 1.3  &  30  & 2.0 & 2.5e-11 & 25.0 & -18.2   &  28.8 & 85.1 & 3.1e-09 & 4.8e-07\\           
		 & 			 & 2015-10-23	& 1.0 &  30  & 2.0 & 2.5e-12 & 50.0 & -14.3  &  24.7 & 118.2 & 6.8e-09 & 1.2e-06\\    
\hline
212571 & B1 III-IV & 2012-11-14 & 1.1 &  60  & 2.5 & 1.0e-11 & 12.5 & -7.7  & 283.7 & 84.1 & 9.3e-10 & 1.5e-07 \\  
       & 	       & 2013-07-24 & 1.1 &  60  & 2.5 & 7.5e-12 & 12.5 & -4.0  & 304.2 & 74.4 & 6.4e-10 & 9.6e-08 \\  
       & 	       & 2015-10-24 & 1.0 &  60  & 2.5 & 1.0e-11 & 12.5 & -10.7 & 209.7 & 83.9 & 1.6e-09 & 2.6e-07 \\
\hline
214748    & B8 Ve & 2012-11-15 & 1.3 &  50  & 3.5 & 2.5e-10 & 12.5 & -4.0 & 131.6 & 28.7 &	2.8e-09	& 2.2e-07  \\   
          &       & 2013-07-24 & 1.3 &  50  & 3.5 & 2.5e-10 & 12.5 & -4.9 & 123.3 & 28.7 &	2.8e-09	& 2.2e-07    \\   
          &       & 2015-07-15 & 1.3 &  50  & 3.5 & 2.5e-10 & 12.5 & -5.7 & 123.3 & 28.7 &	2.8e-09	& 2.2e-07   \\   
          &       & 2015-10-24 & 1.3 &  50  & 3.5 & 2.5e-10 & 12.5 & -5.7 & 135.7 & 28.7 &	2.8e-09	& 2.2e-07    \\   
\hline
217891   & B6 Ve  & 2012-11-13 & 1.0 &  40  & 2.0 & 5.0e-11 & 50.0 & -21.1   & 0  & 94.1 &	1.7e-08	& 2.9e-06 \\   
         &        & 2013-07-25 & 1.0 &  40  & 2.0 & 5.0e-11 & 50.0 & -22.8   & 0 & 94.1 &	1.7e-08	& 2.9e-06   \\   
\hline
219688$^{a}$   &  B5 V & 2015-10-24 &   1.0 & 50  & 3.0 & 2.5e-12 & 12.5 & 2.6 & 0 & - & - & - \\ 
\hline 
221507$^{a}$   & B9.5 IIIpHgMnSi    & 2013-07-24 & 1.0 & 89 & 3.0 & 2.5e-12 & 6.0 & 2.5  & 0 & - & - & - \\  
               &                    & 2015-07-15 & 1.0 & 89 & 3.0 & 2.5e-12 & 6.0 & 3.5  & 0 & - & - & - \\  
               &                    & 2015-10-23 & 1.0 & 89 & 3.0 & 2.5e-12 & 6.0 & 4.1   & 0 & - & - & - \\ 
 \hline
224686   & B8 Ve  & 2012-11-13 & 1.0 &  80  & 2.0 & 2.5e-10 & 6.0 & -2.0 & 275.4 & 28.7 &	2.8e-9	& 2.8e-07  \\  
\enddata
\tablenotetext{a  }{     Absorption profiles}
\tablenotetext{pf  }{     Poor fit}
\tablenotetext{-  }{      Not agreement model}
\tablecomments{The information displayed in this table are for the best (visual inspection) and representative ($\mathcal{F} /\mathcal{F}_{min} \leq $ 1.25) models of each observation. \\
Values of the representative models are only for emission profiles without a poor fit.\\
The Spectral Type (Sp.T) is obtained from Simbad database. \\
Blue and red (indicated next to the date) refer to the blue or red peak fit, respectively.}
\end{deluxetable*}

\bibliographystyle{aasjournal}
\bibliography{References}

\newpage
\appendix

\section{Emission profiles} \label{ApA}
Observed emission line profiles from our program stars (black lines) shown with the best-fit model (red dashed lines). Variable stars in our sample are indicated an asterisk symbol beside the star name.

\begin{figure}
          \gridline{
                    \fig{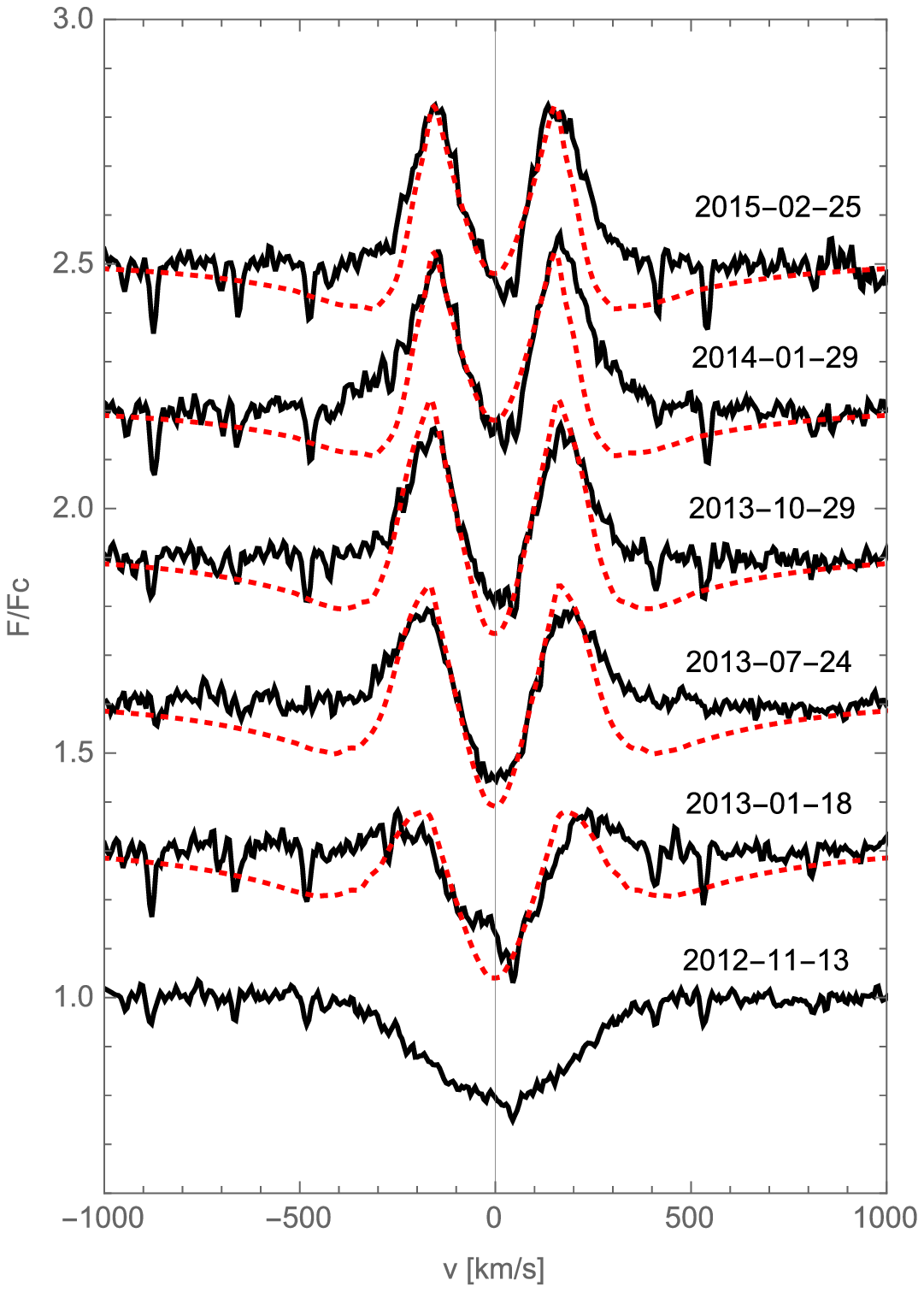}{0.2\textheight}{HD10144*}
                    \fig{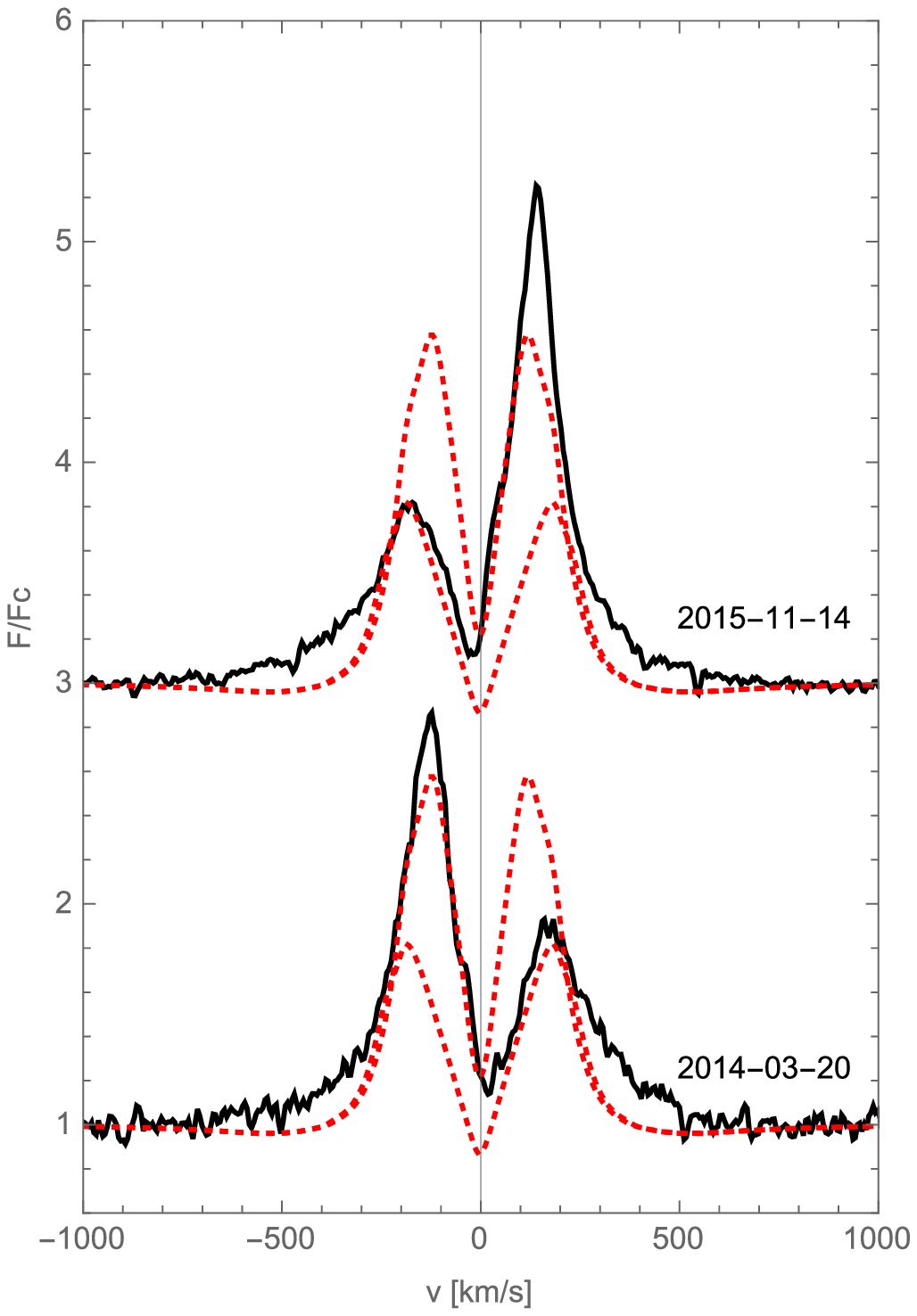}{0.2\textheight}{HD35165*}
                    \fig{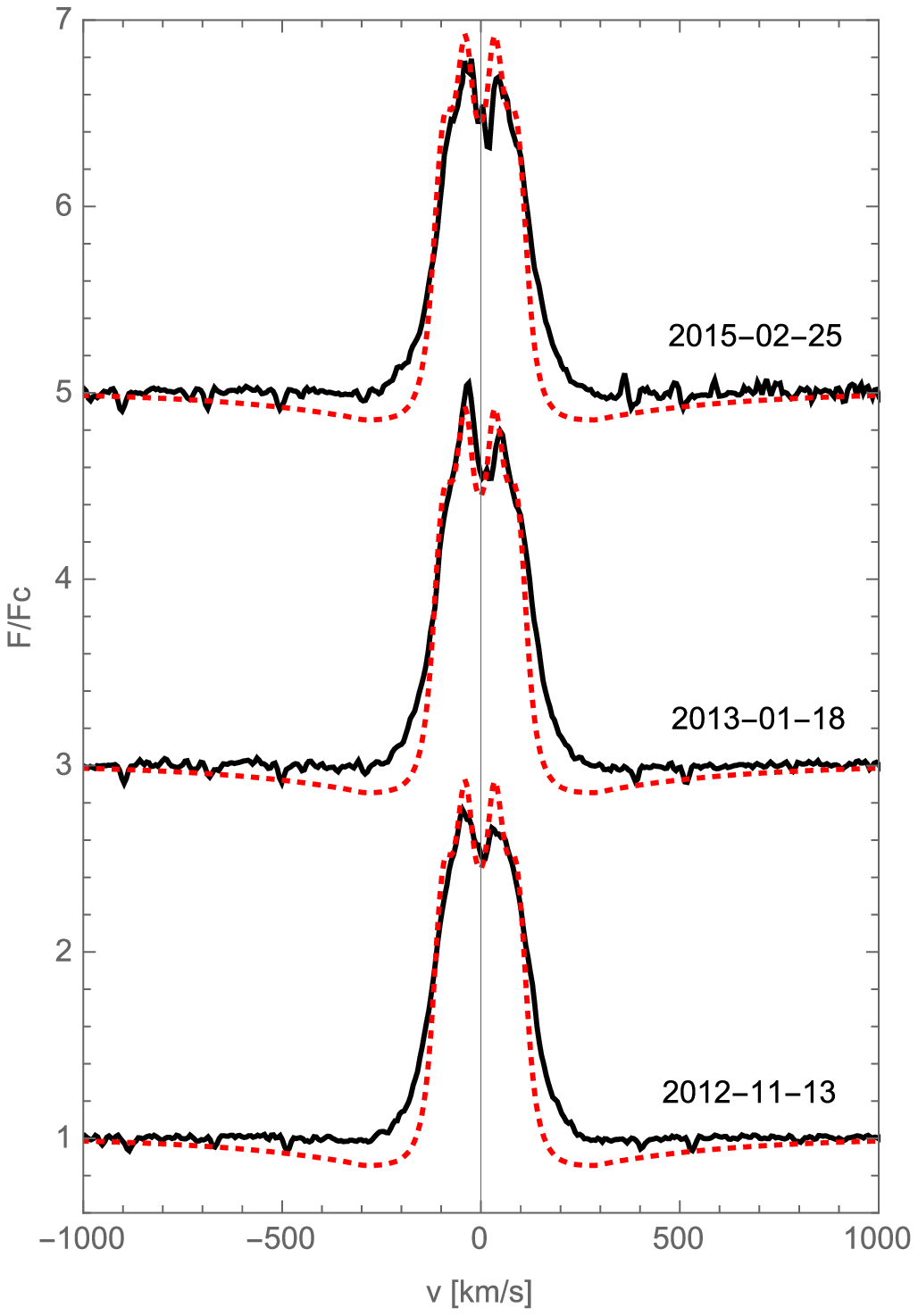}{0.2\textheight}{HD37795}
          }
          
          \gridline{
                    \fig{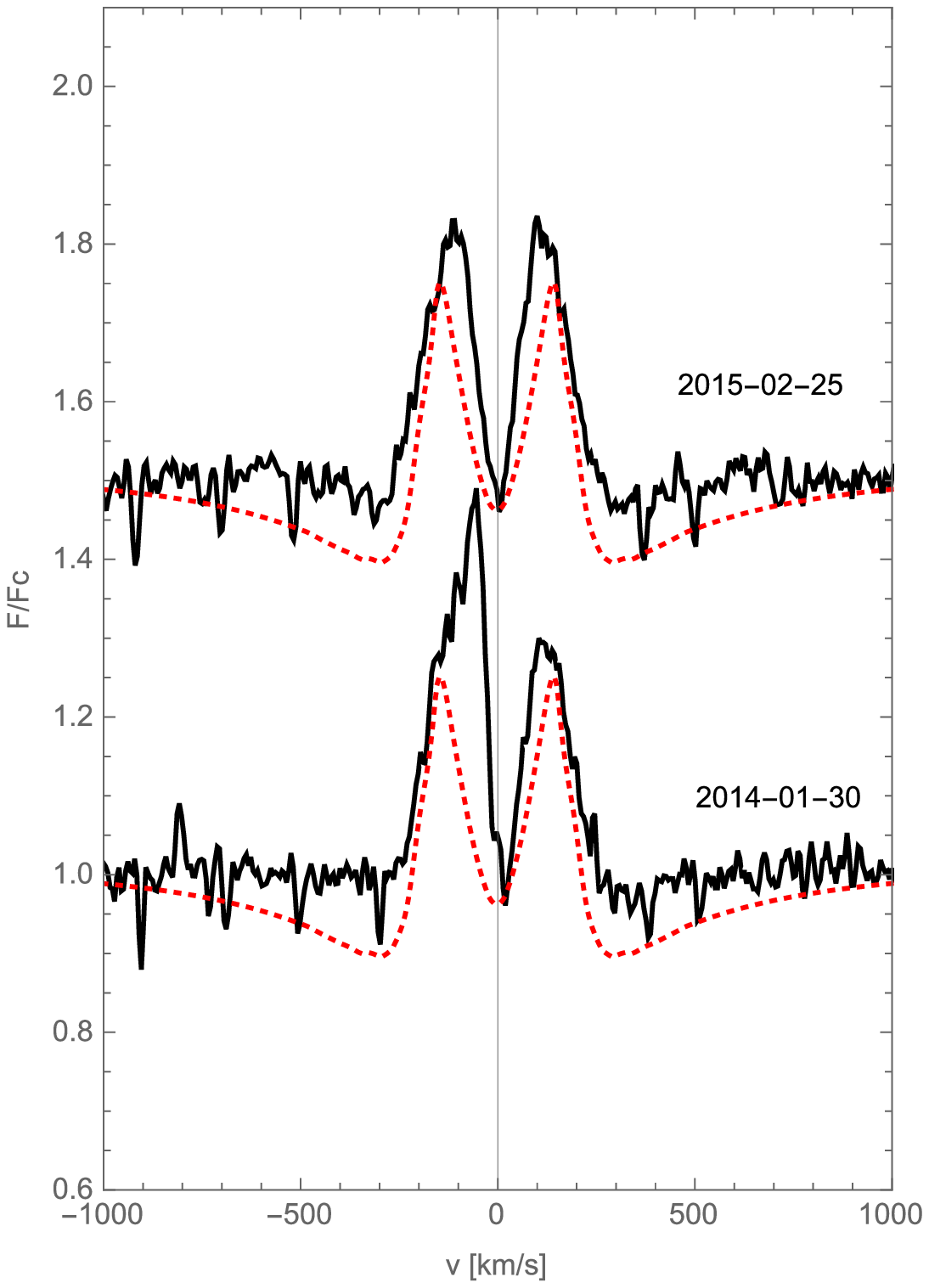}{0.2\textheight}{HD42167}
                    \fig{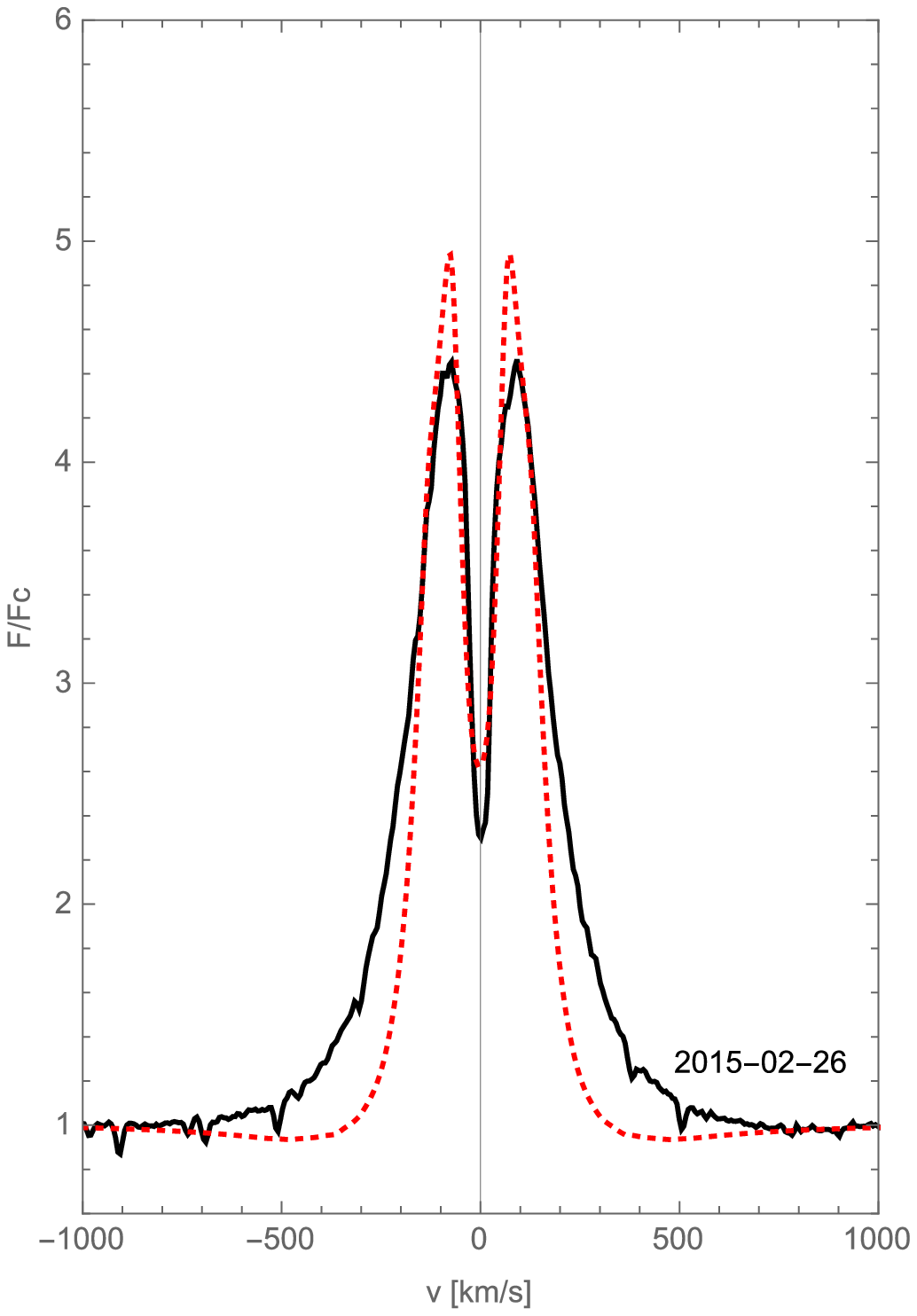}{0.2\textheight}{HD45725}
                    \fig{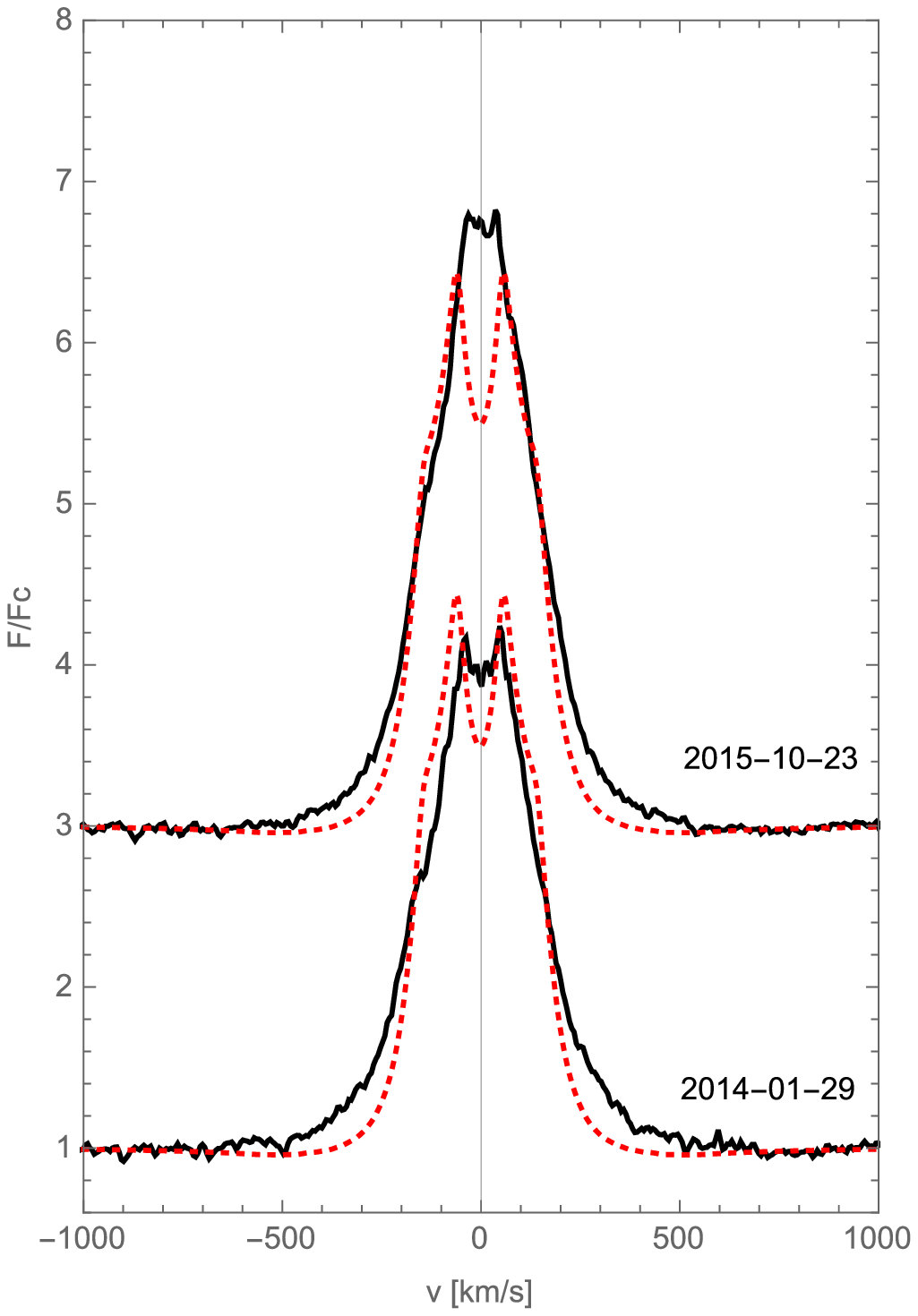}{0.2\textheight}{HD48917}
           }
        
          \gridline{
                    \fig{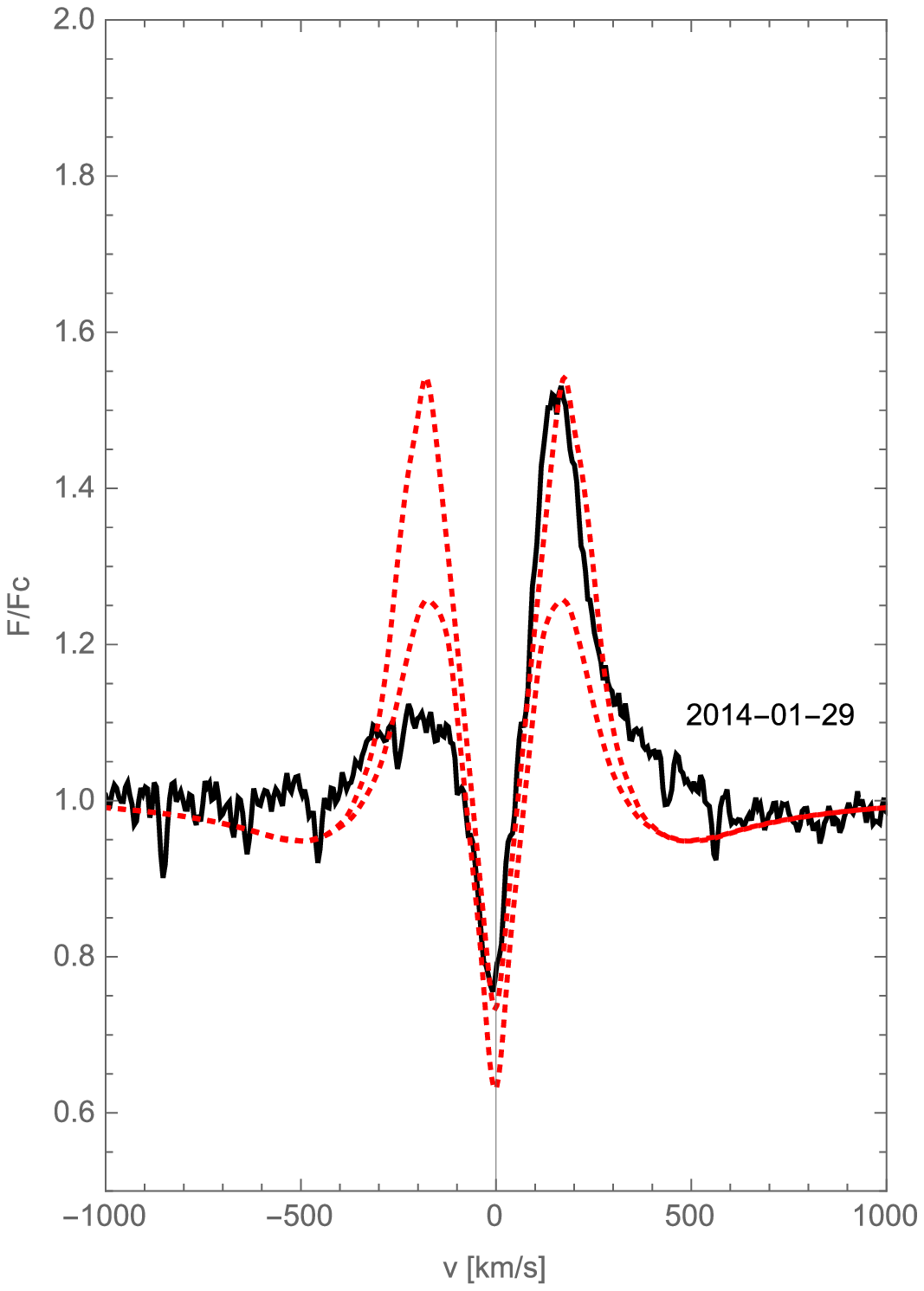}{0.2\textheight}{HD56014*}
                    \fig{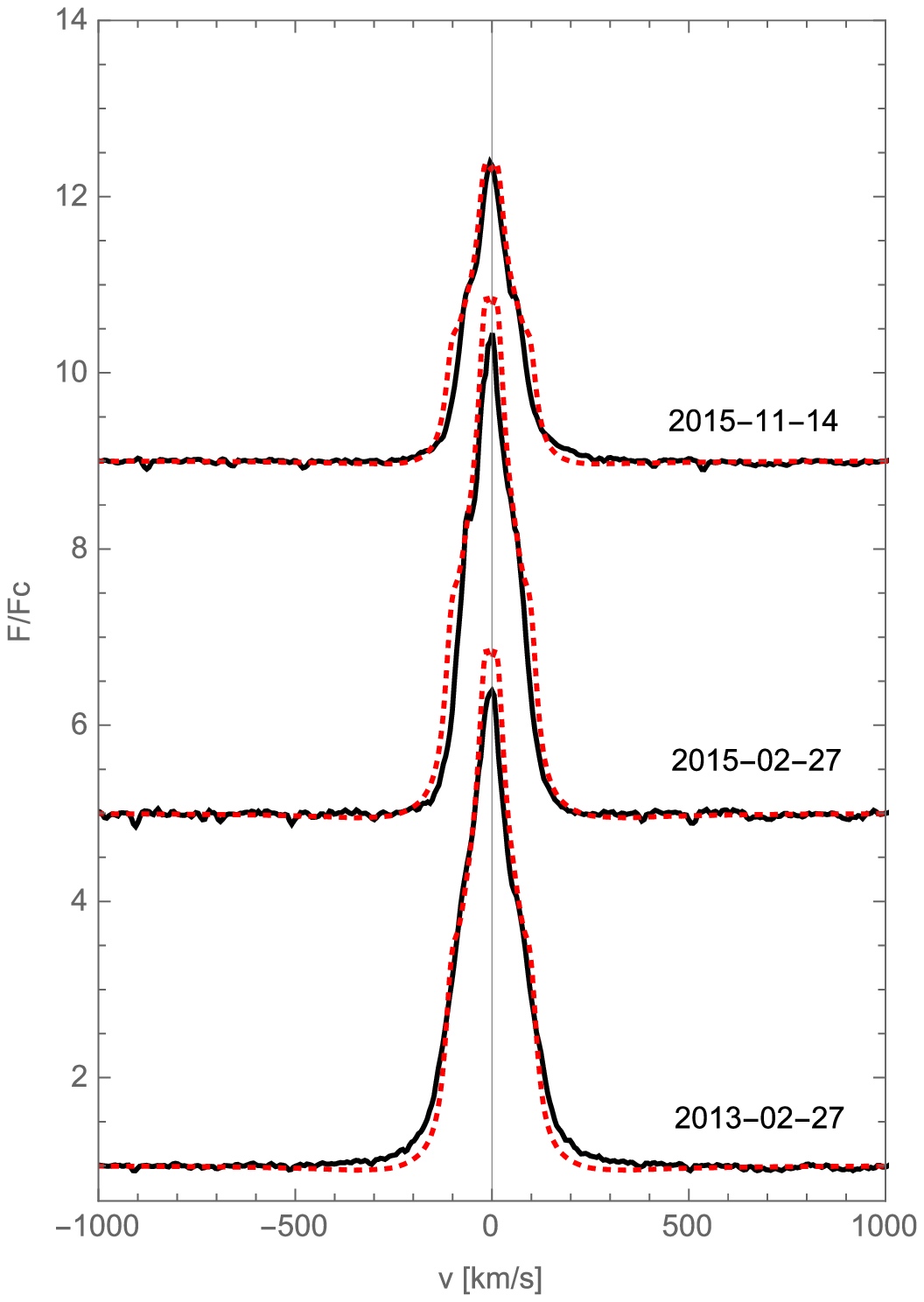}{0.2\textheight}{HD56139*}
                    \fig{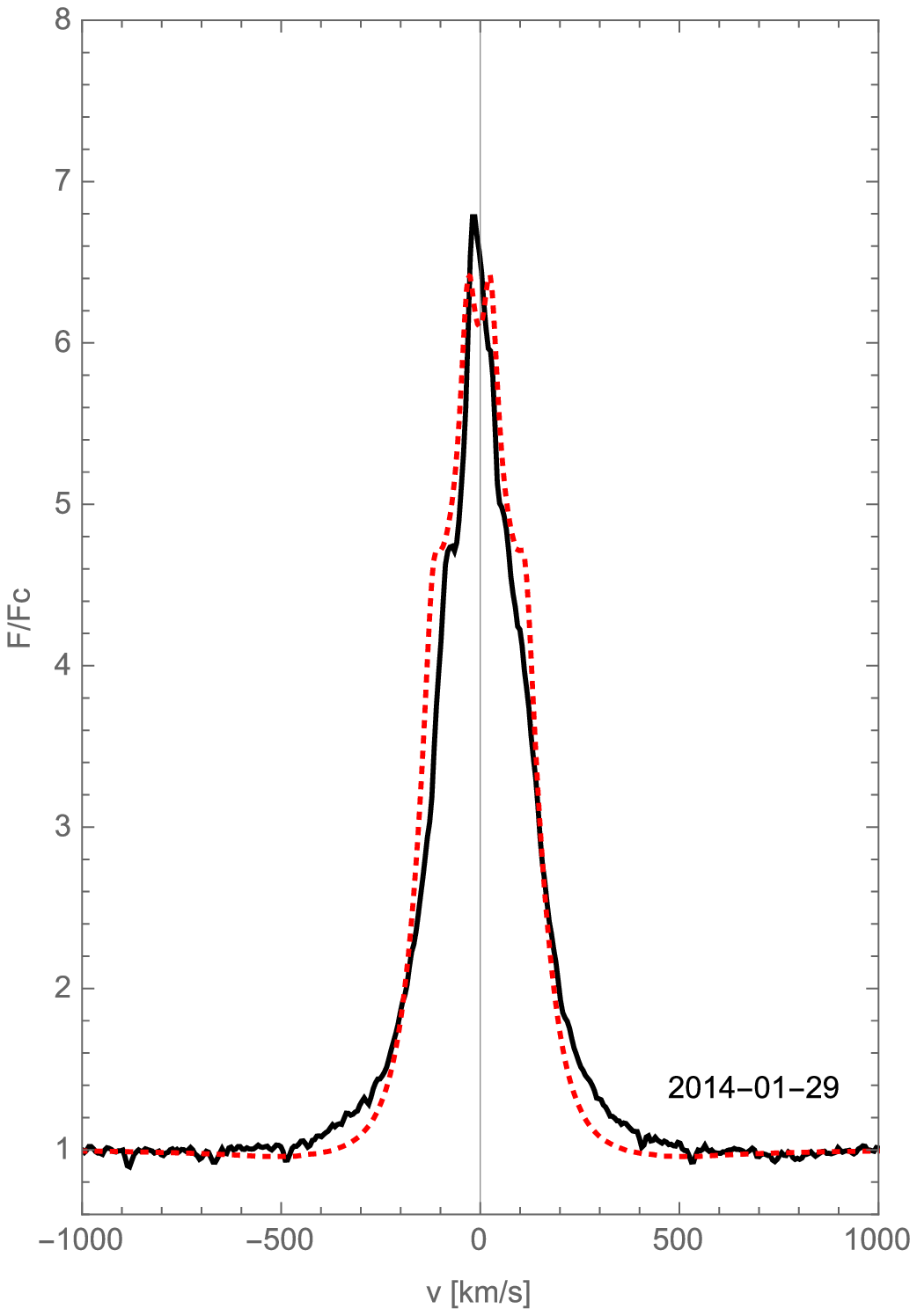}{0.2\textheight}{HD57150}
           }
           
\end{figure}

\begin{figure}
          \gridline{
                    \fig{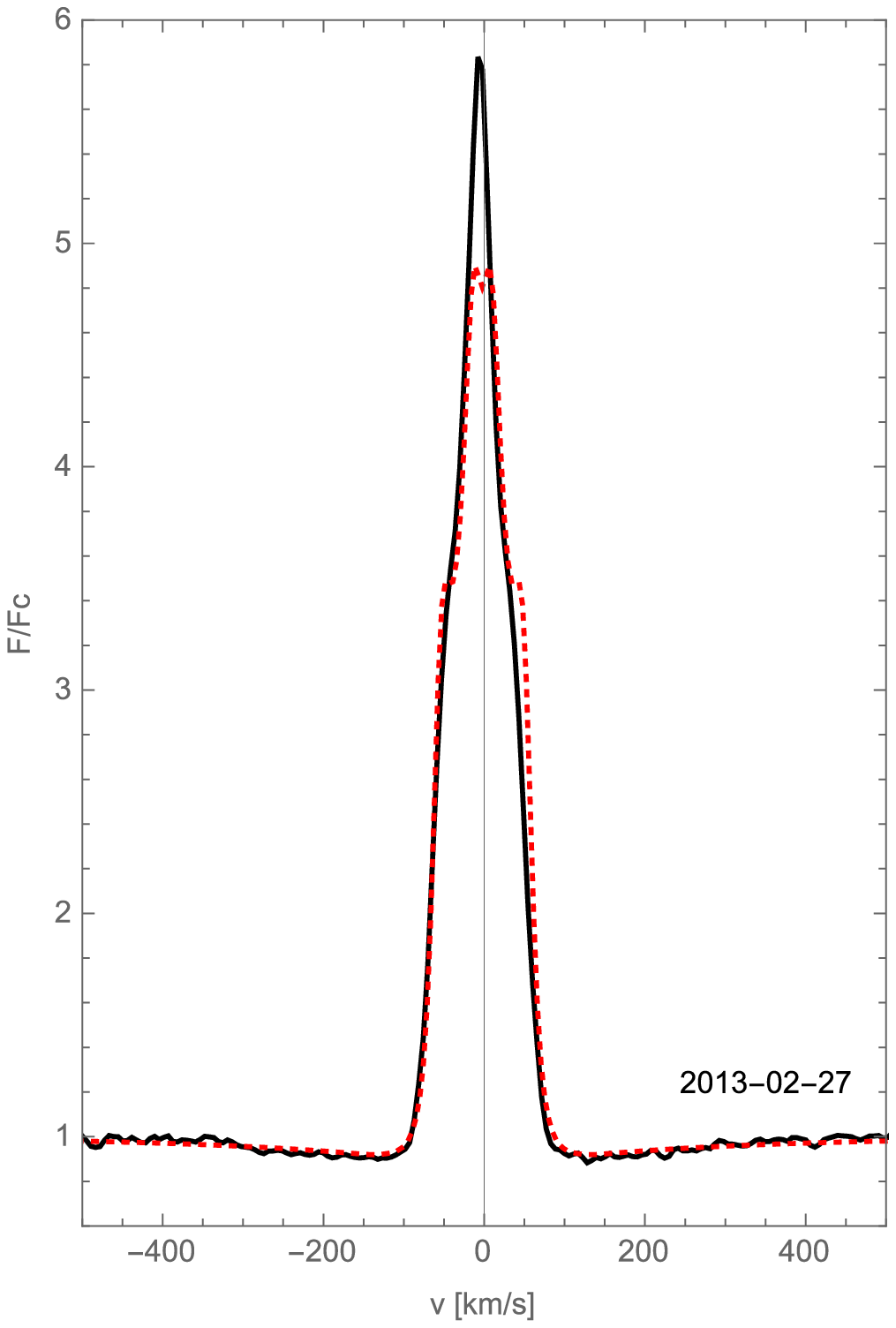}{0.2\textheight}{HD58343}
                    \fig{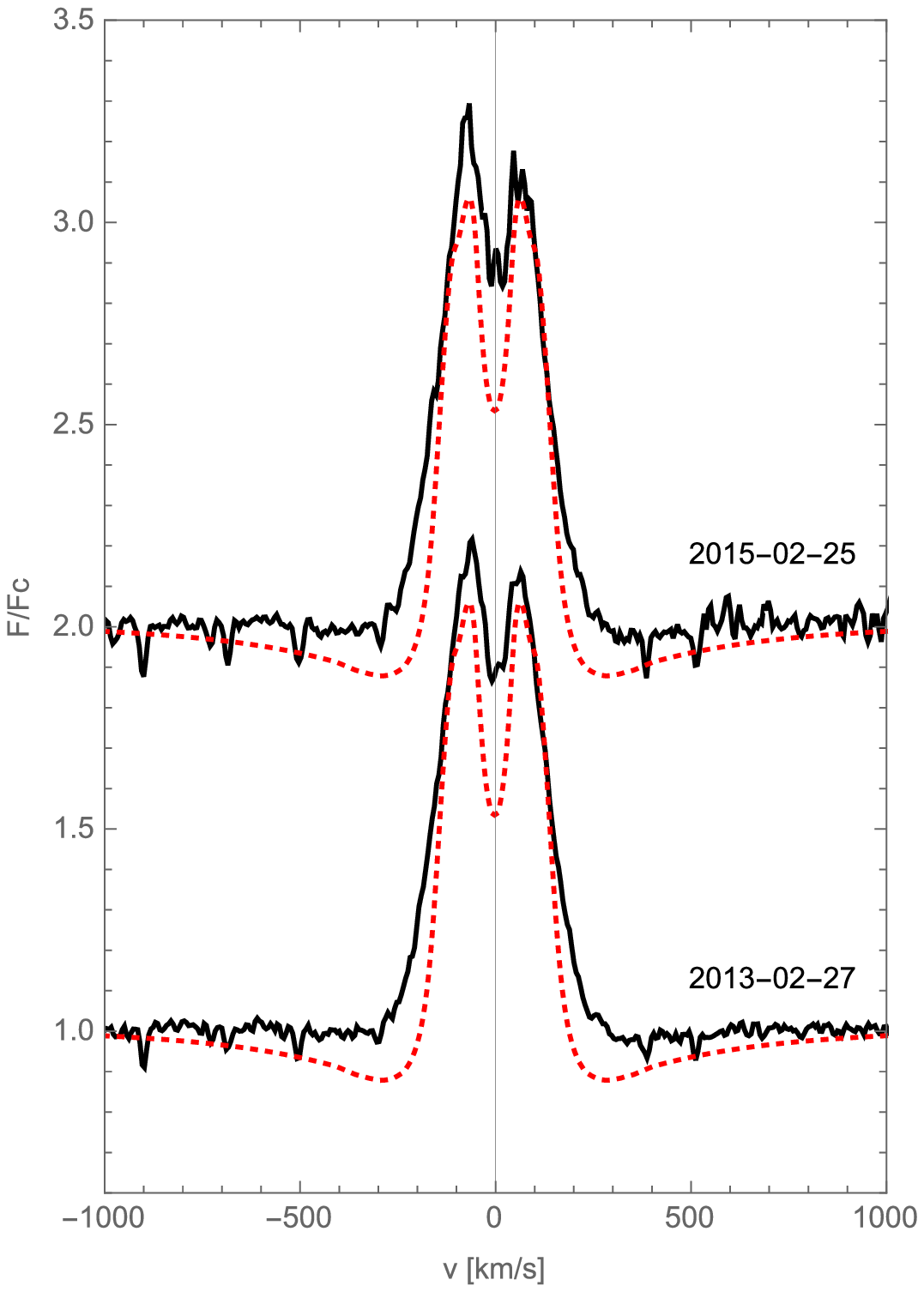}{0.2\textheight}{HD58715}
                    \fig{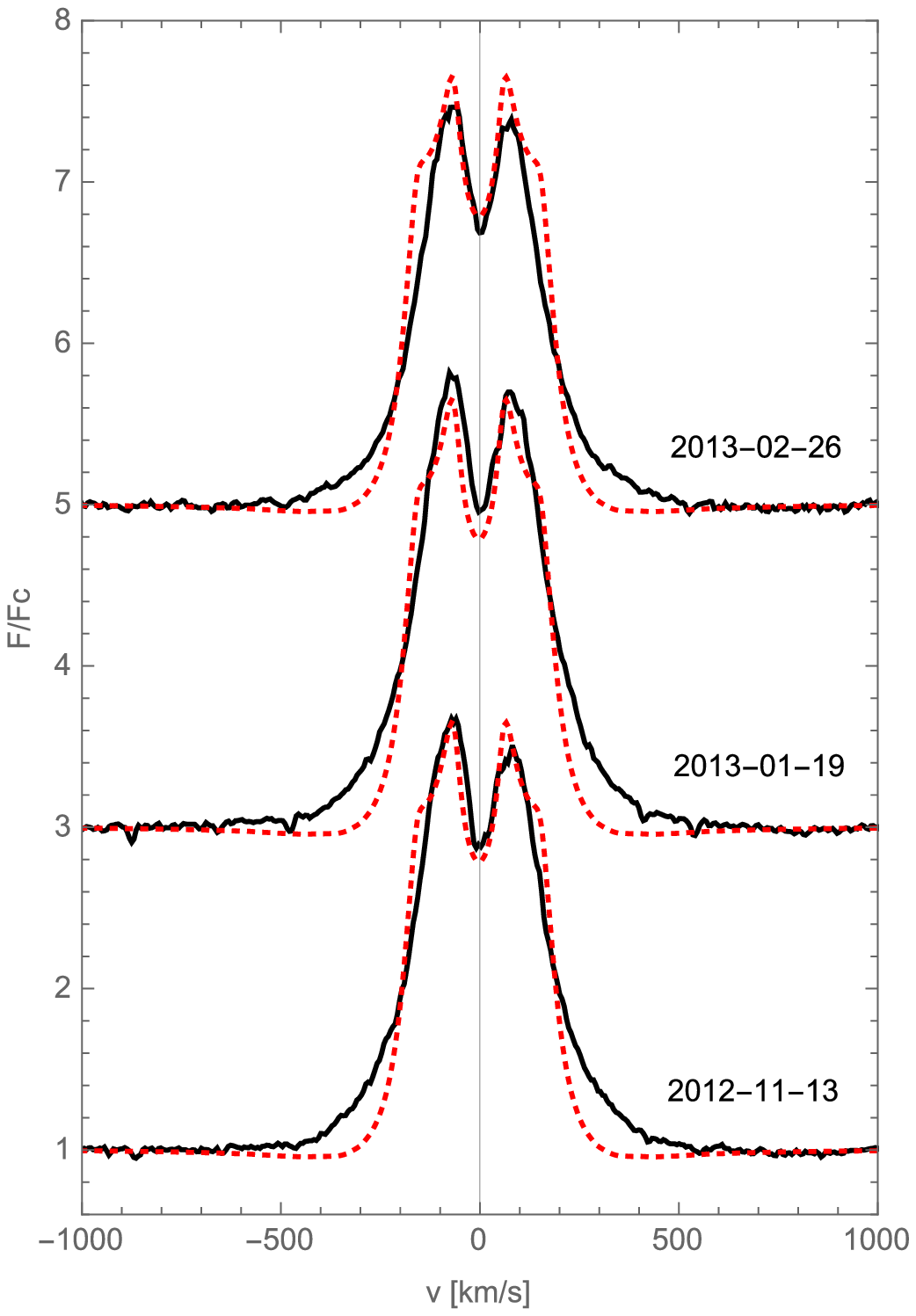}{0.2\textheight}{HD60606}
          }
          
          \gridline{
                    \fig{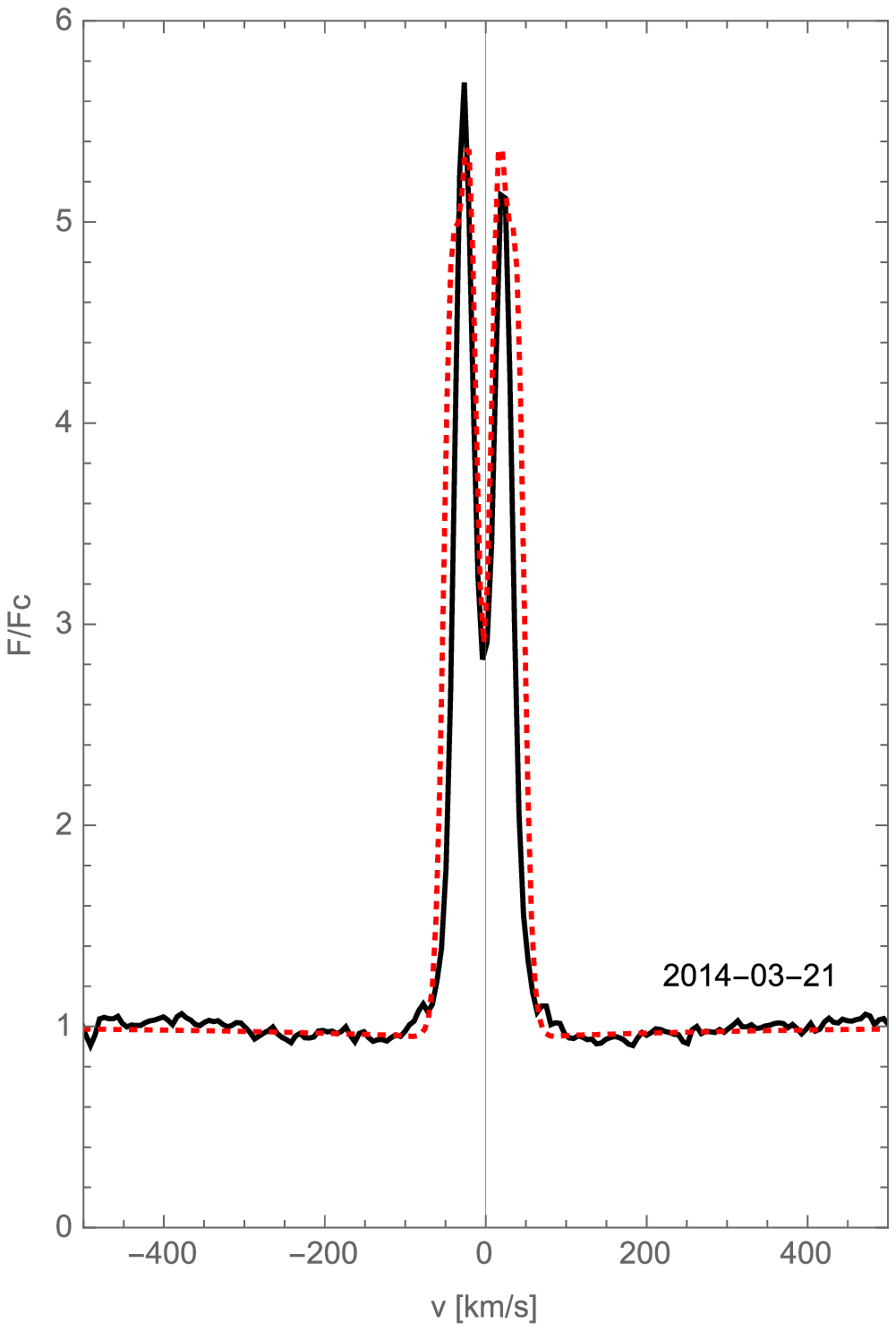}{0.2\textheight}{HD68423}
                    \fig{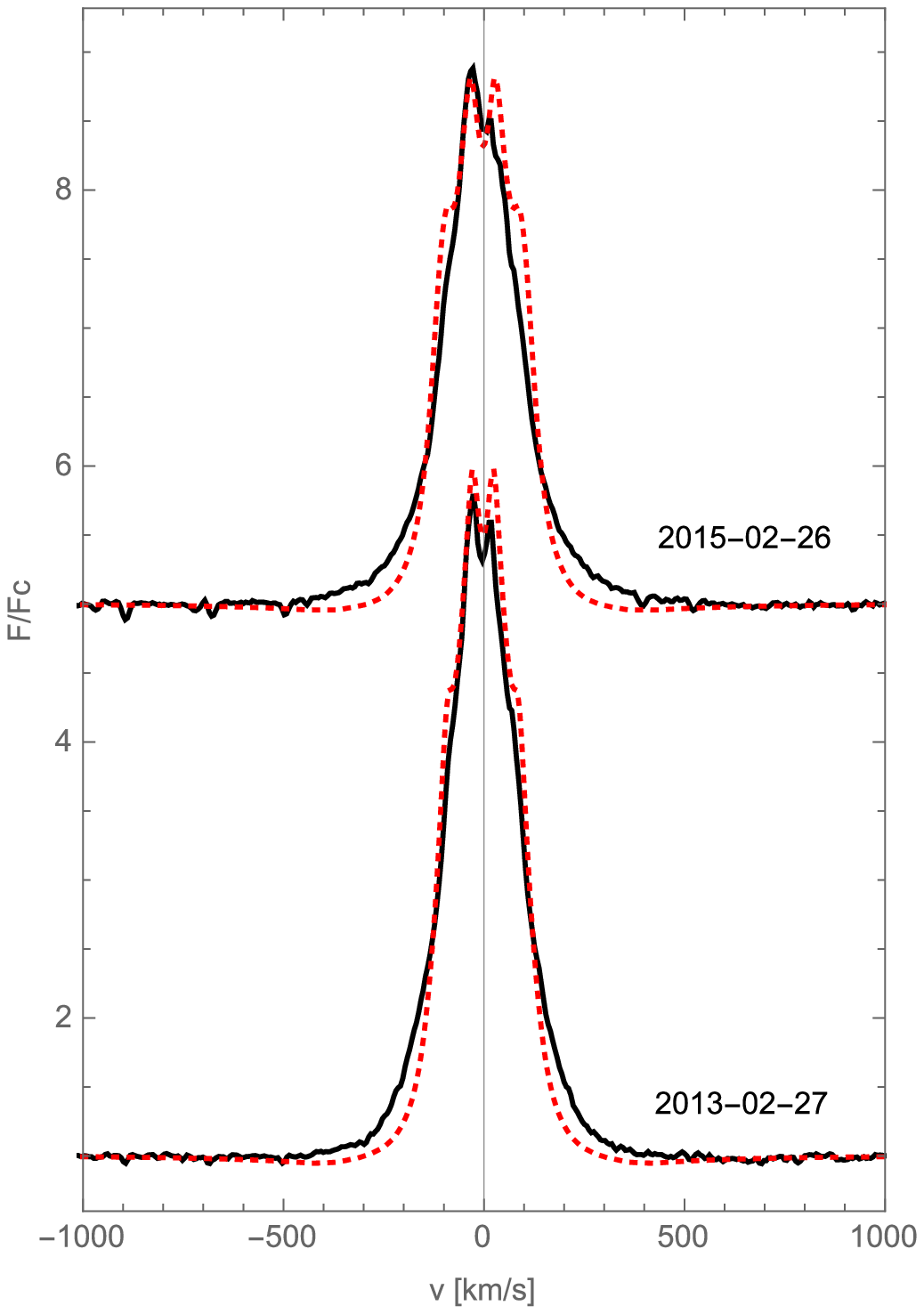}{0.2\textheight}{HD68980*}
                    \fig{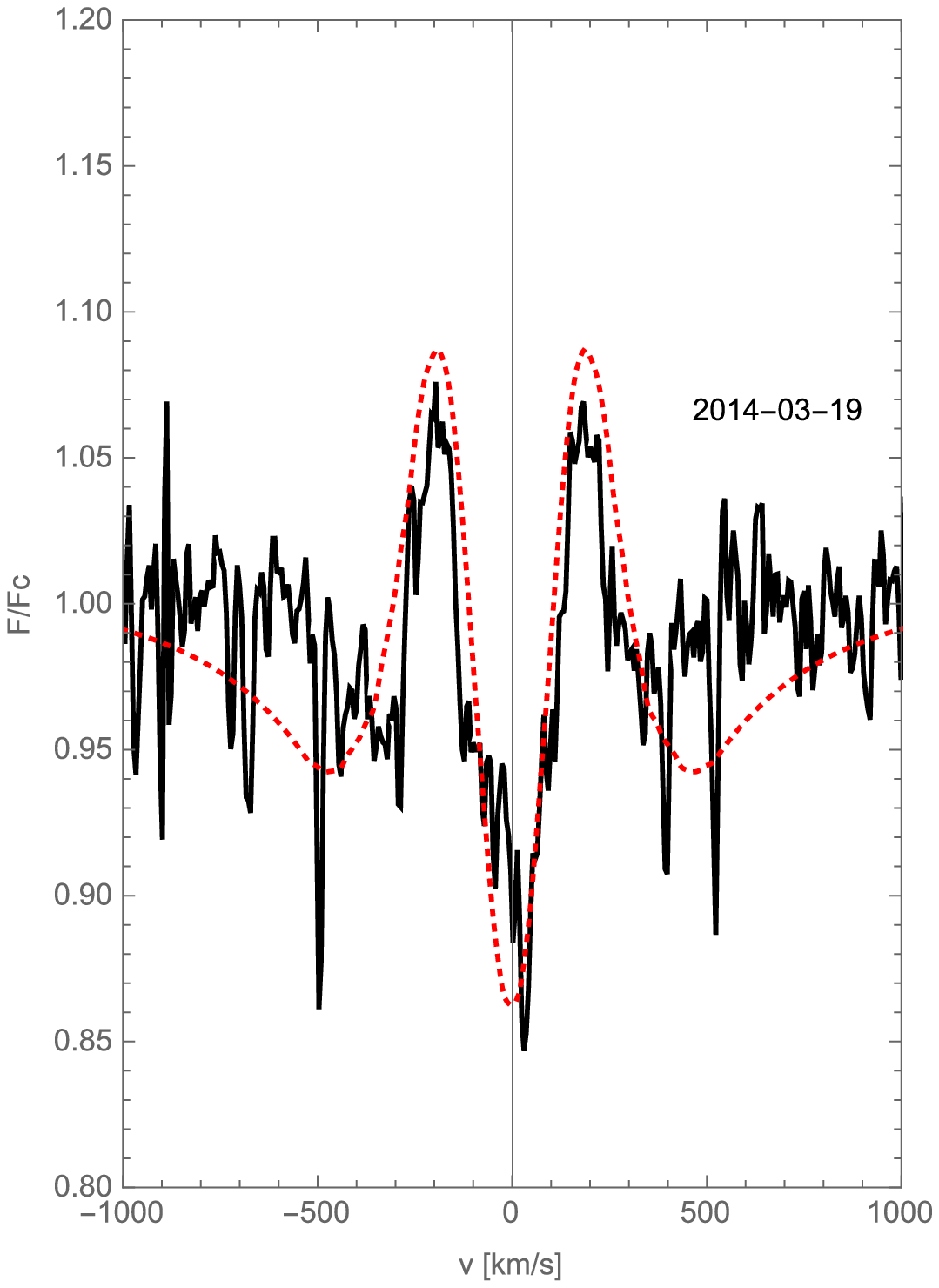}{0.2\textheight}{HD75311}
          }
         
         \gridline{
                    \fig{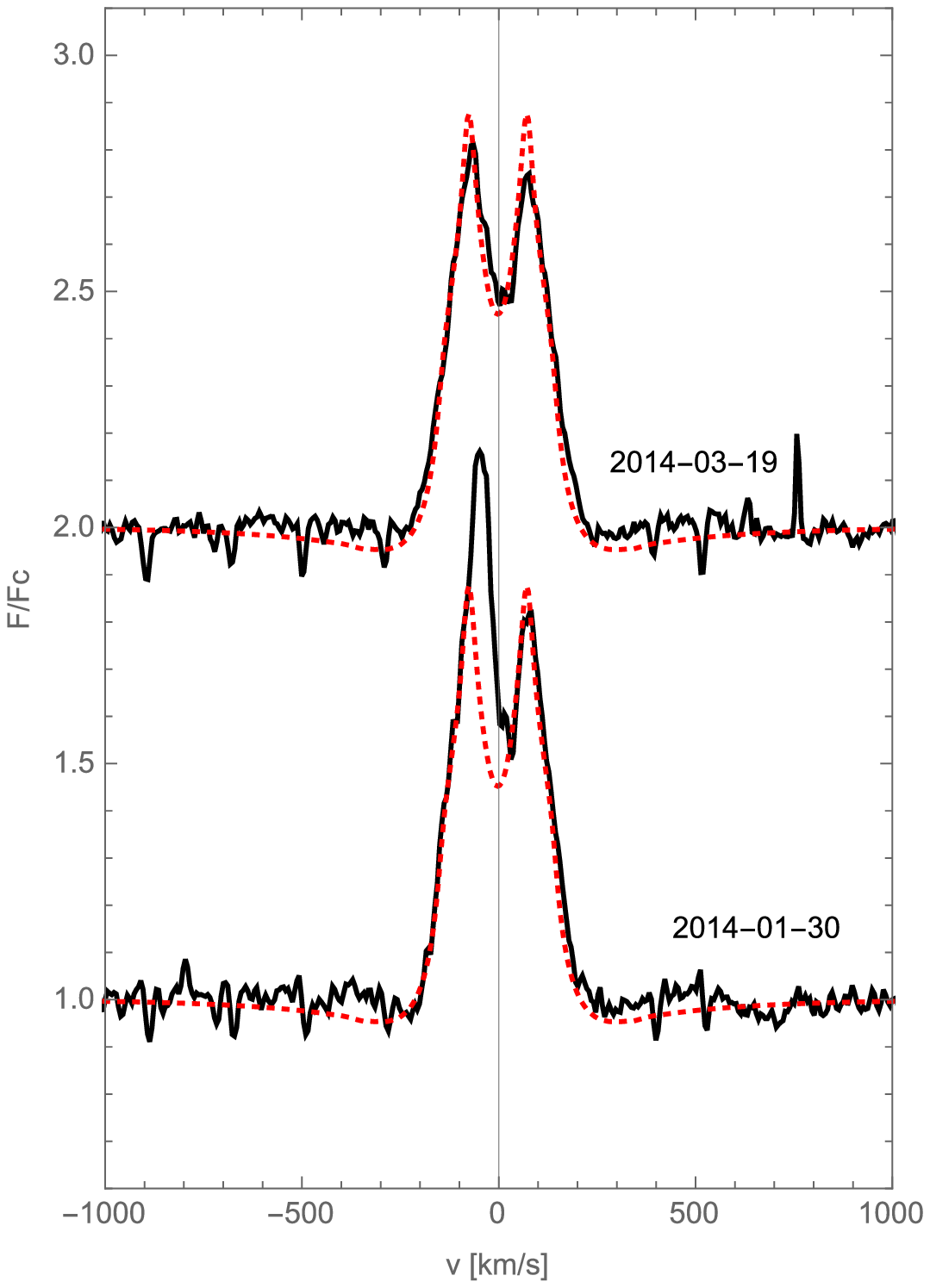}{0.2\textheight}{HD78764}
                   \fig{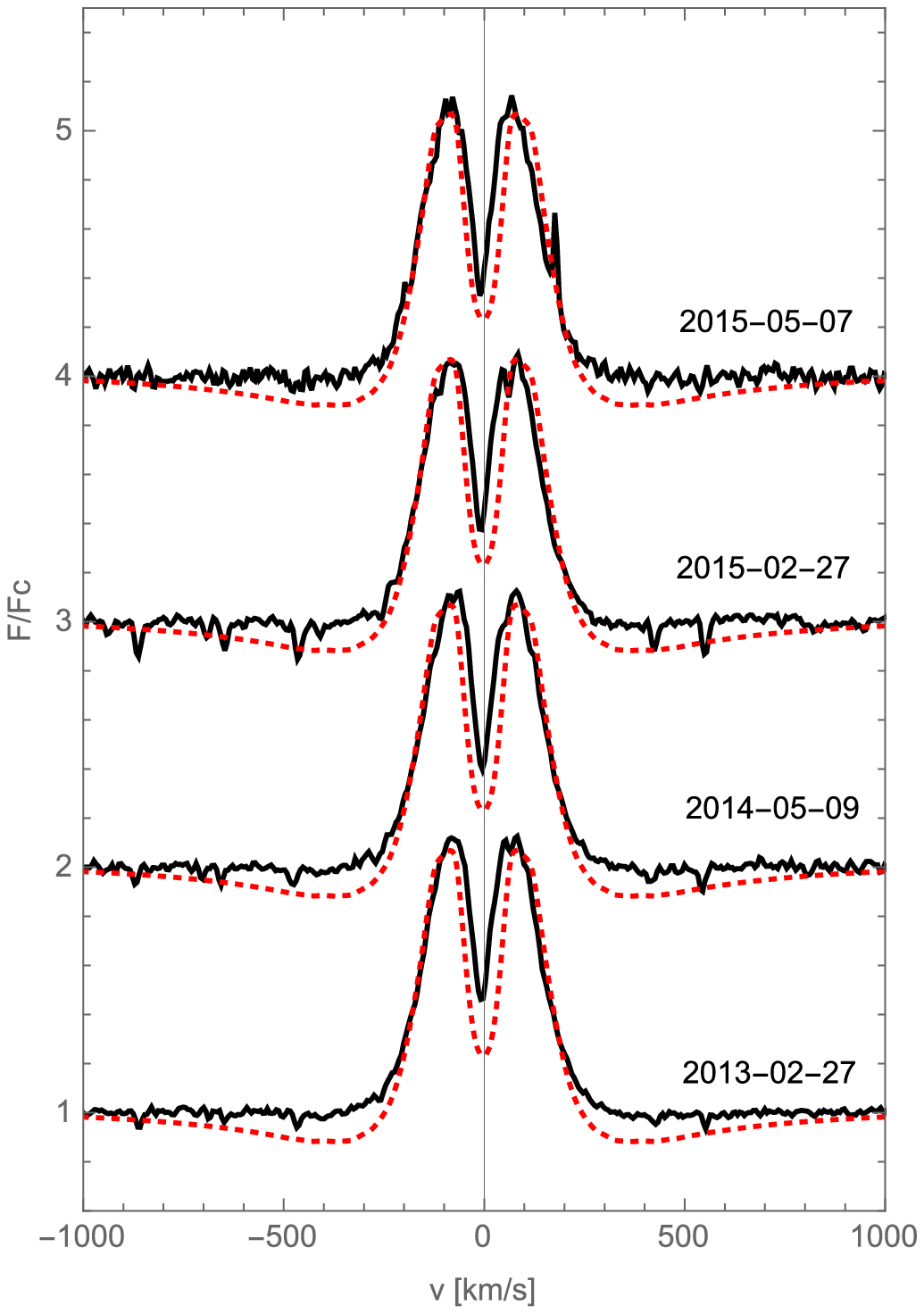}{0.2\textheight}{HD89080}
                   \fig{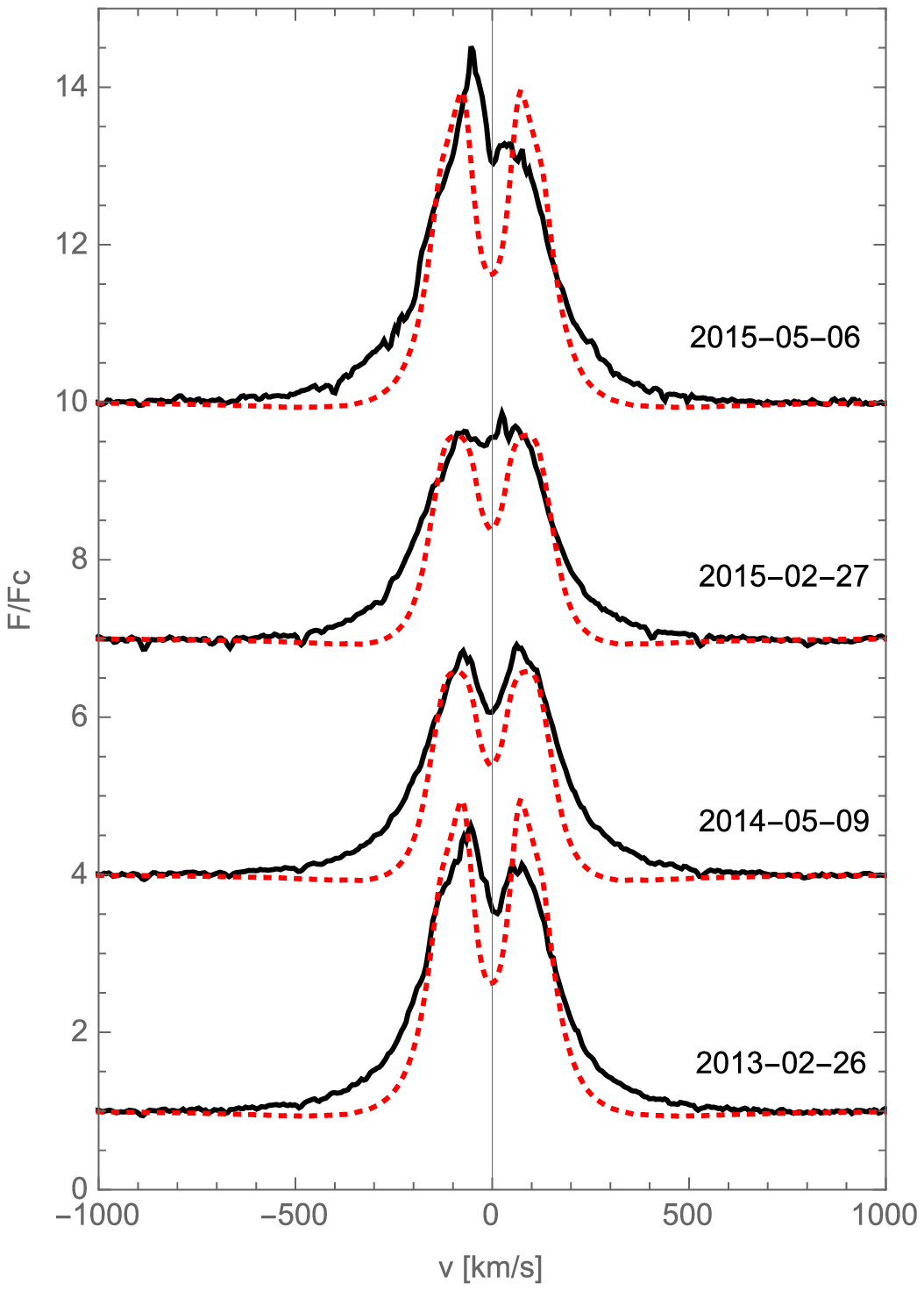}{0.2\textheight}{HD91465*}
          }

\end{figure}

\begin{figure}
            \gridline{
                    \fig{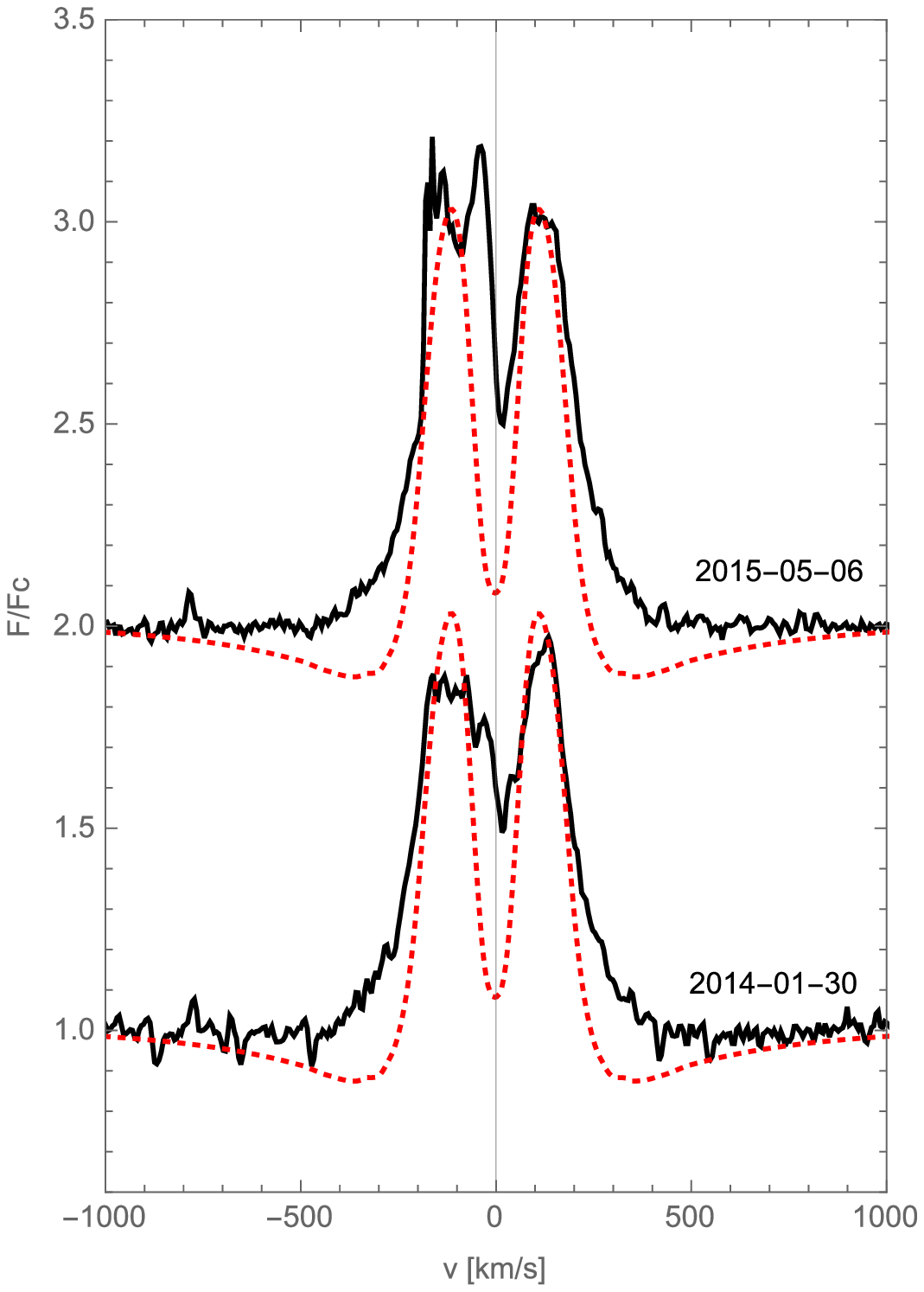}{0.2\textheight}{HD93563}
                    \fig{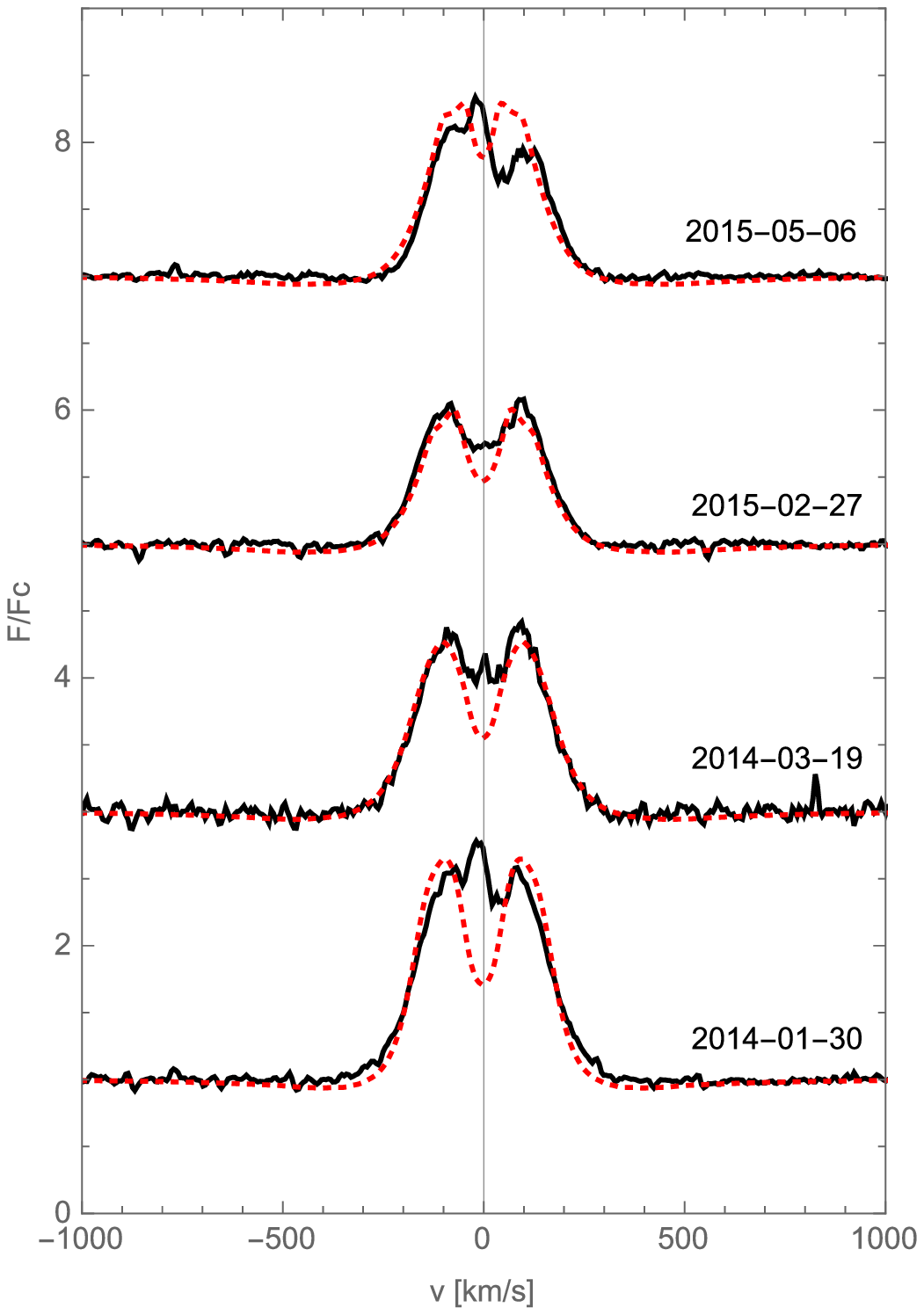}{0.2\textheight}{HD102776*}
                    \fig{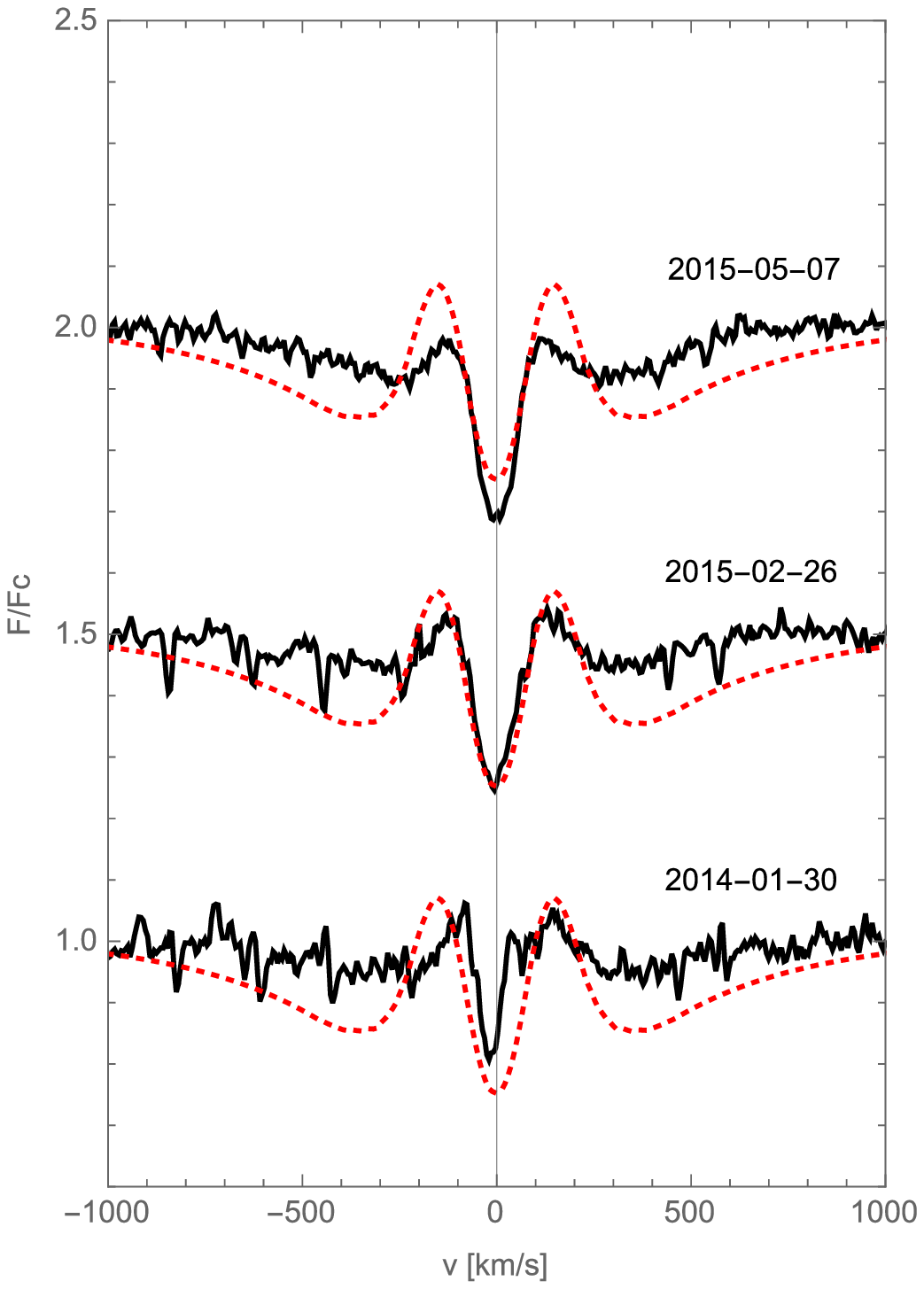}{0.2\textheight}{HD103192}
          }
           
           \gridline{
                     \fig{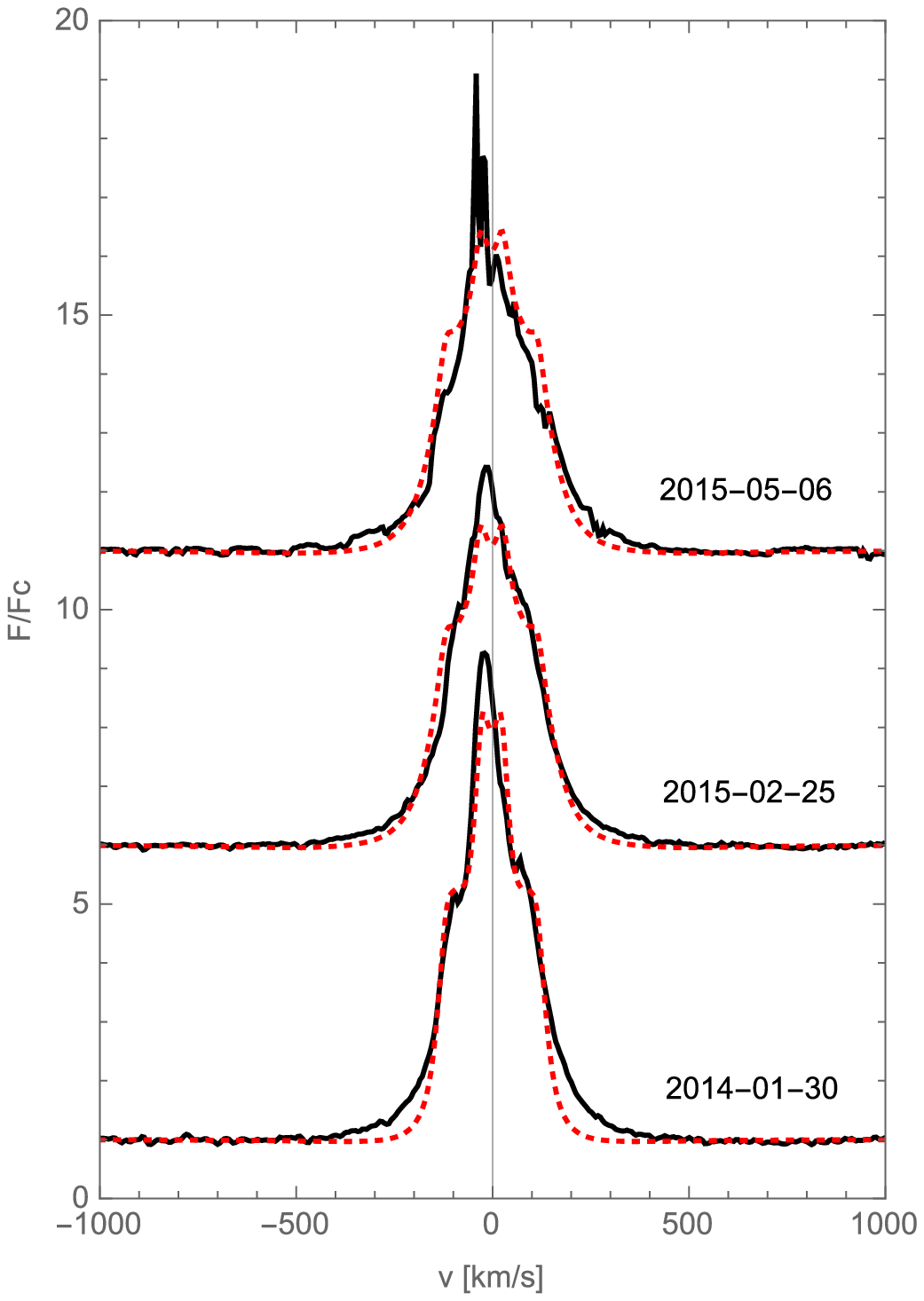}{0.2\textheight}{HD105435*}
                     \fig{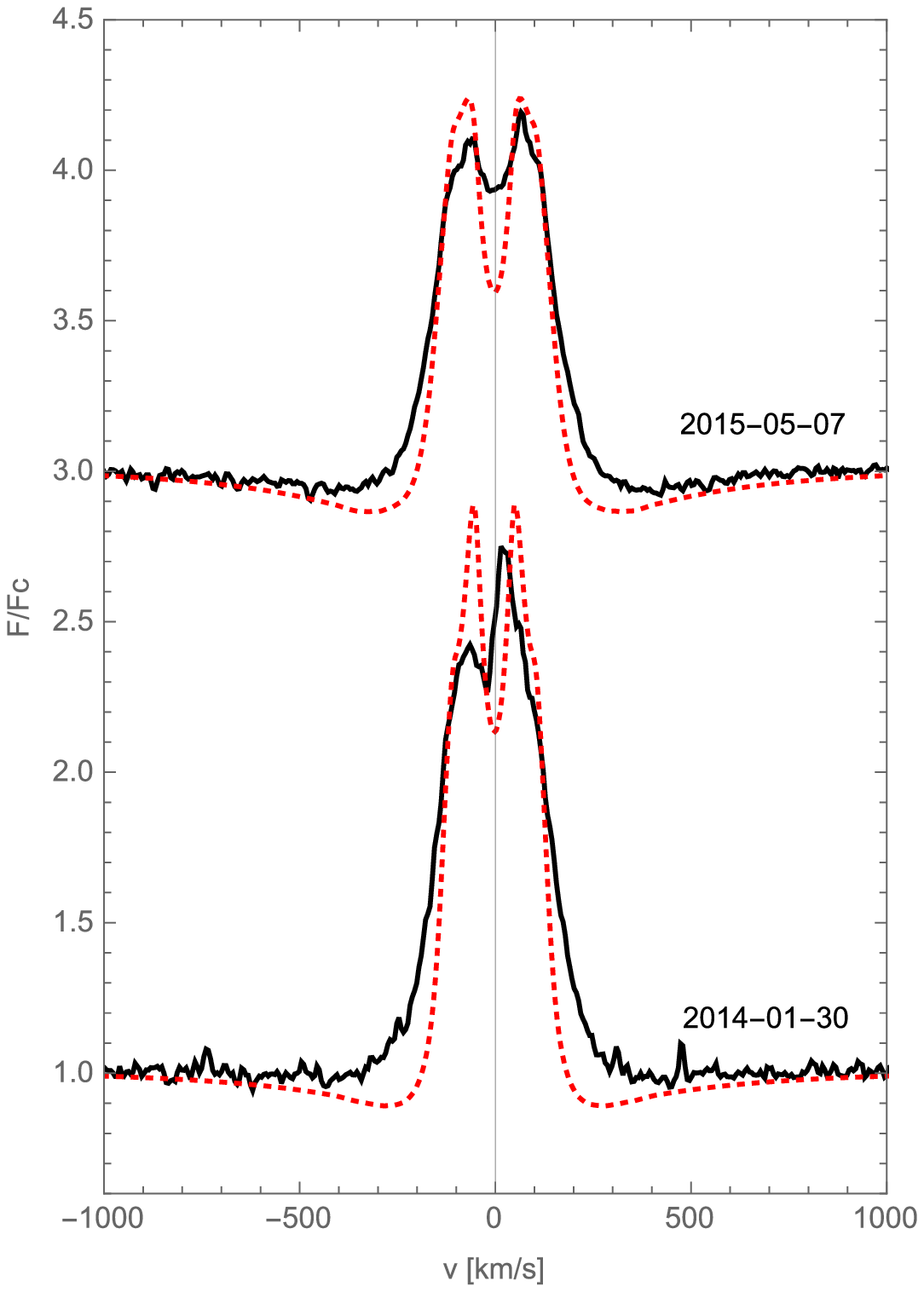}{0.2\textheight}{HD107348*}
                     \fig{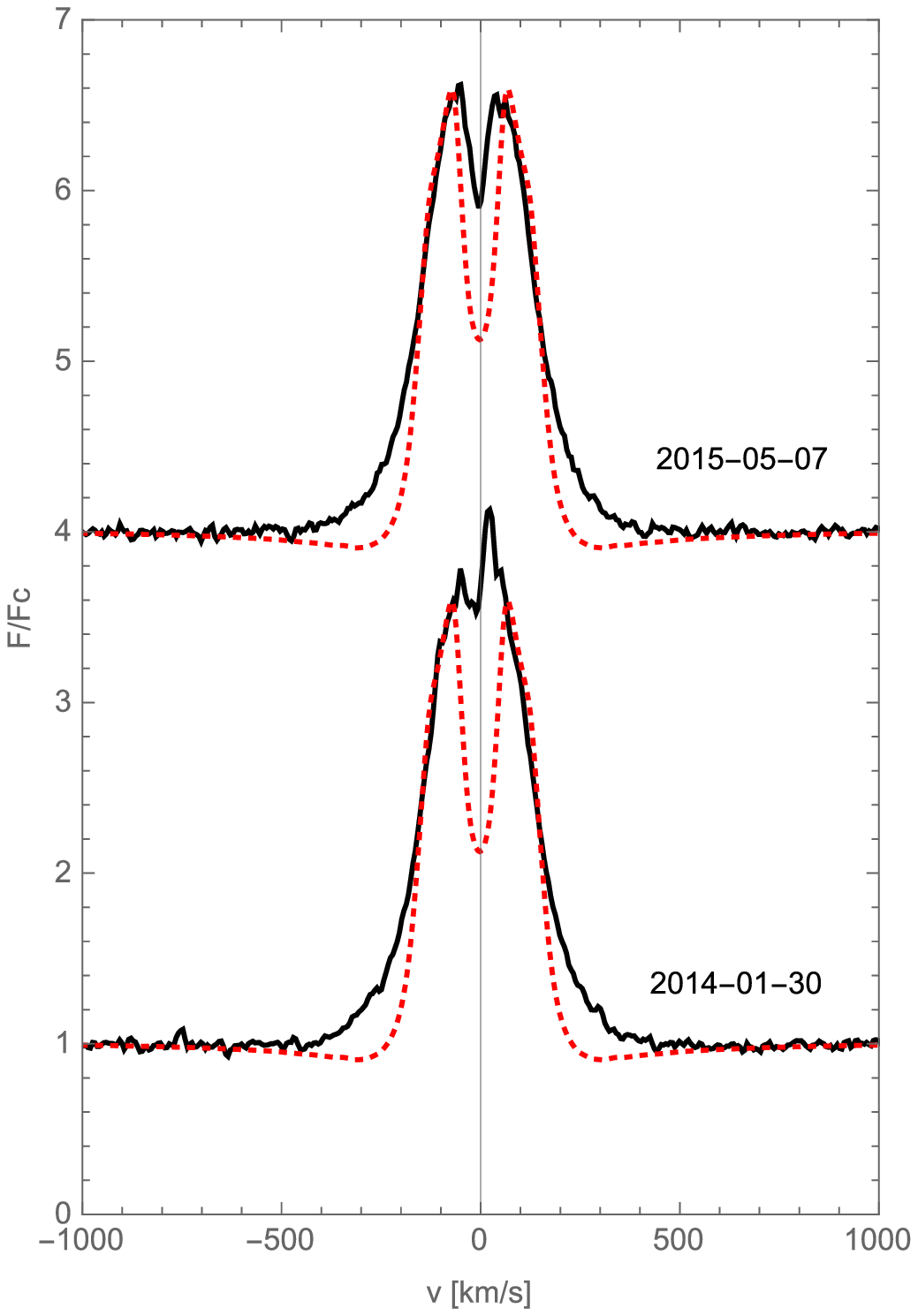}{0.2\textheight}{HD110335}
          }
         
         \gridline{
                   \fig{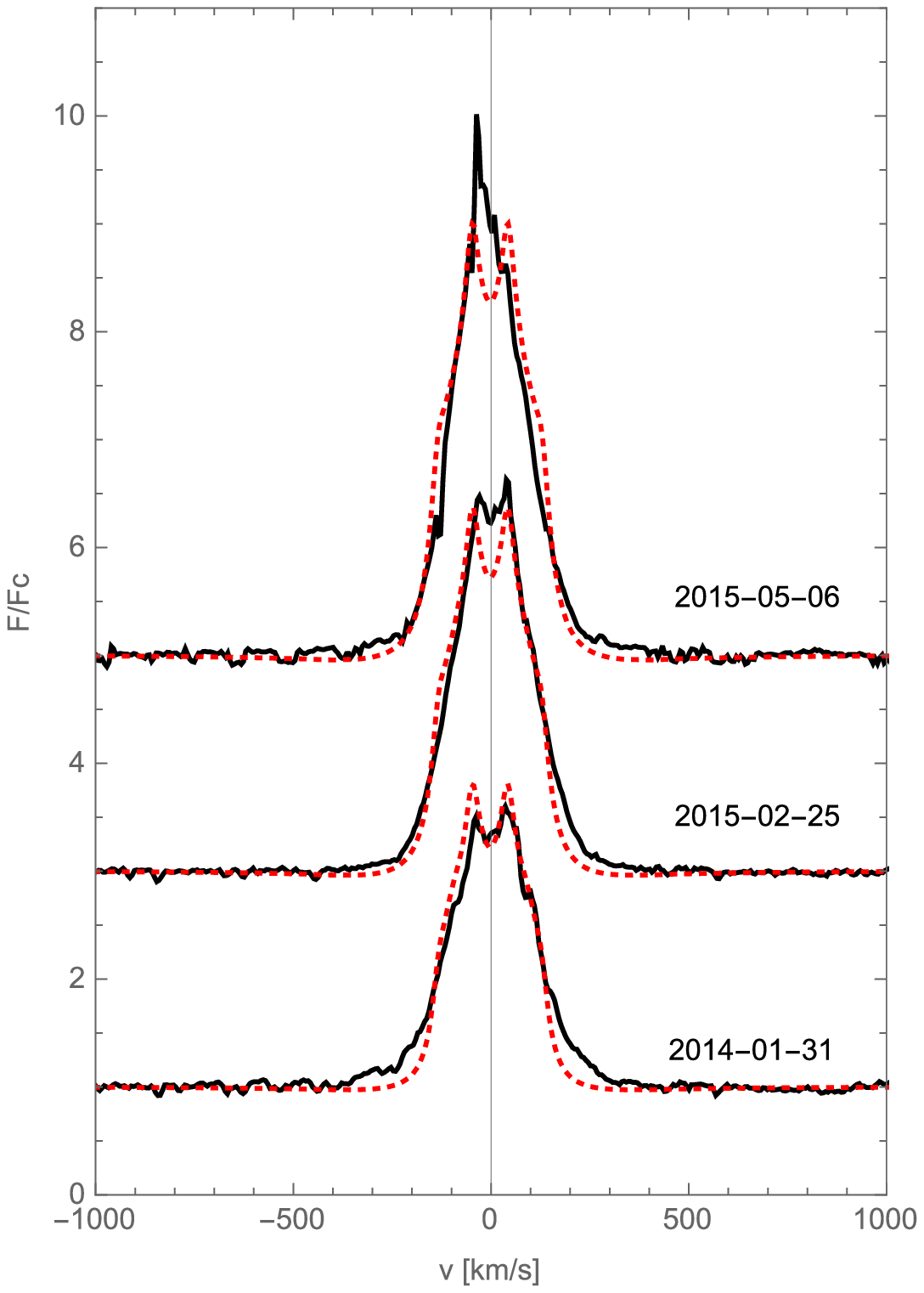}{0.2\textheight}{HD120324*}
                   \fig{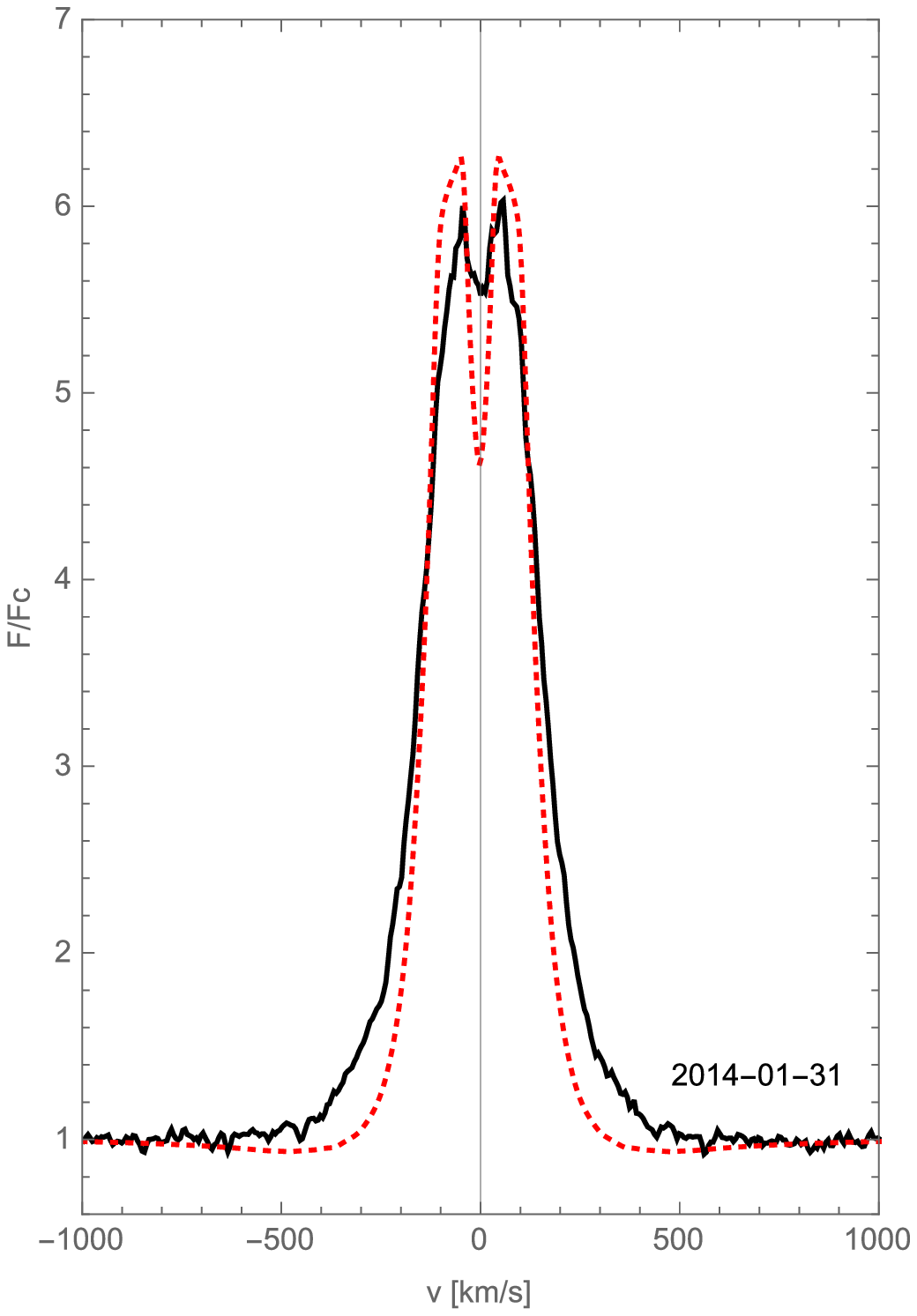}{0.2\textheight}{HD124367}
                   \fig{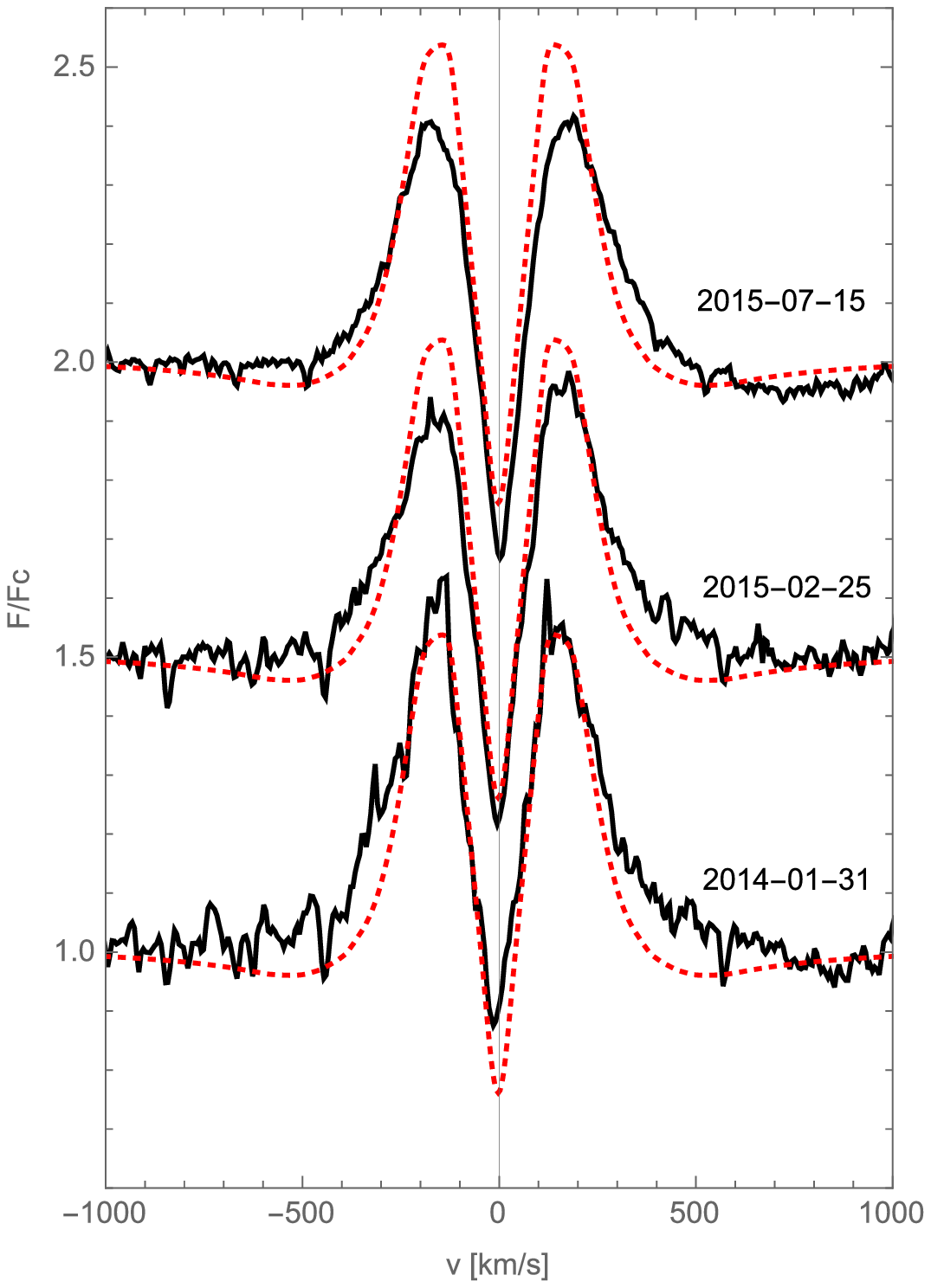}{0.2\textheight}{HD127972}
          }
          
\end{figure}

\begin{figure}
          \gridline{\fig{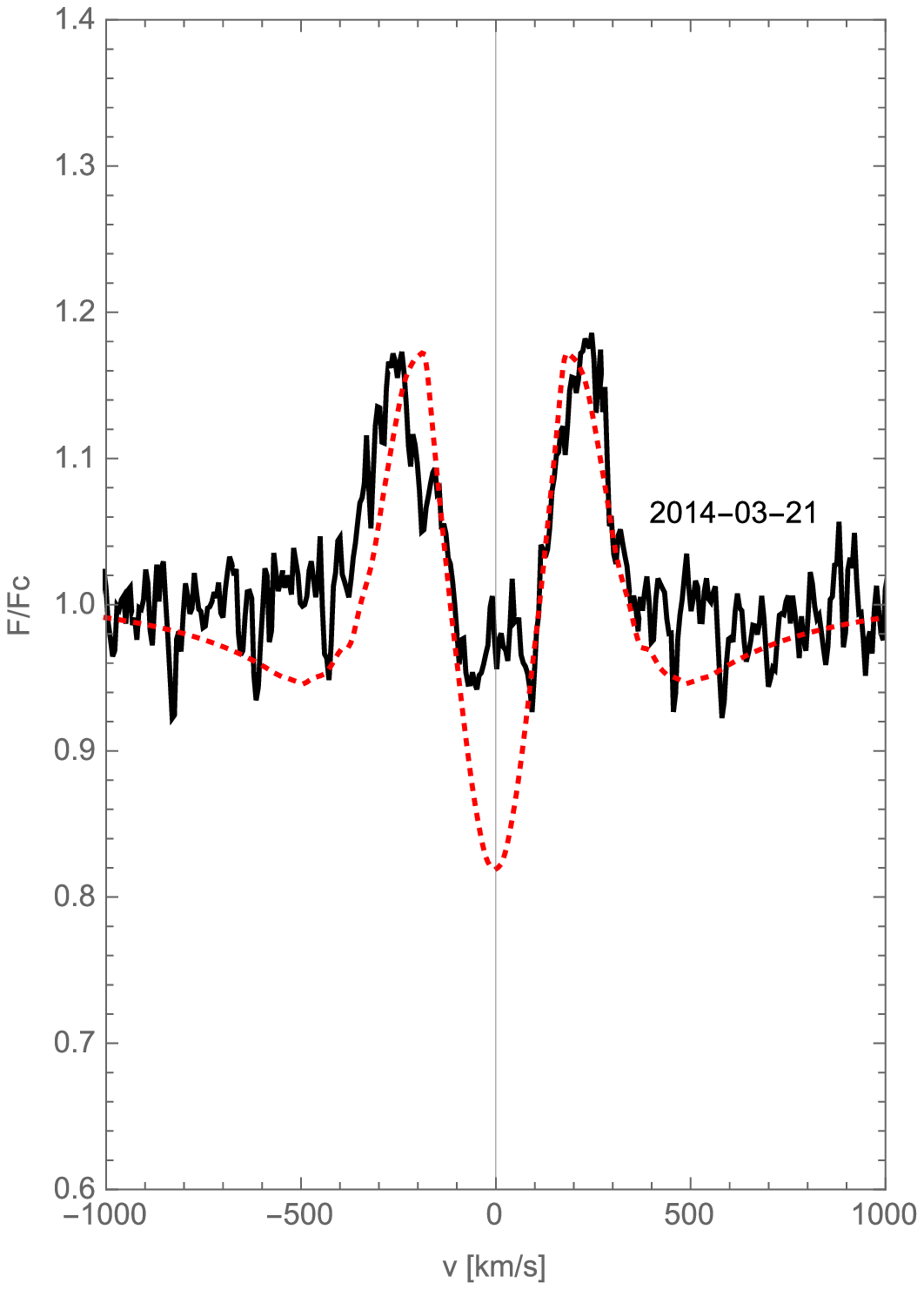}{0.2\textheight}{HD131492}
                    \fig{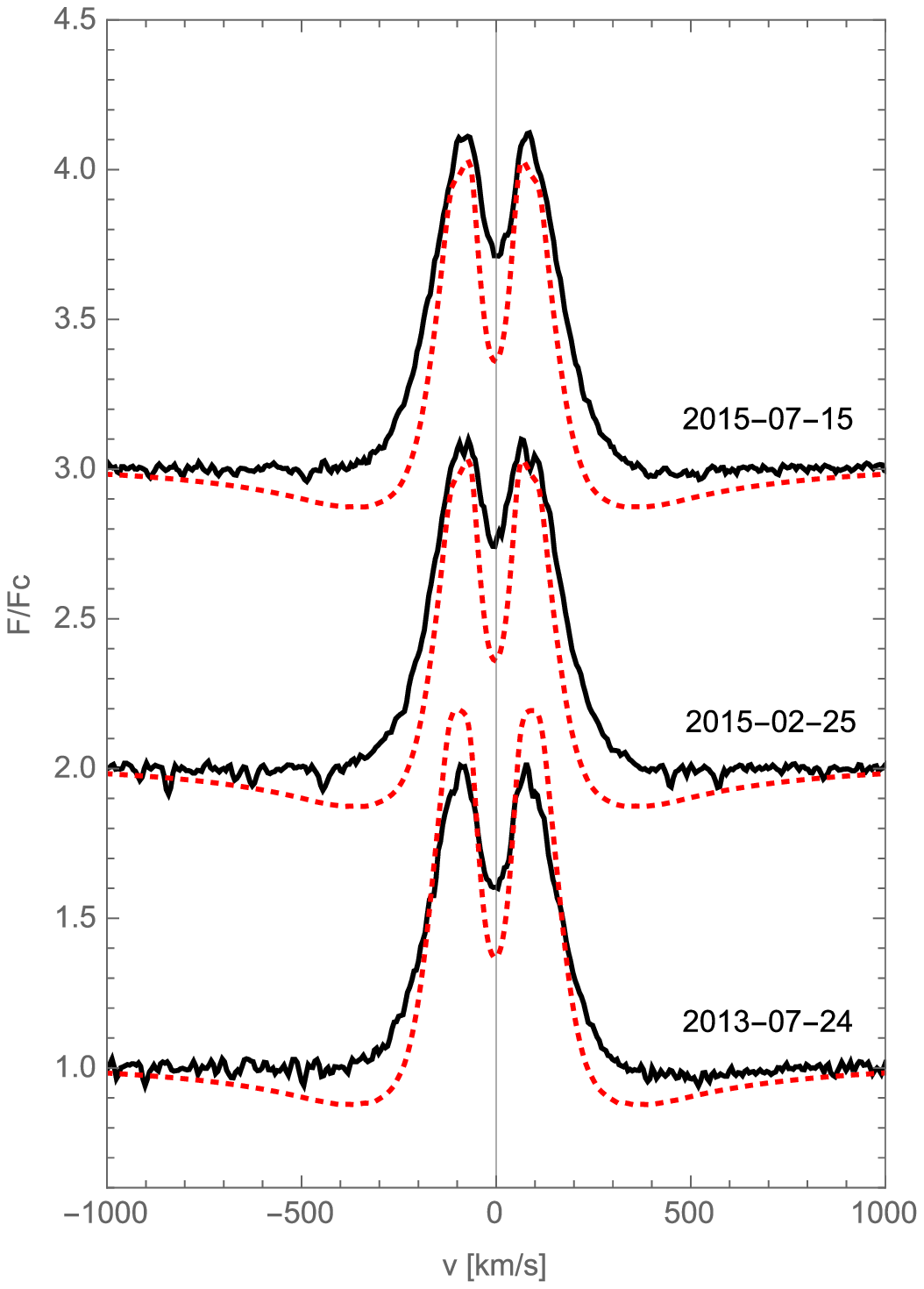}{0.2\textheight}{HD135734*}
                    \fig{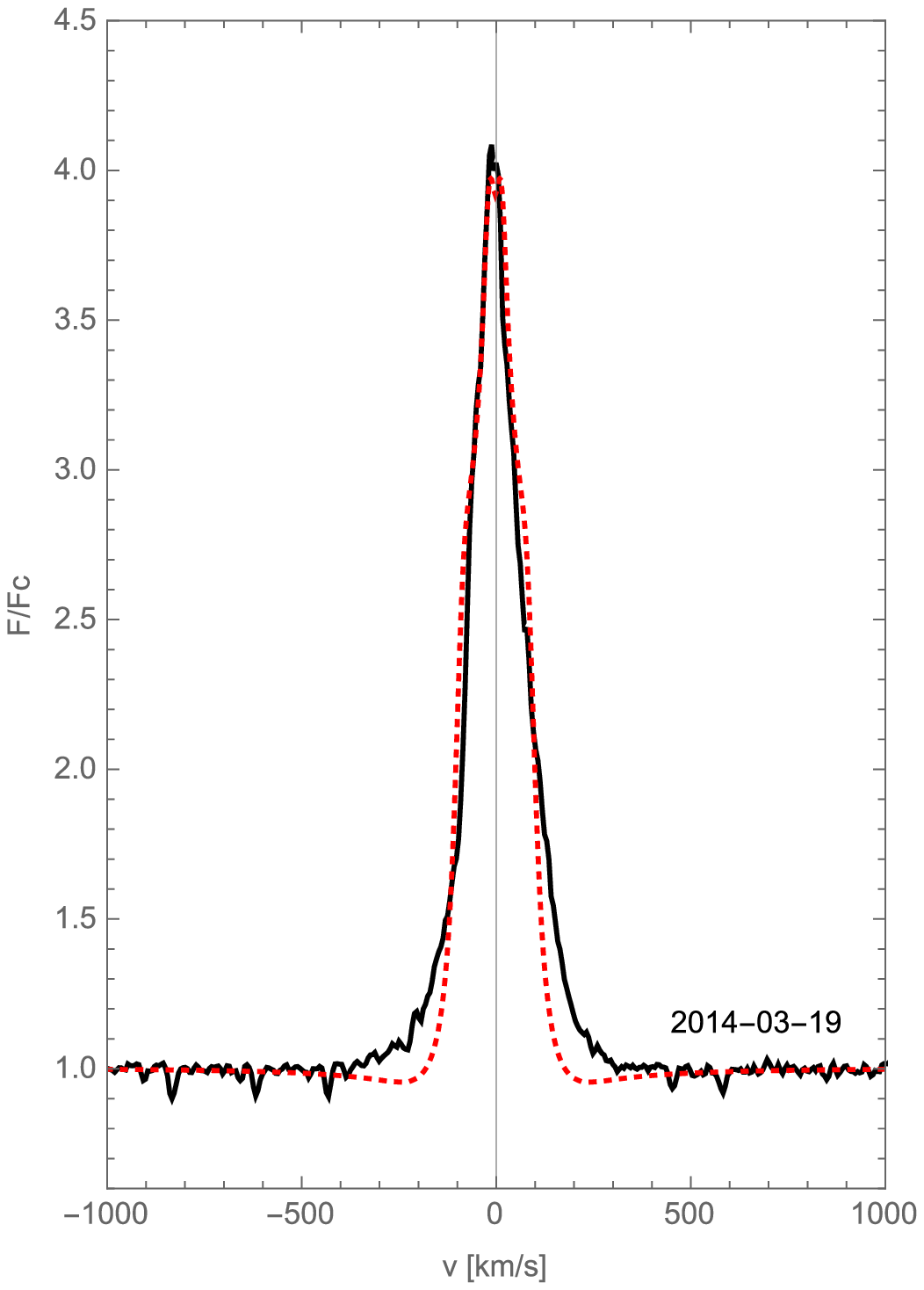}{0.2\textheight}{HD143275}
          }

          \gridline{\fig{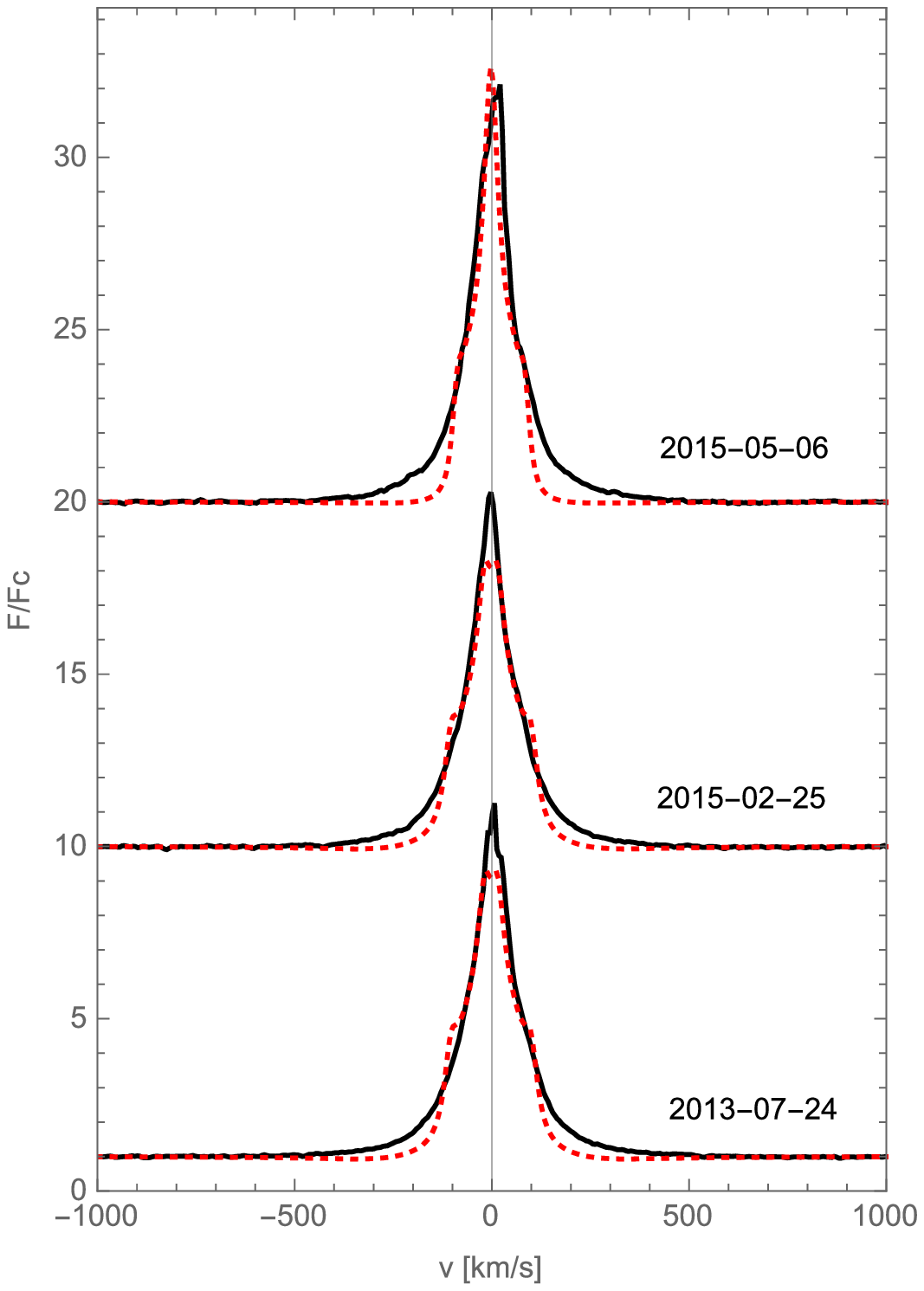}{0.2\textheight}{HD148184}
                    \fig{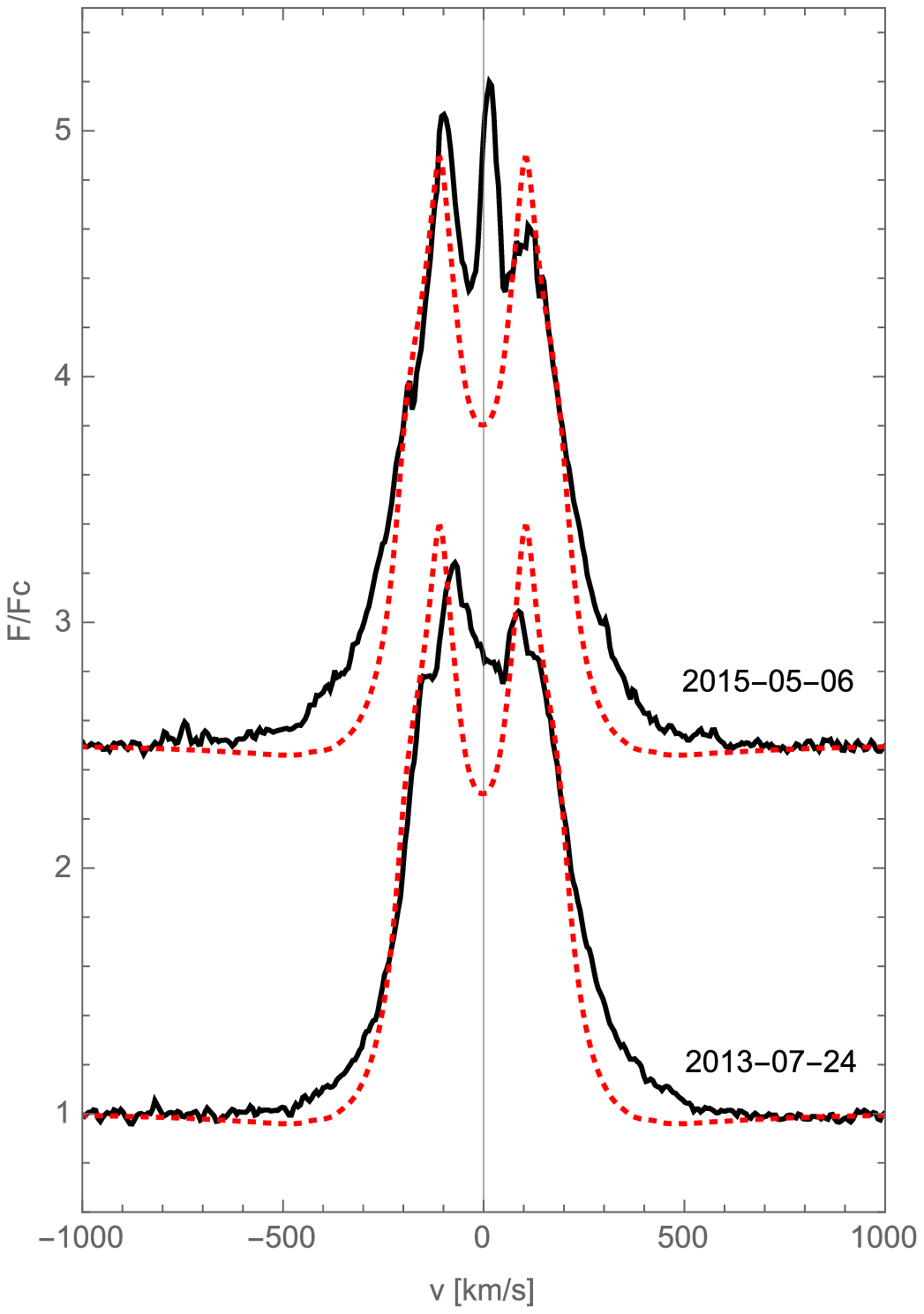}{0.2\textheight}{HD157042}
                    \fig{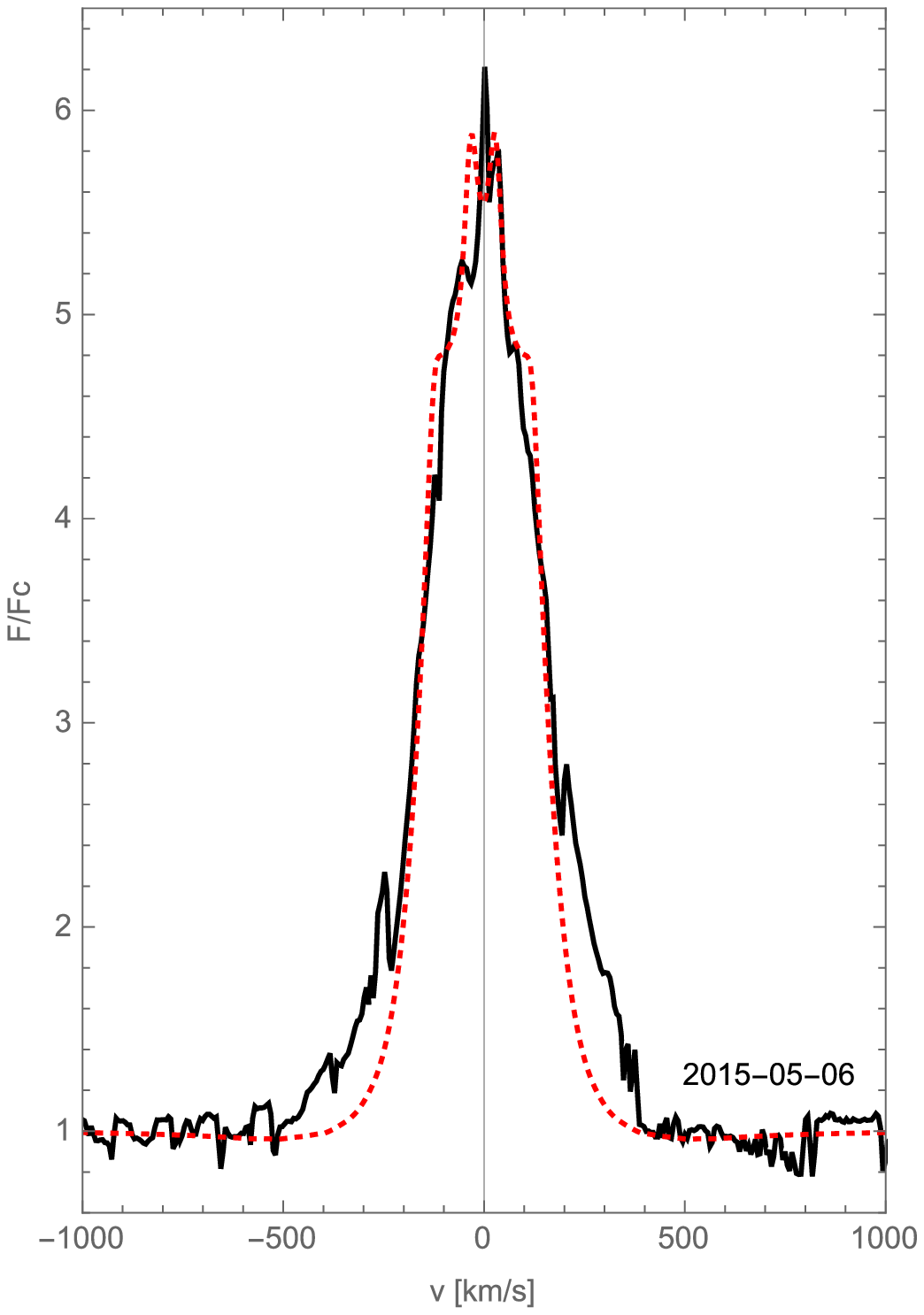}{0.2\textheight}{HD158427}
          }
          
           \gridline{\fig{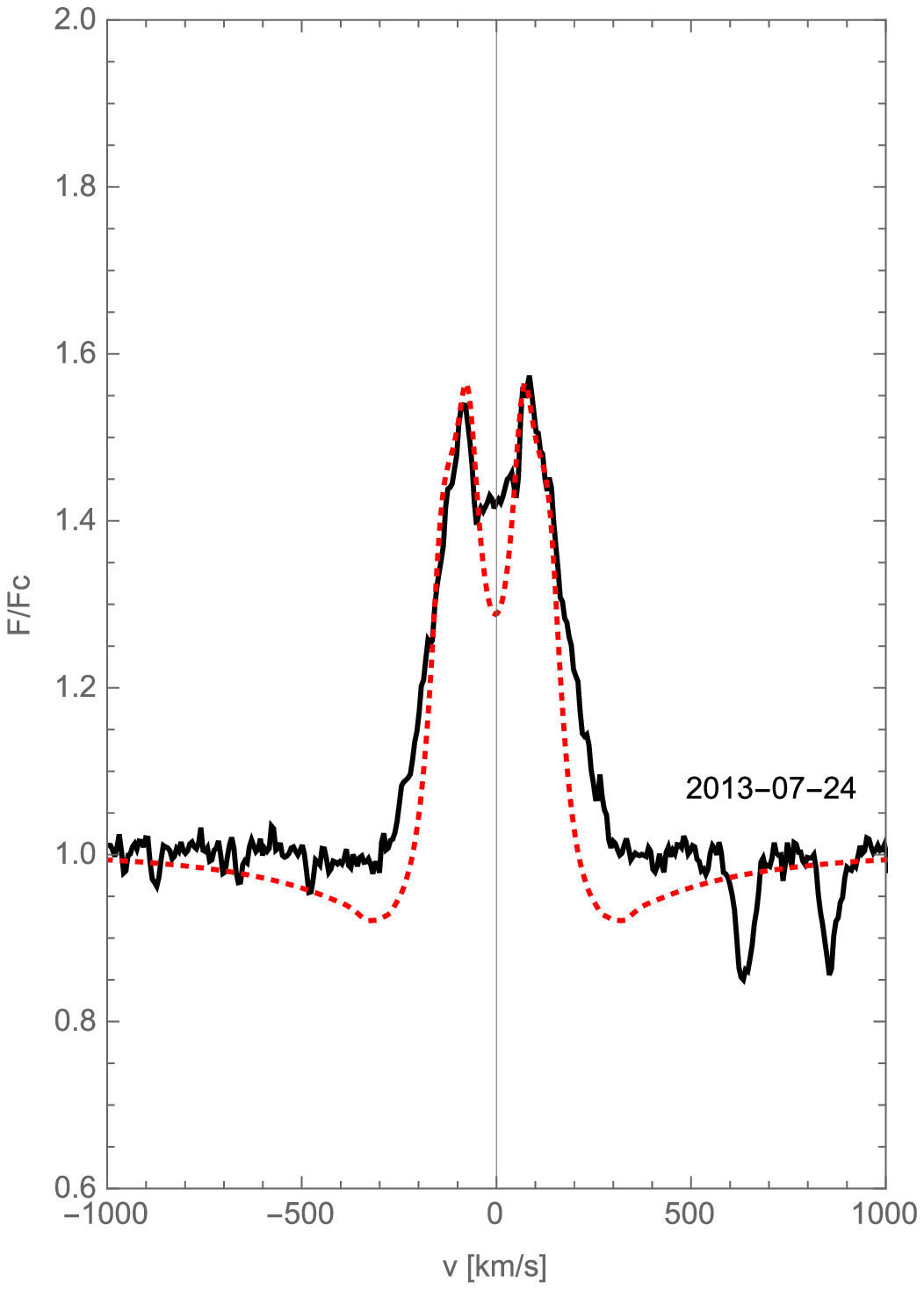}{0.2\textheight}{HD167128}
                     \fig{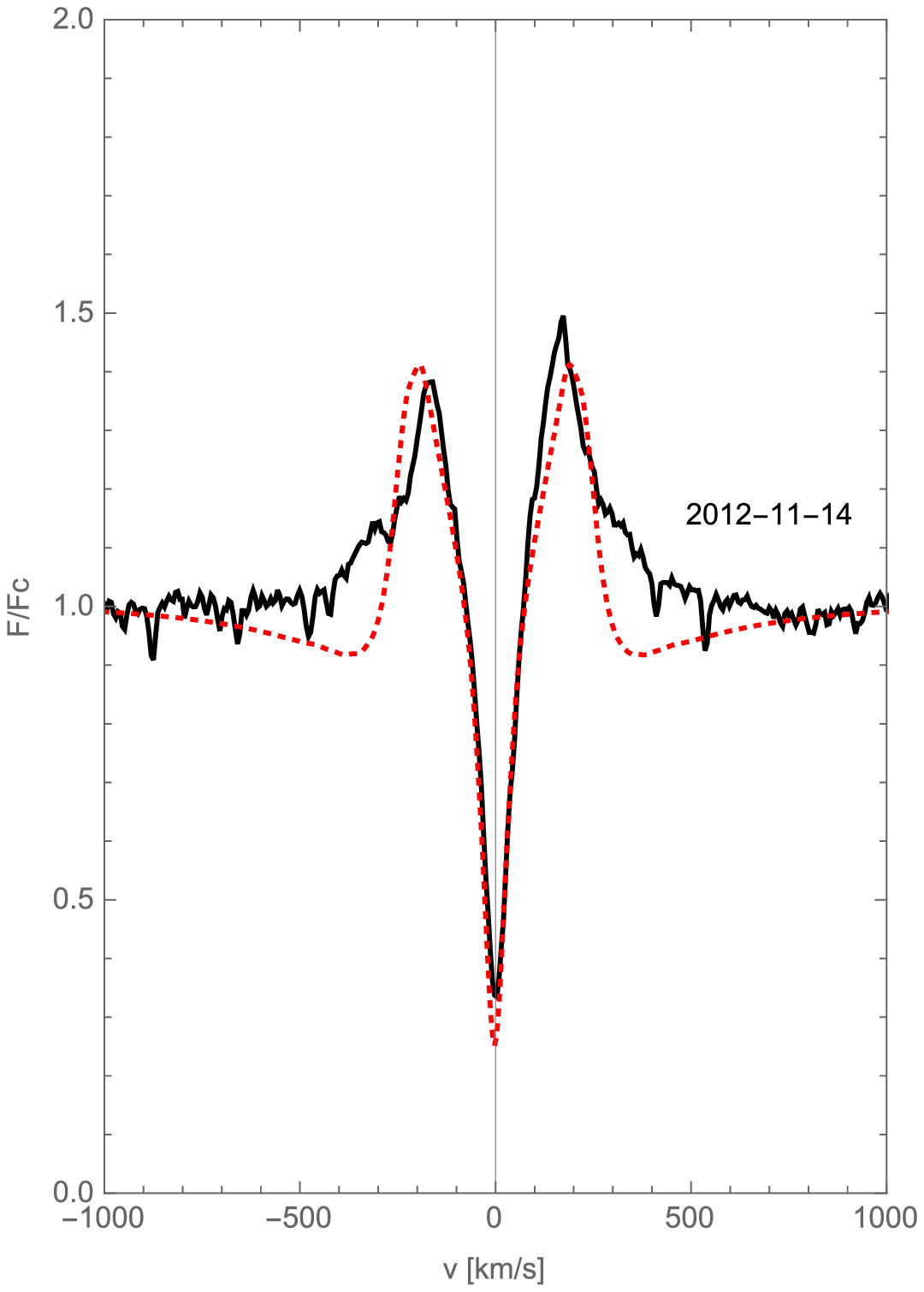}{0.2\textheight}{HD205637}
                     \fig{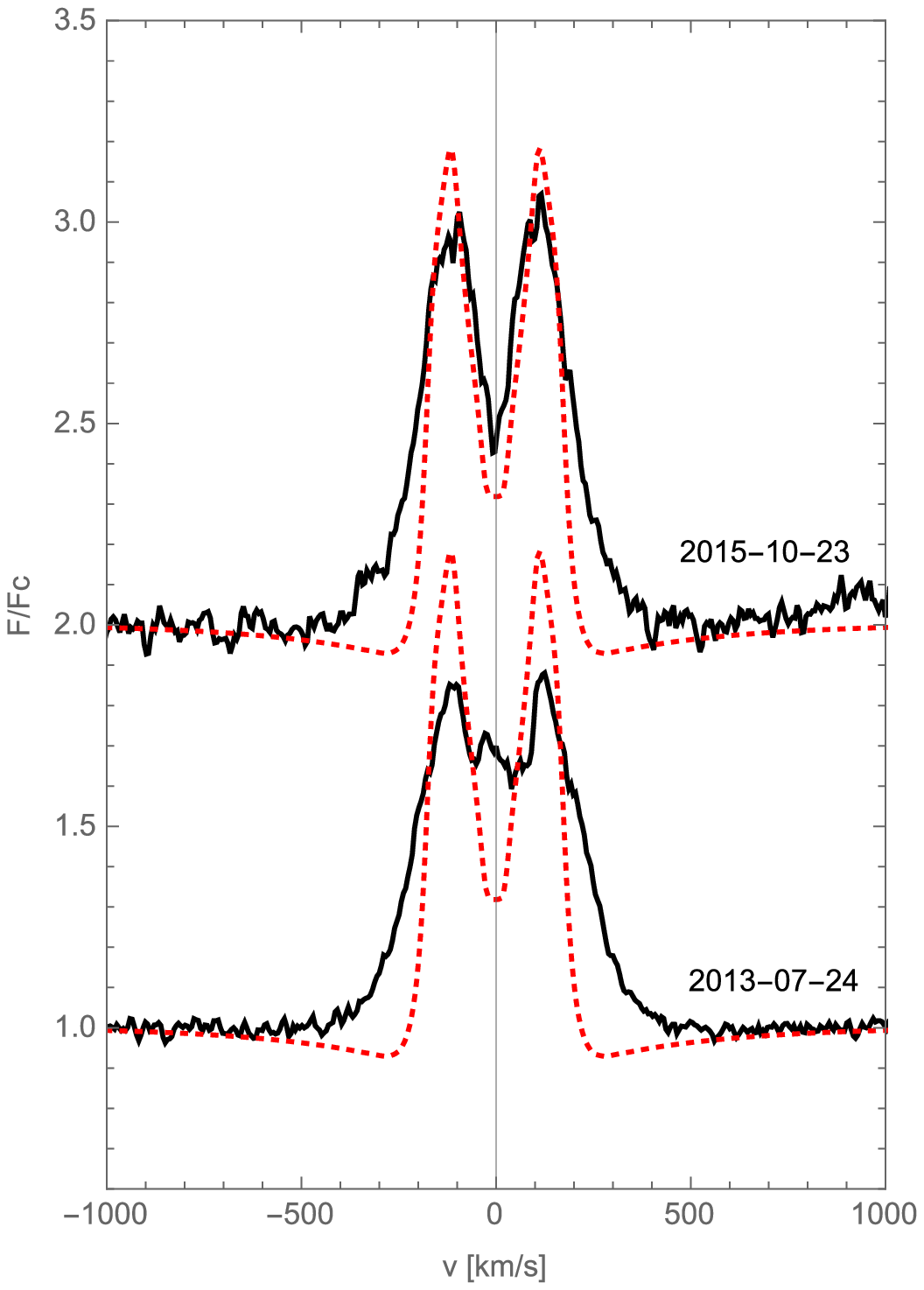}{0.2\textheight}{HD209014}
          }
          
\end{figure}

\begin{figure}
          \gridline{\fig{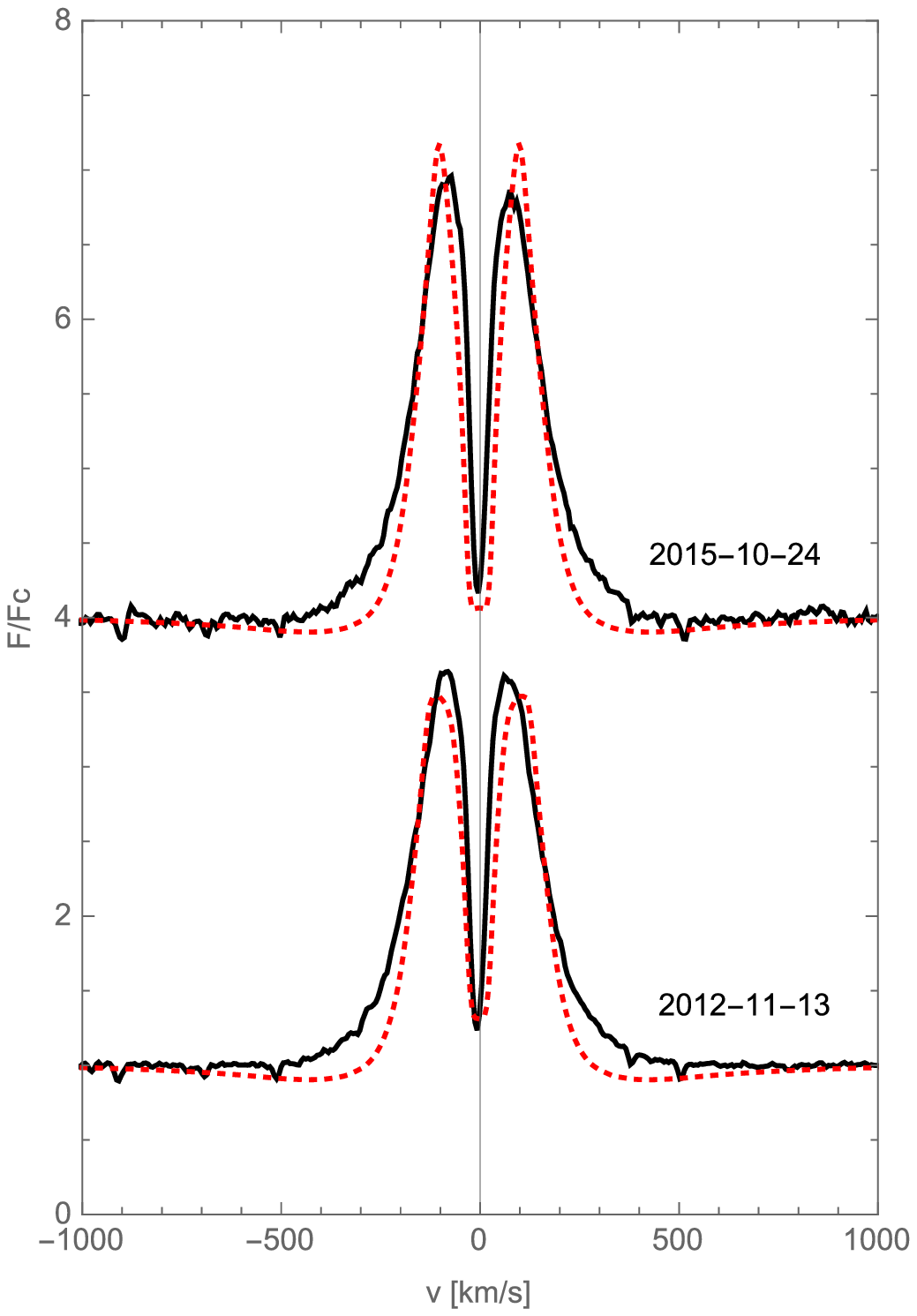}{0.2\textheight}{HD209409*}
                    \fig{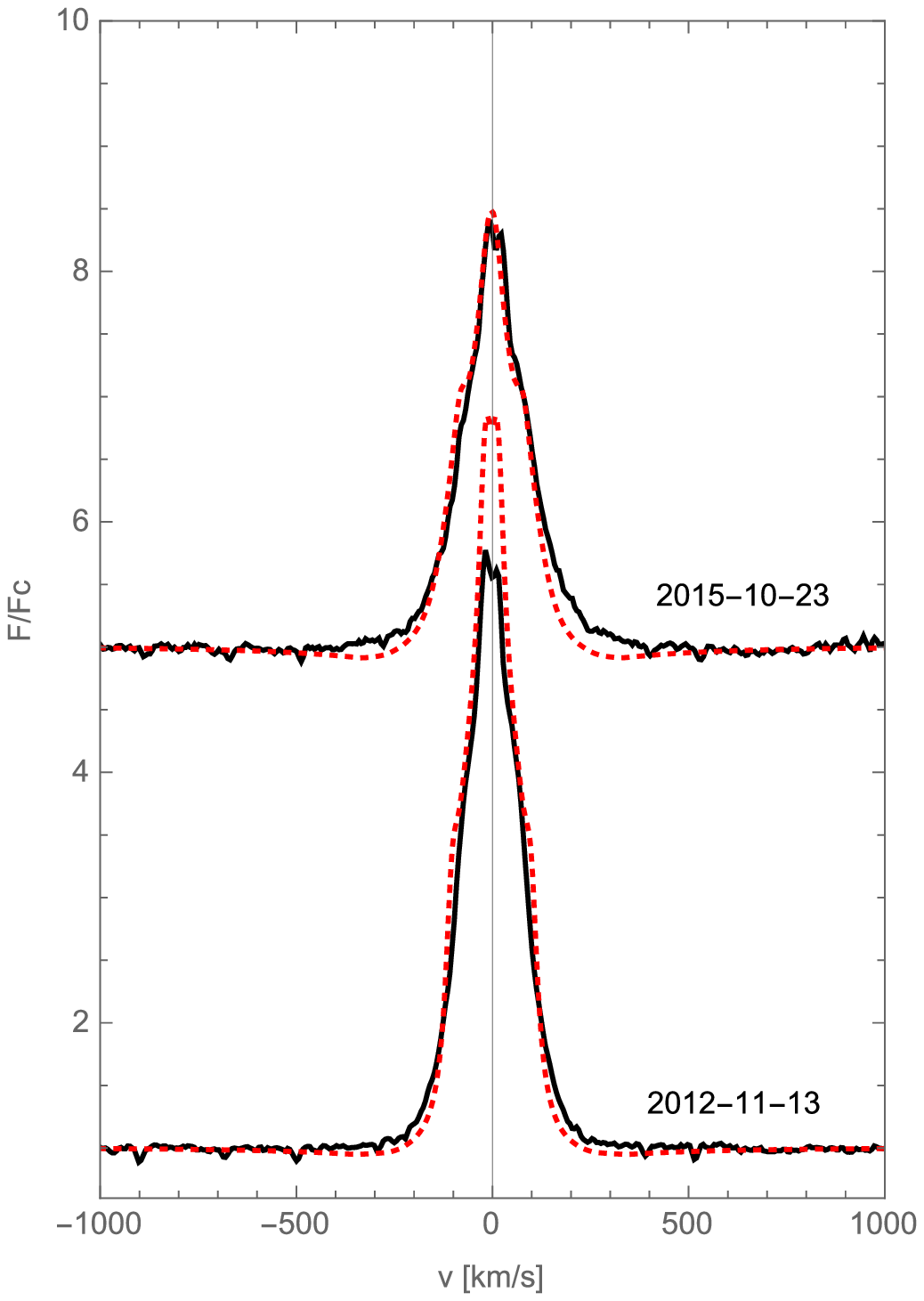}{0.2\textheight}{HD212076*}
                    \fig{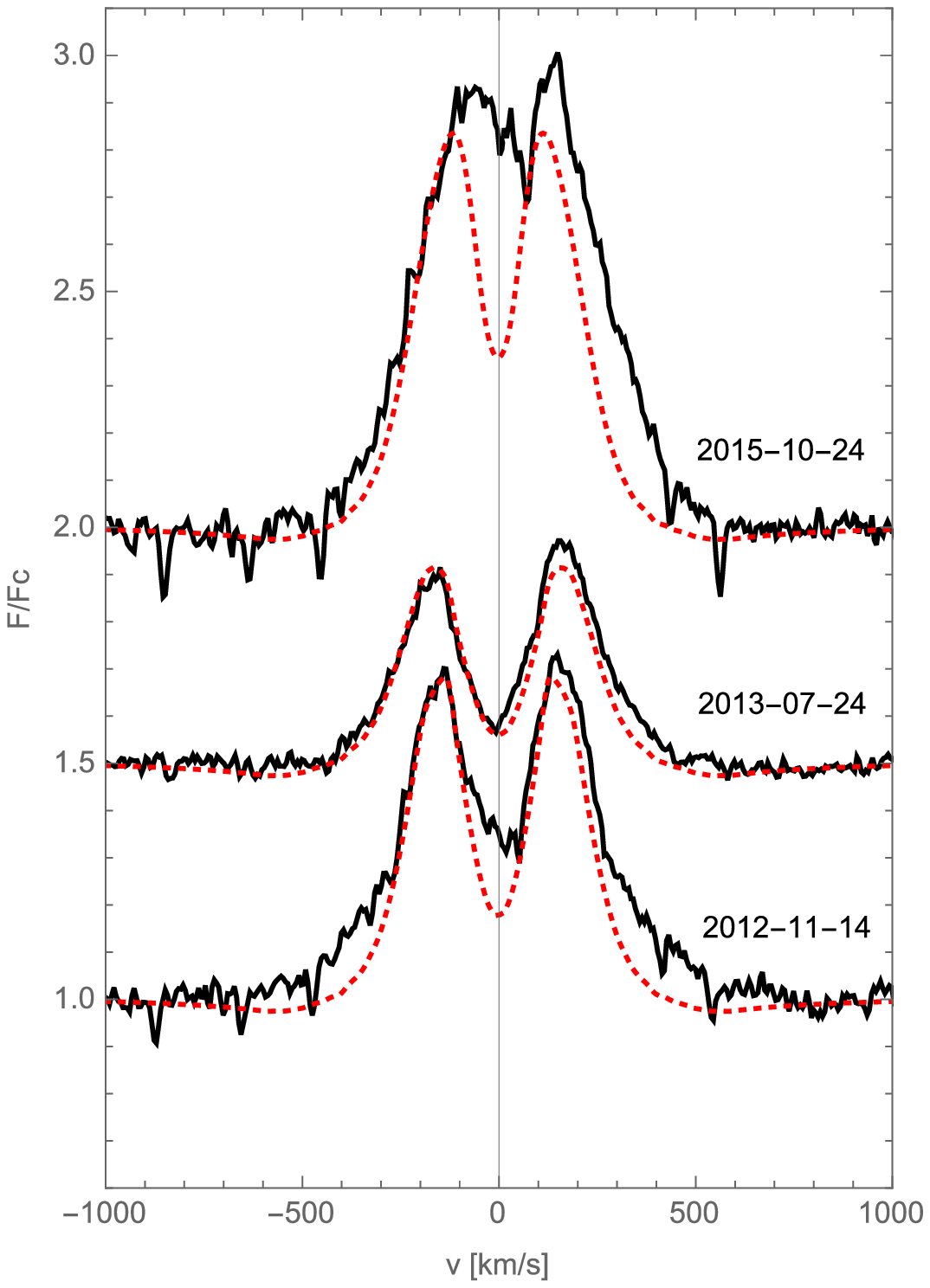}{0.2\textheight}{HD212571*}
          }
          
          \gridline{\fig{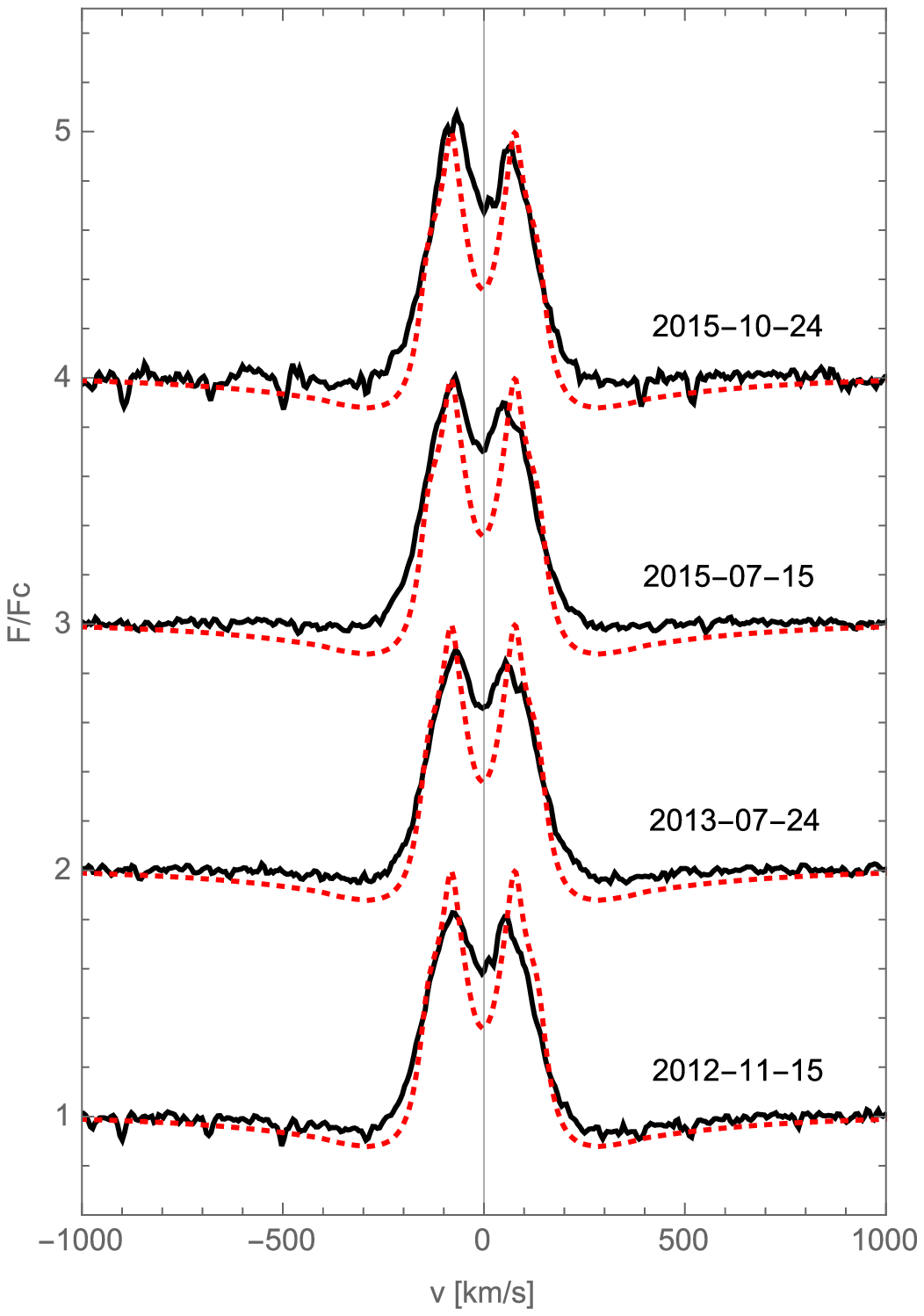}{0.2\textheight}{HD214748}
                    \fig{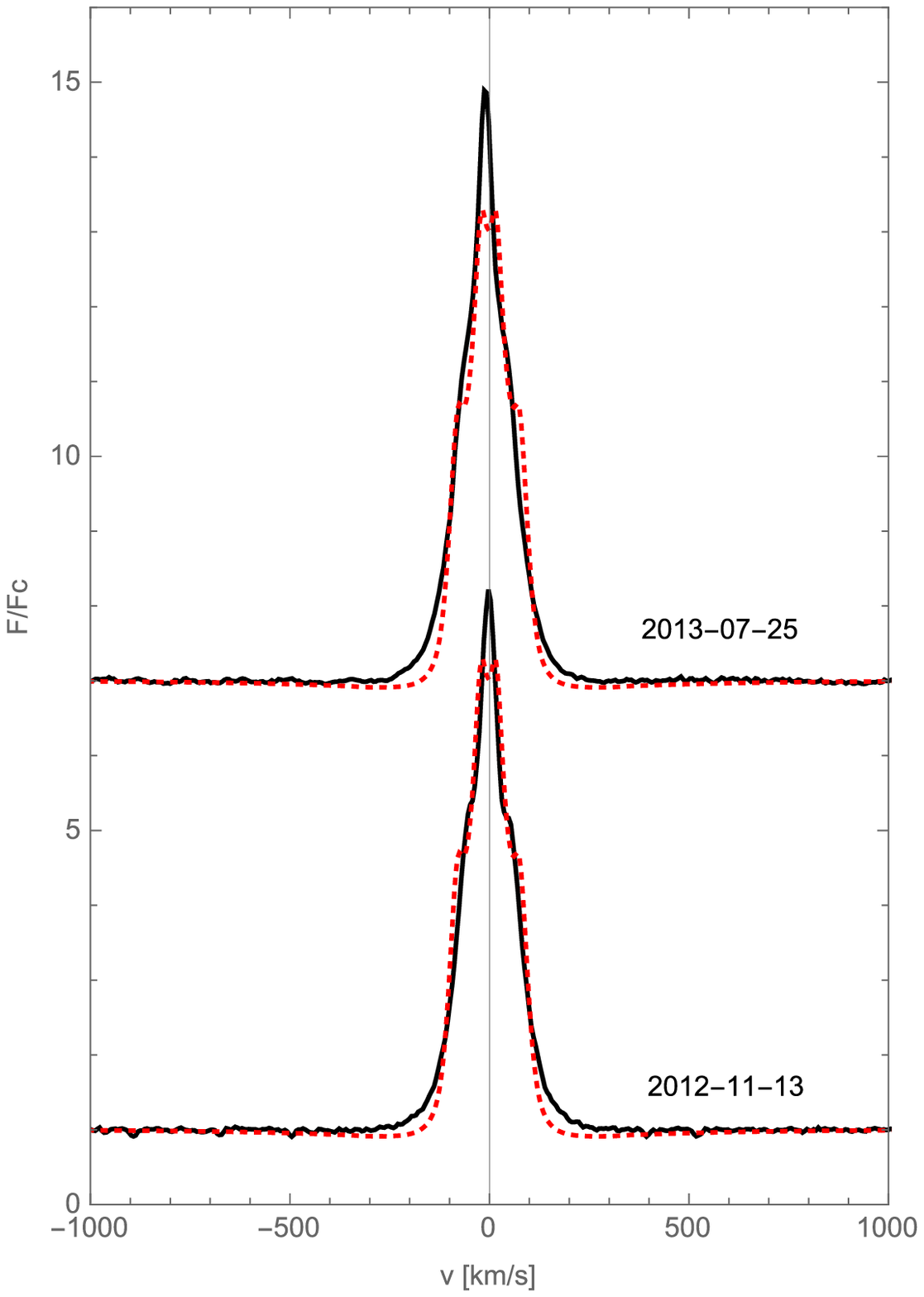}{0.2\textheight}{HD217891}   
                    \fig{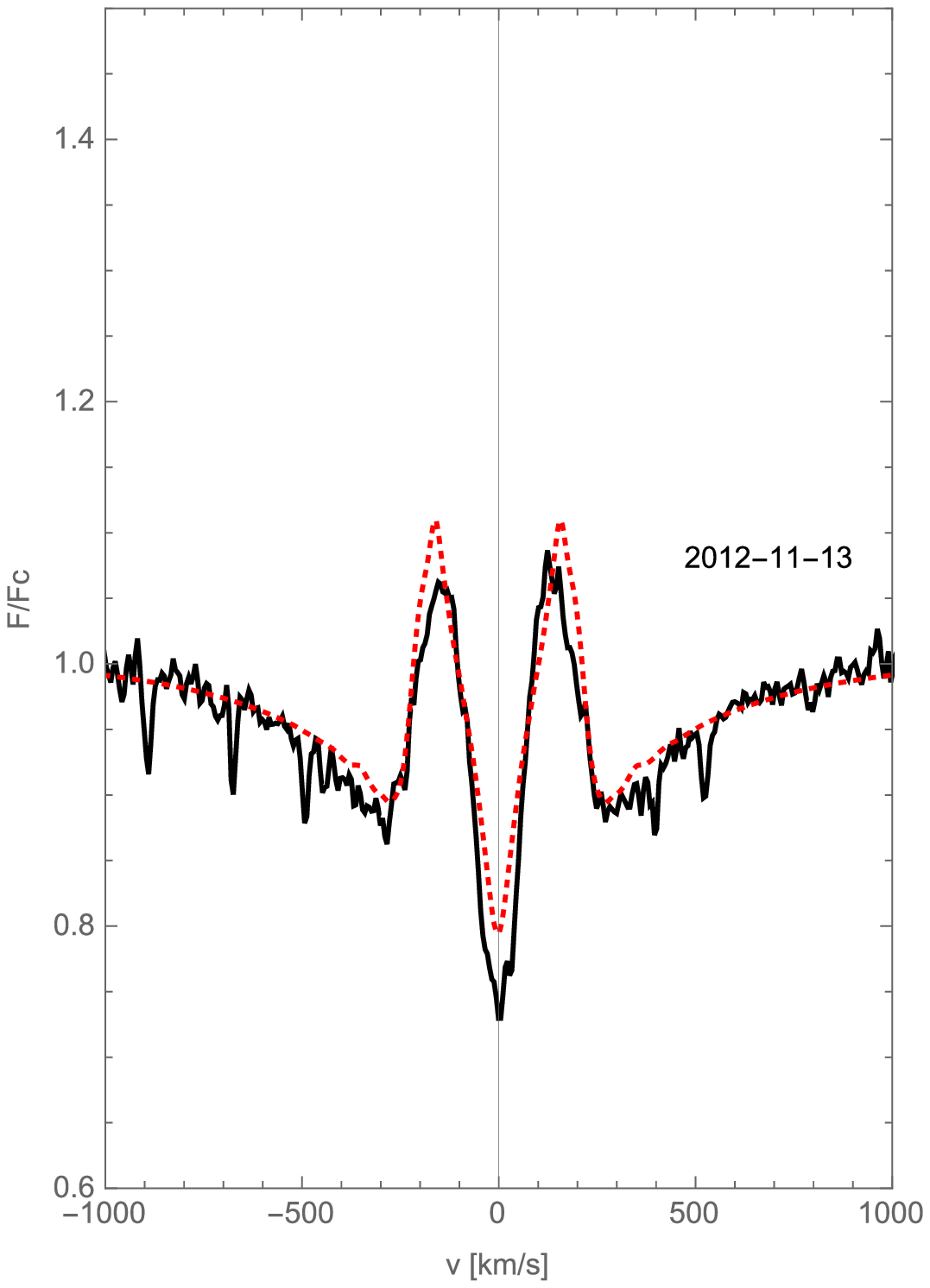}{0.2\textheight}{HD224686} 
          }
          
\end{figure}

\newpage
\section{Absorption profiles} \label{ApB}
The same as Appendix~\ref{ApA} except for absorption profiles.
\begin{figure}

           \gridline{\fig{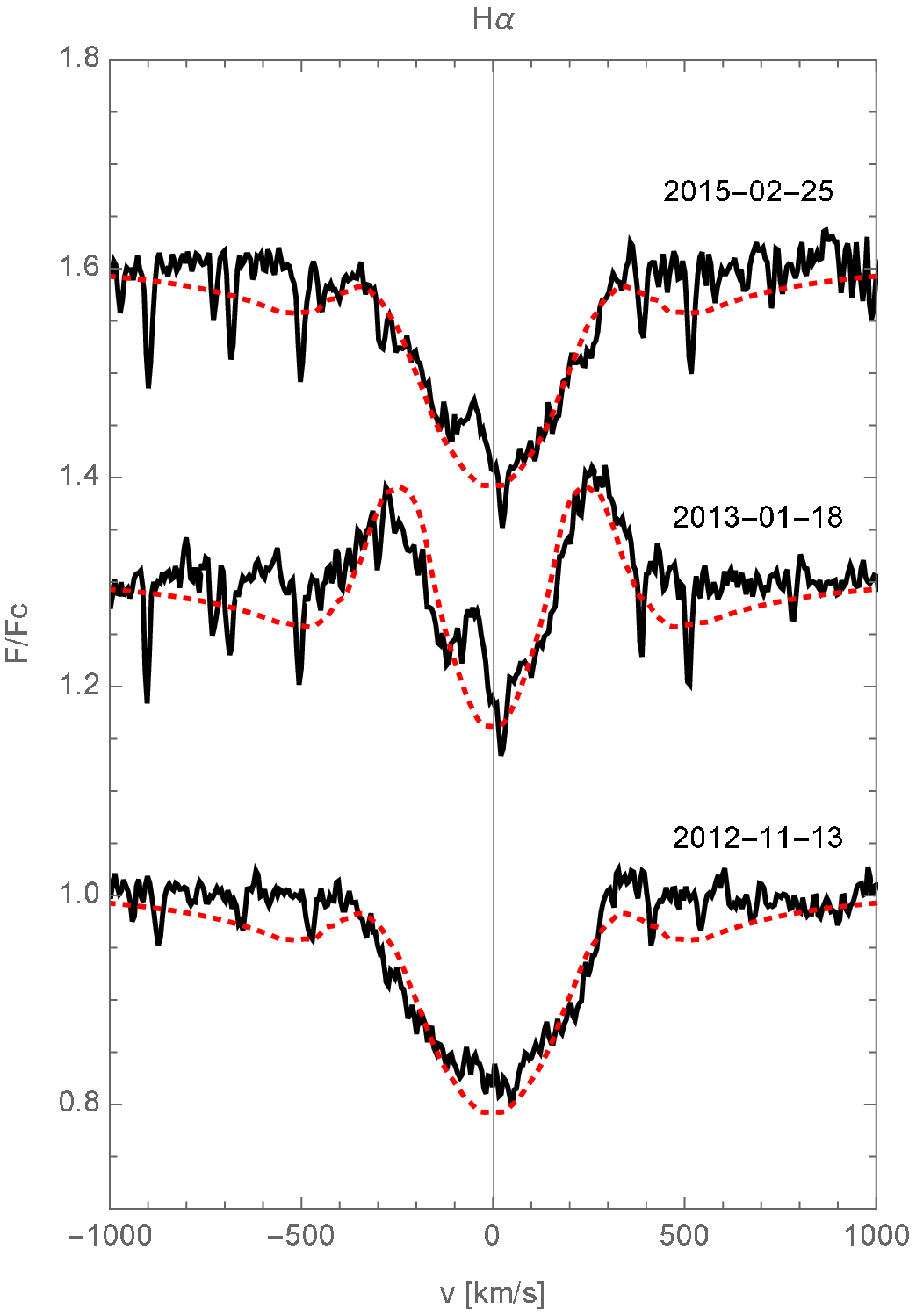}{0.2\textheight}{HD33328*}
                     \fig{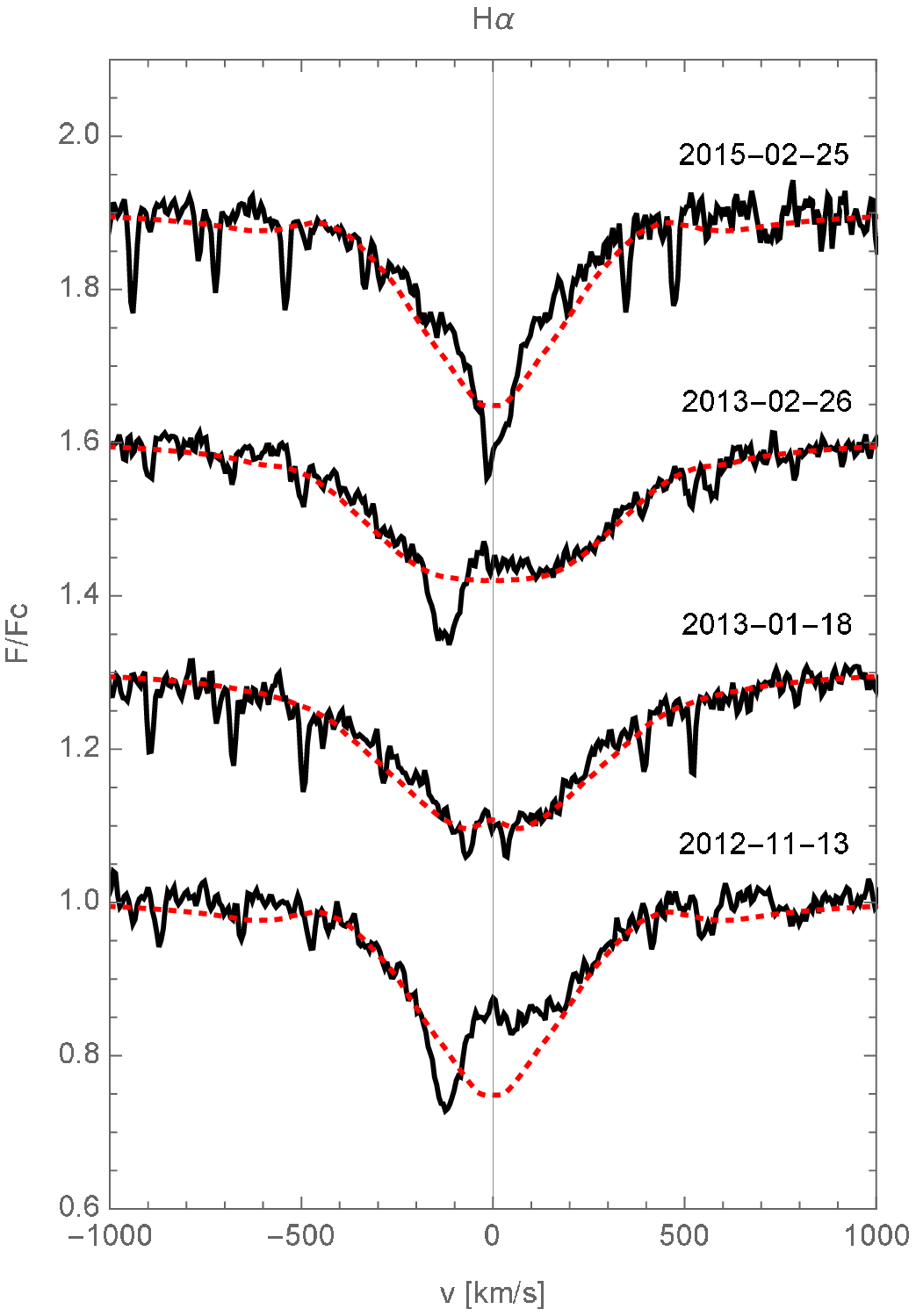}{0.2\textheight}{HD35411*}
                     \fig{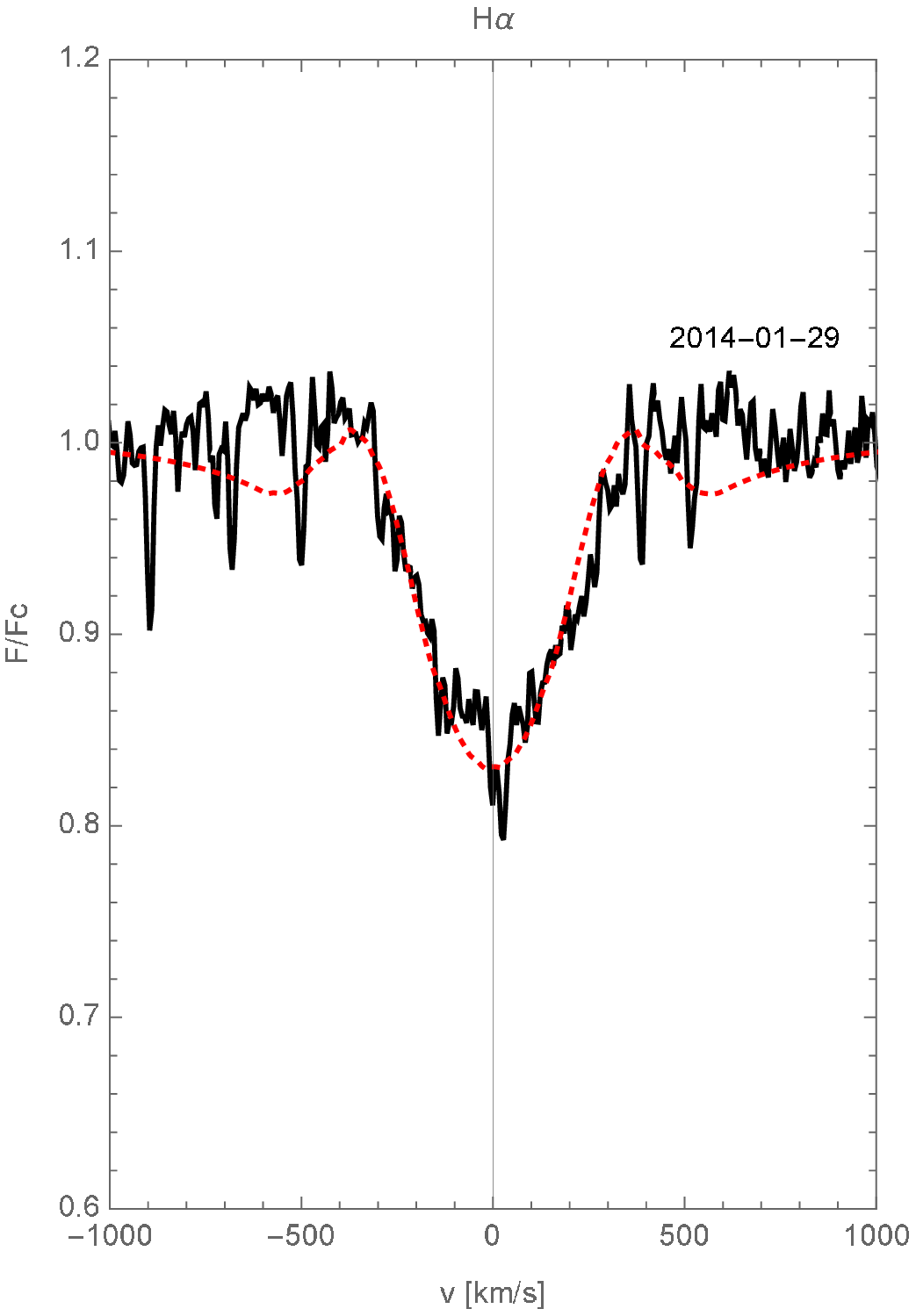}{0.2\textheight}{HD52918}
          }
          
          \gridline{\fig{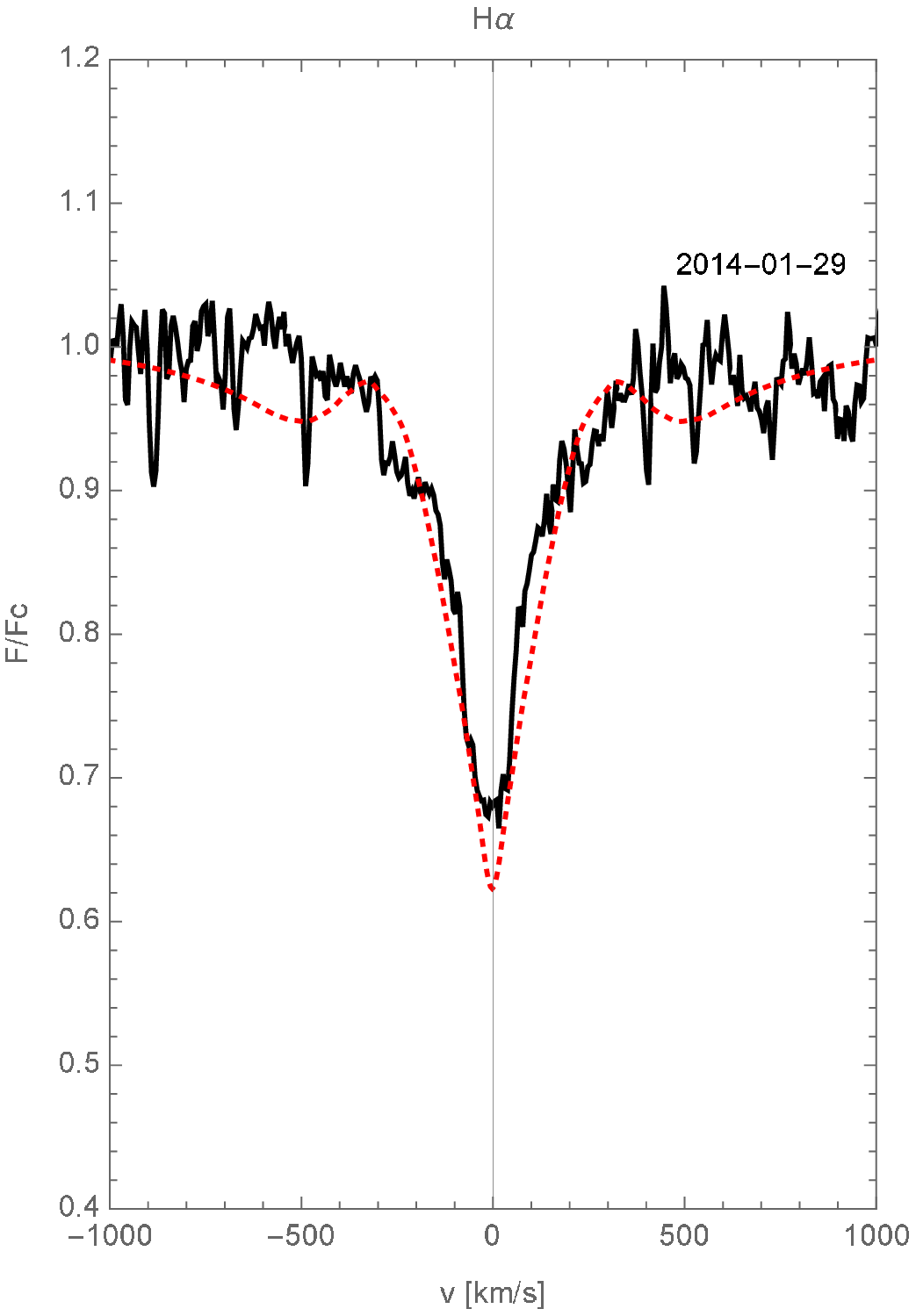}{0.2\textheight}{HD57219}
                    \fig{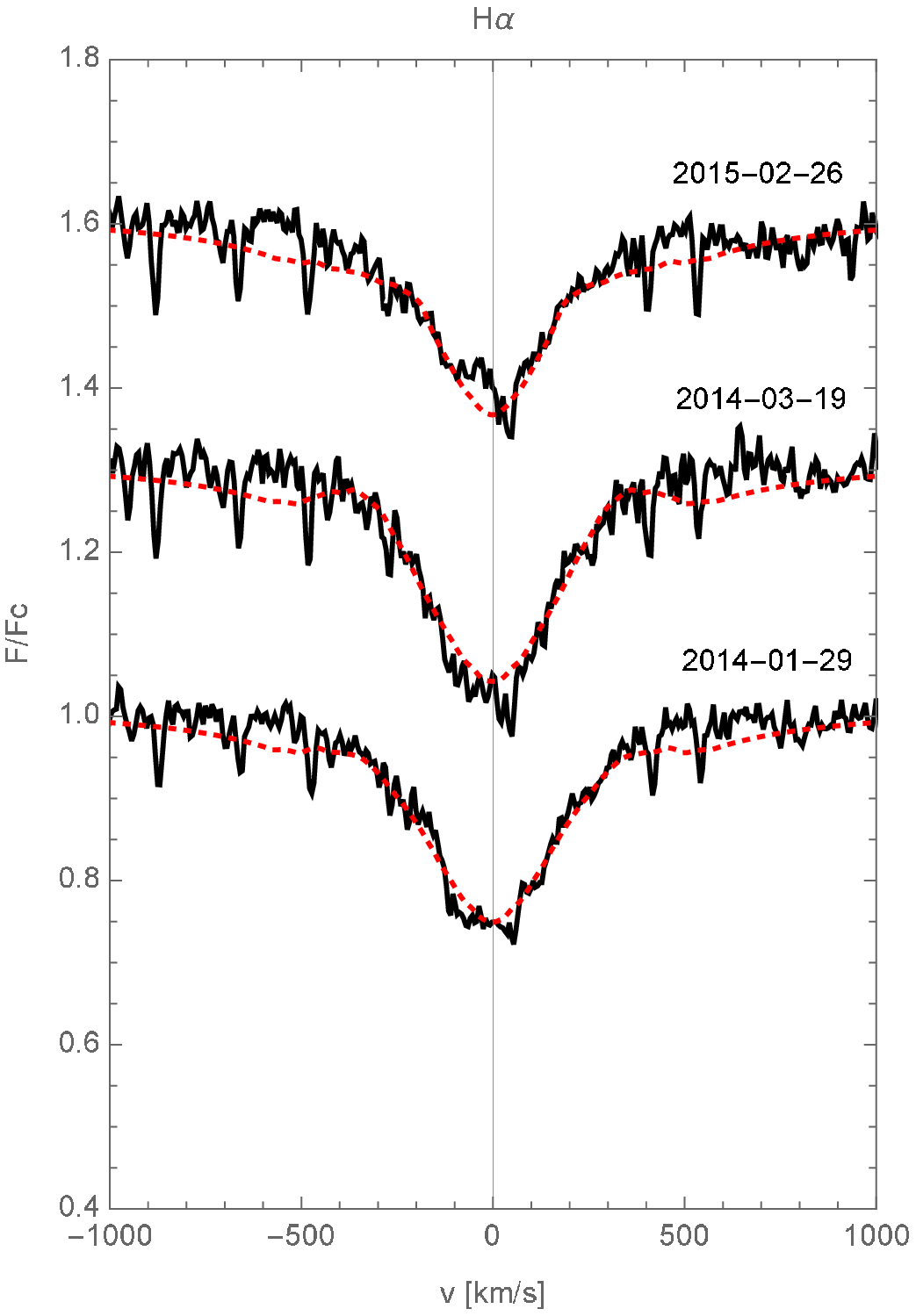}{0.2\textheight}{HD71510*}
                    \fig{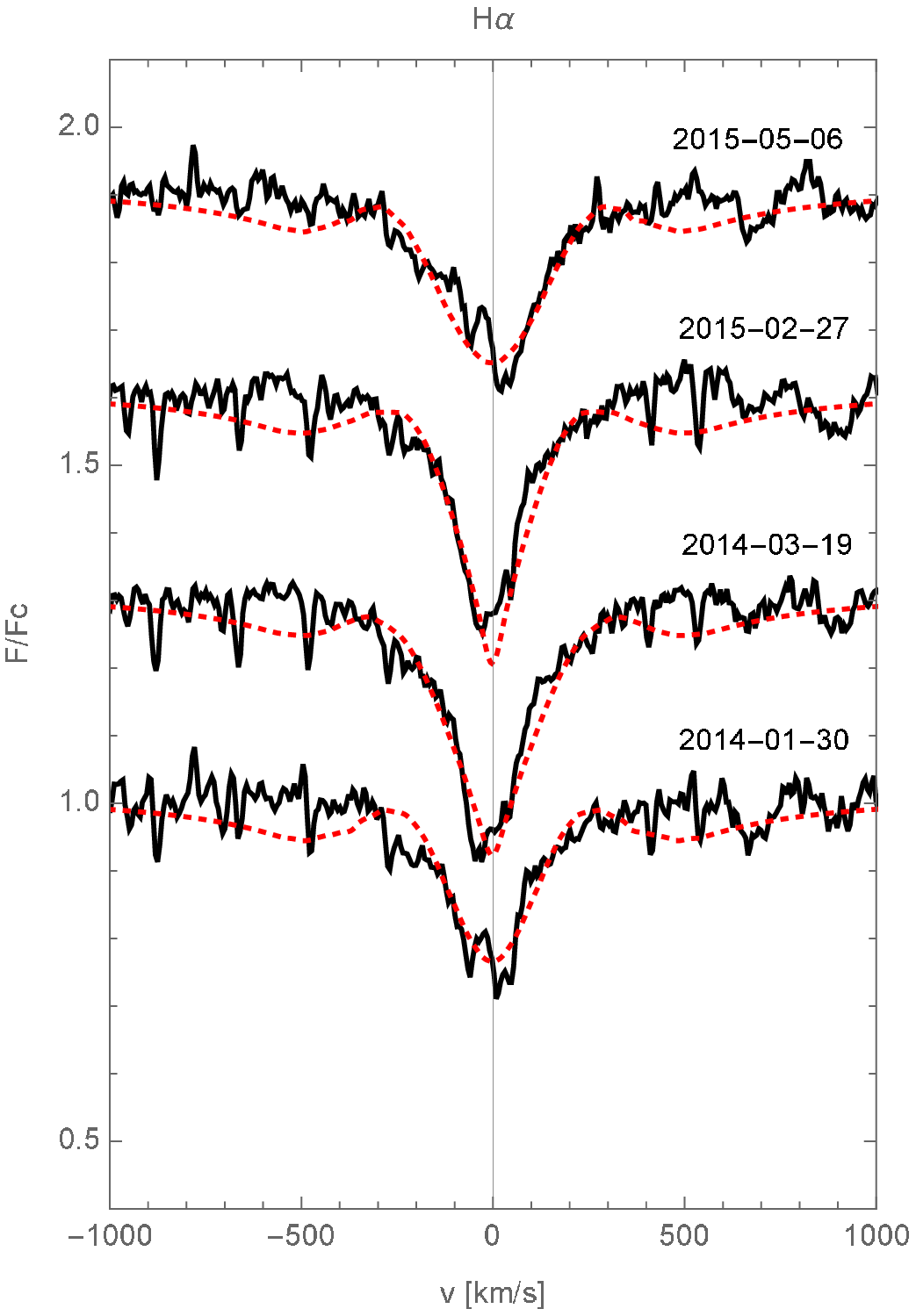}{0.2\textheight}{HD89890*}
          }
           
          \gridline{\fig{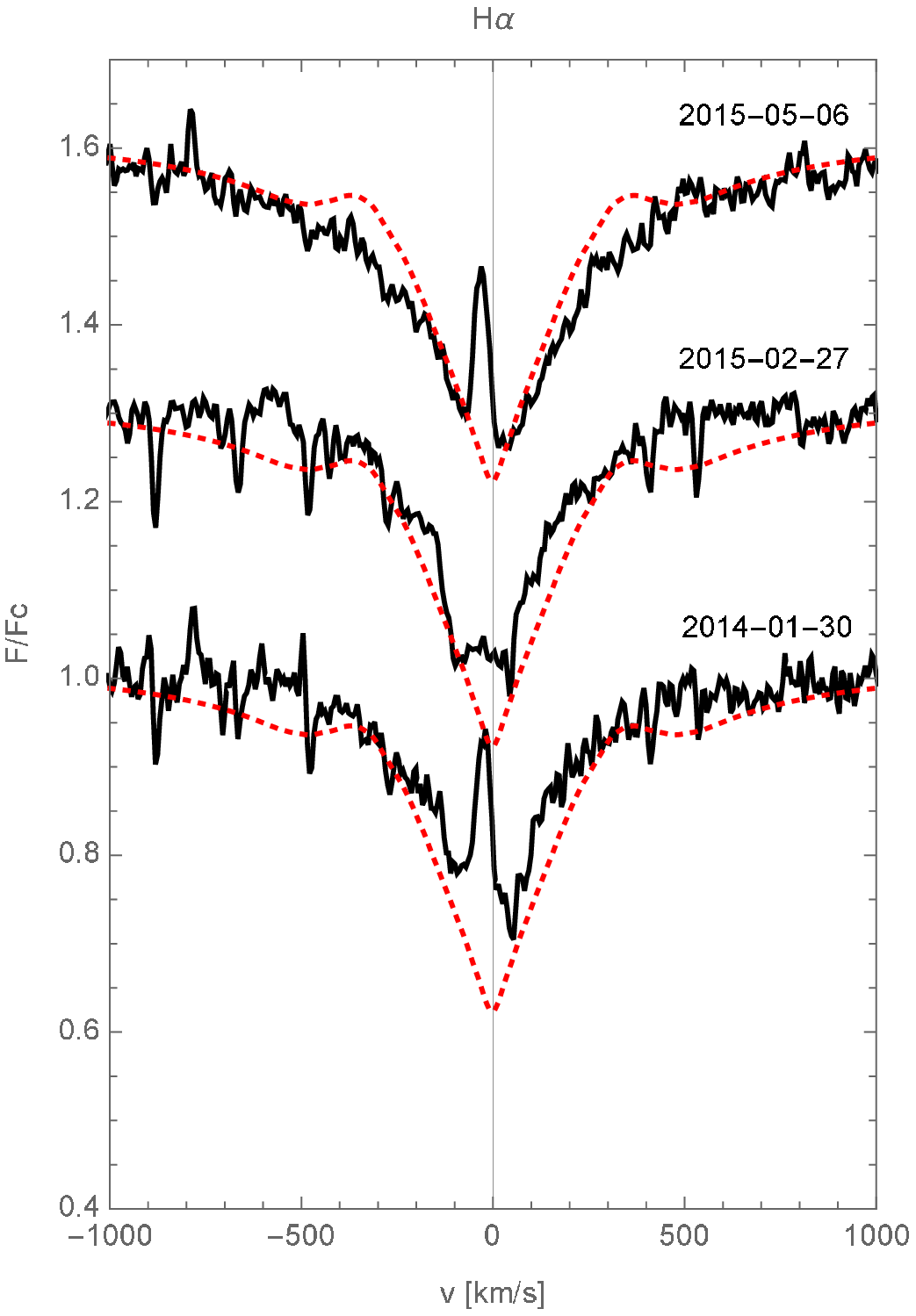}{0.2\textheight}{HD92938*}
                    \fig{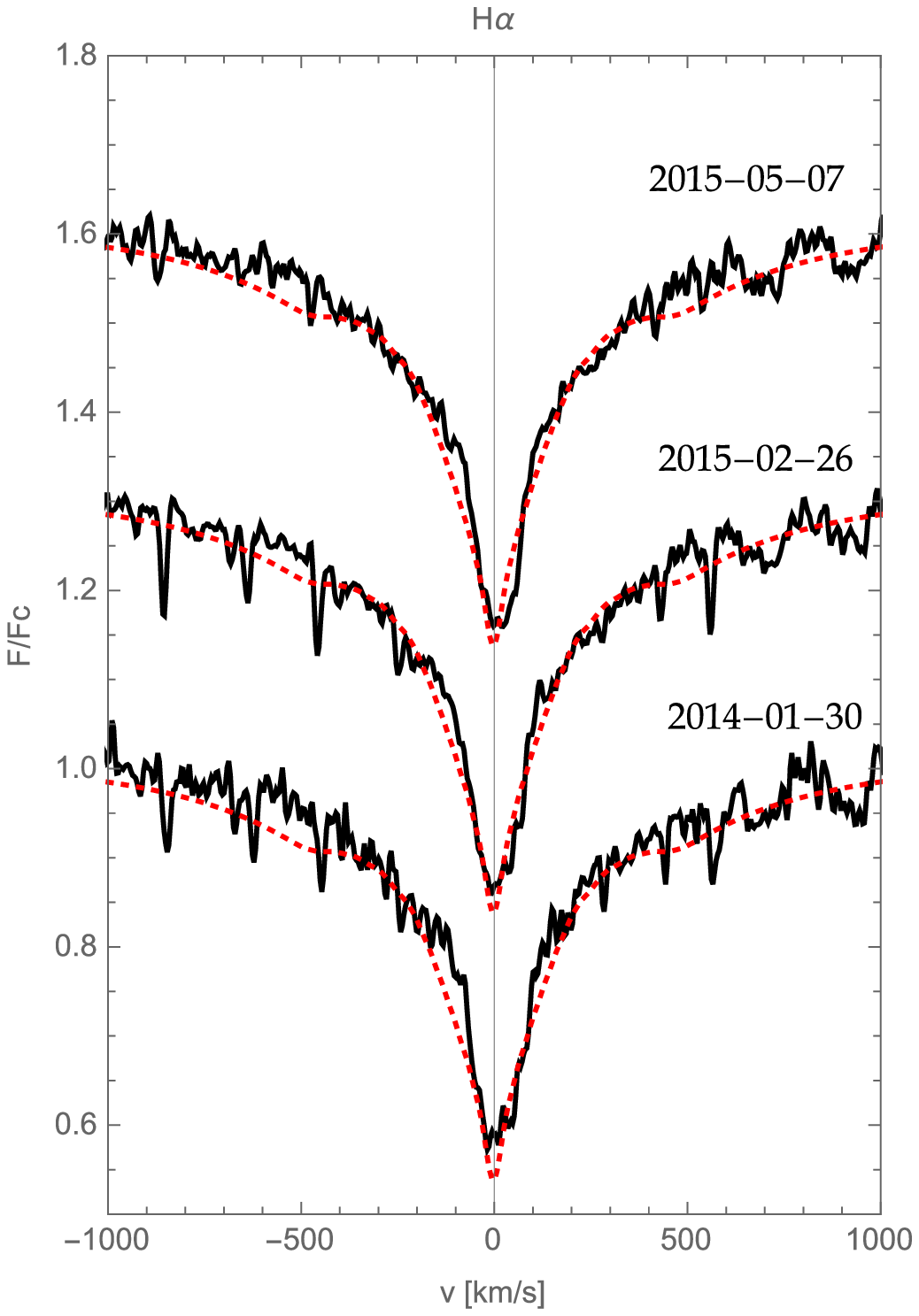}{0.2\textheight}{HD105382*}
                    \fig{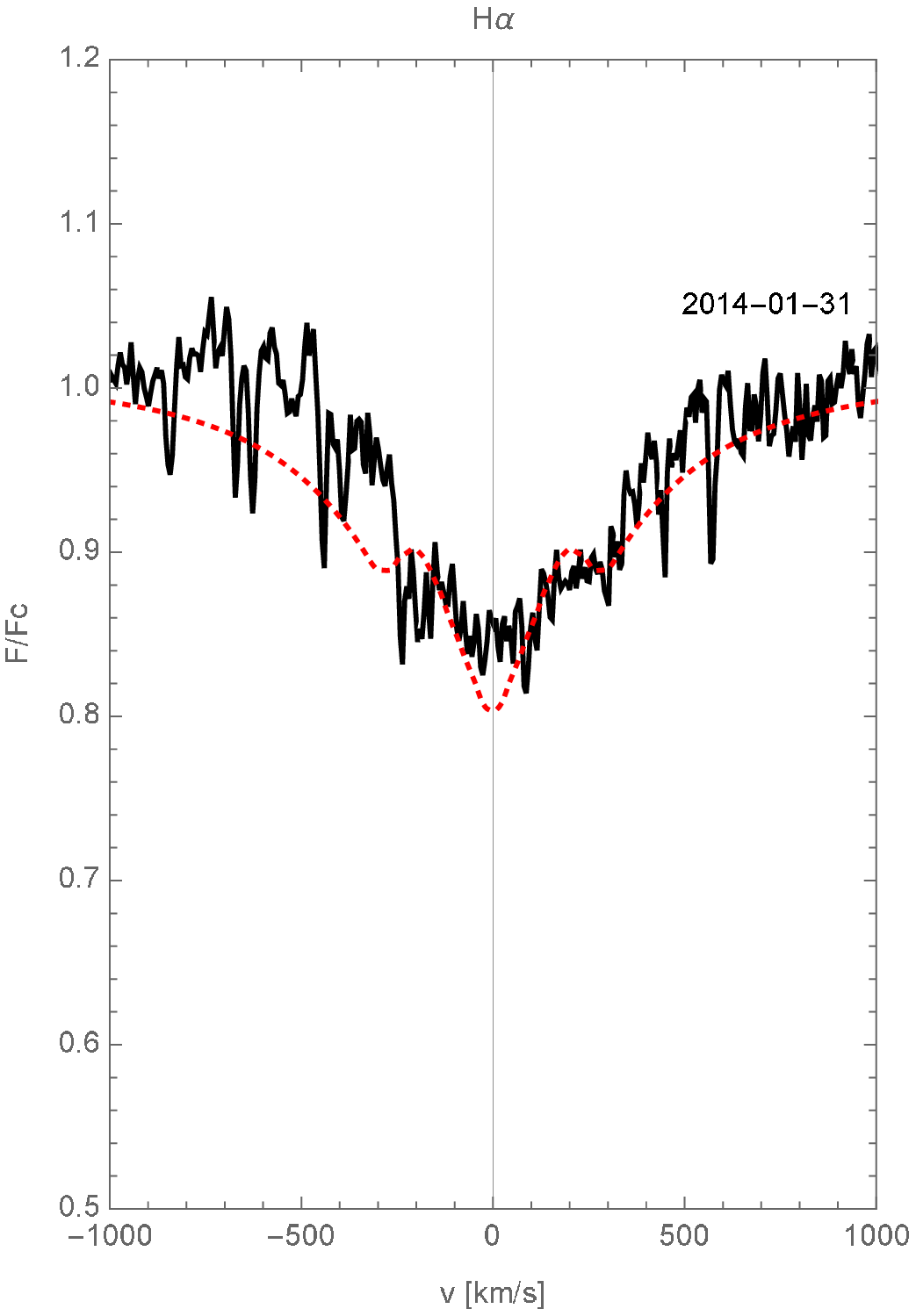}{0.2\textheight}{HD112078}
         }
          
\end{figure}

\begin{figure}

          \gridline{\fig{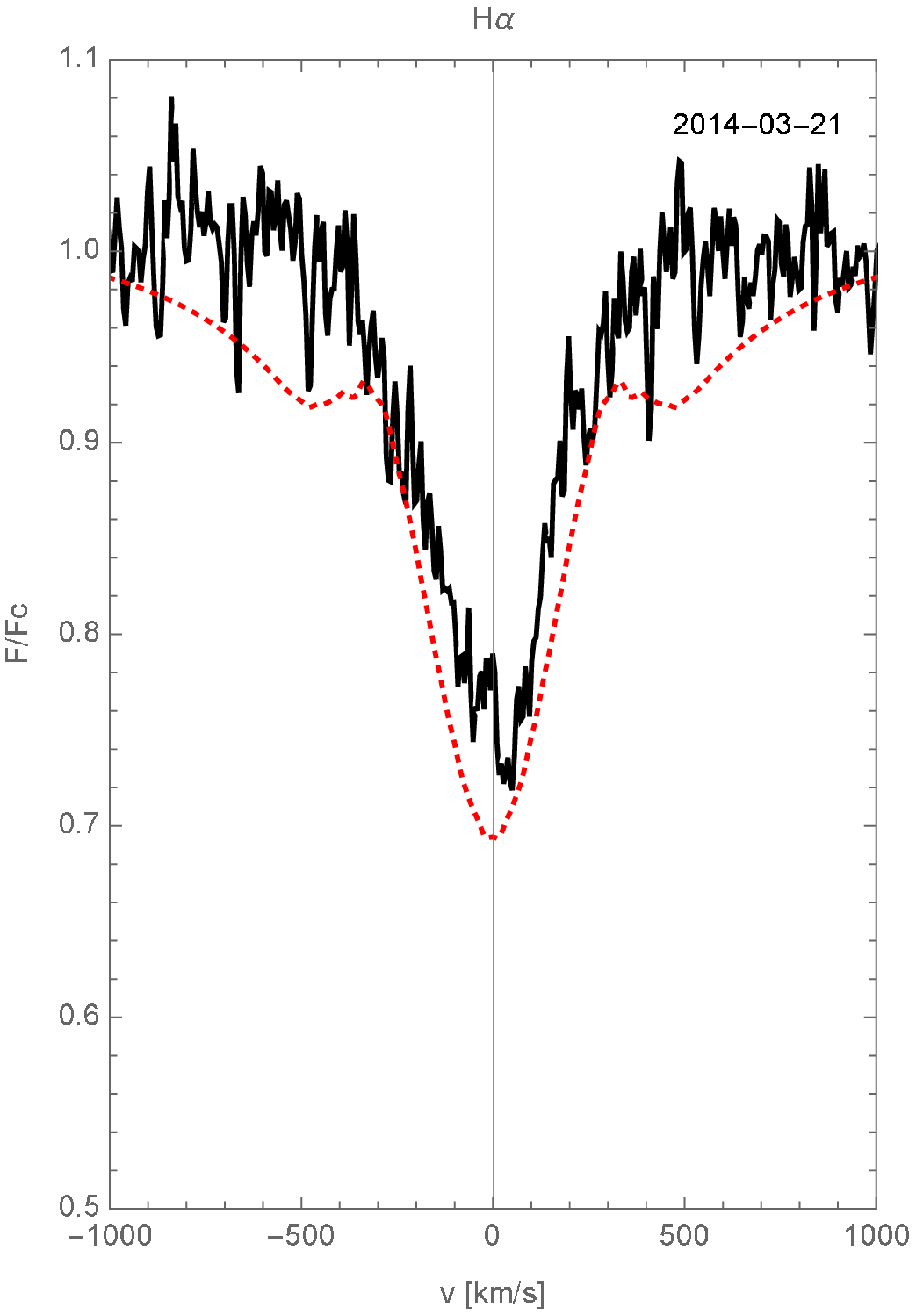}{0.2\textheight}{HD124195}
                    \fig{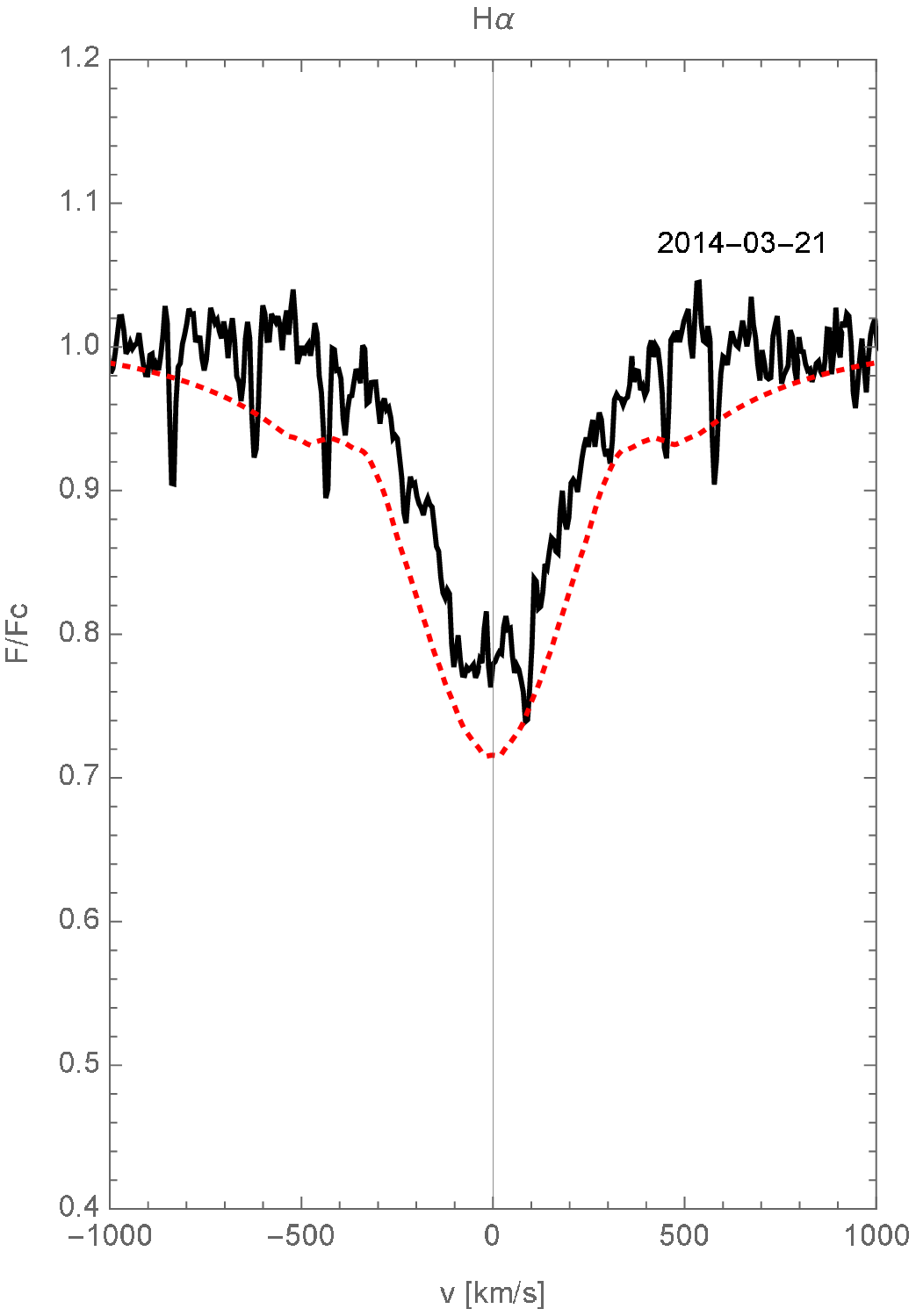}{0.2\textheight}{HD124771}
                    \fig{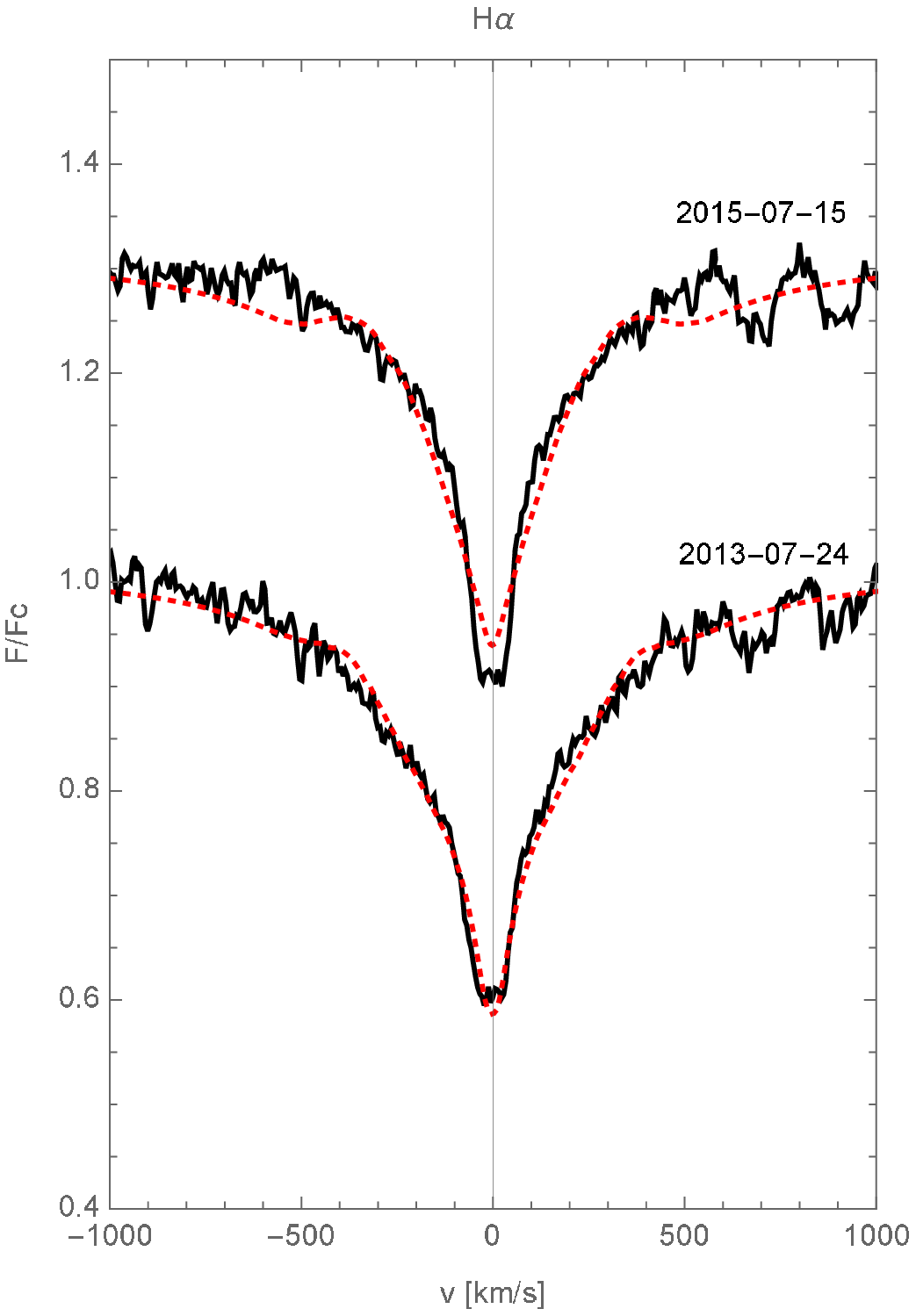}{0.2\textheight}{HD138769*}
           }
          
          \gridline{\fig{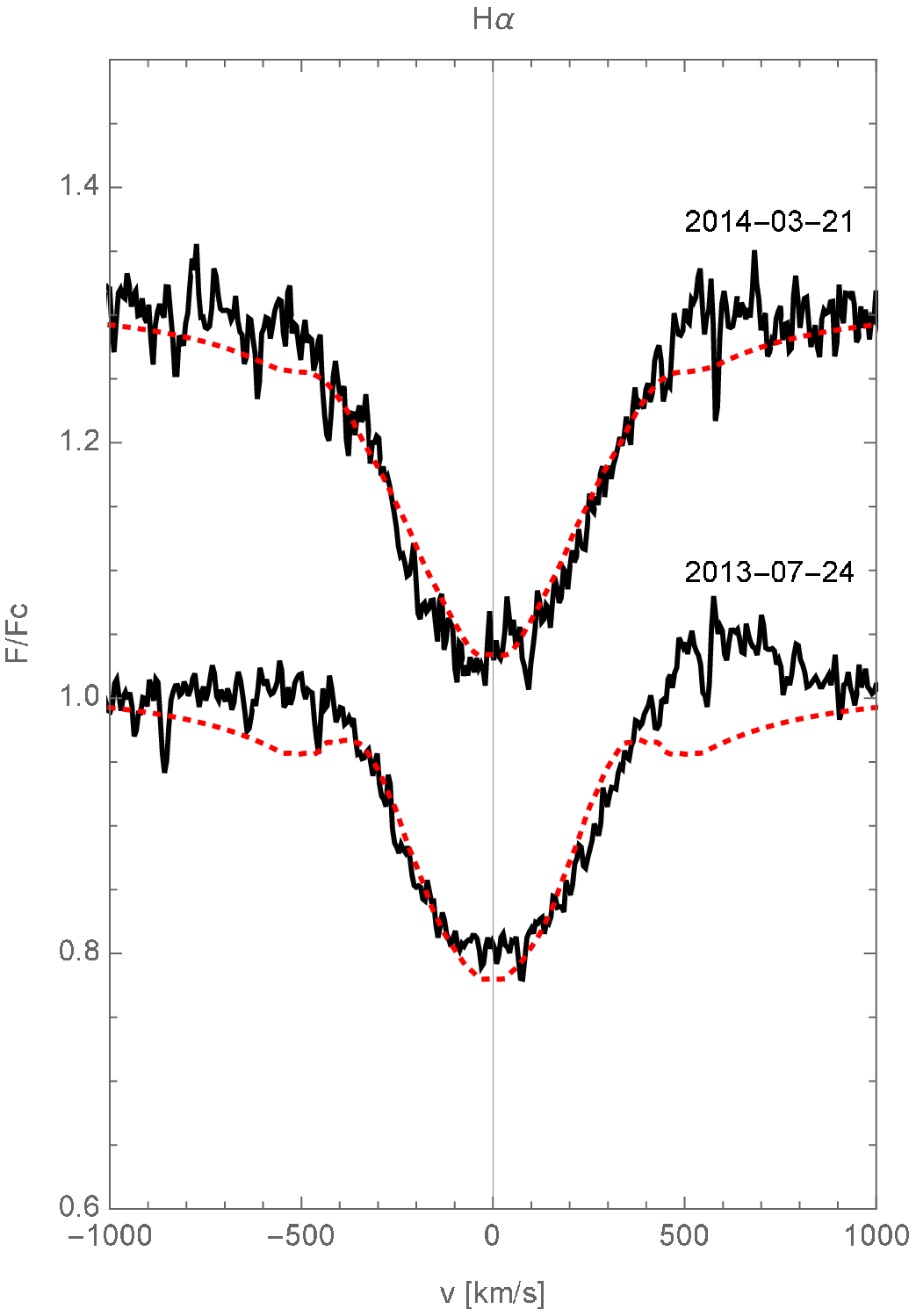}{0.2\textheight}{HD142184*}
                    \fig{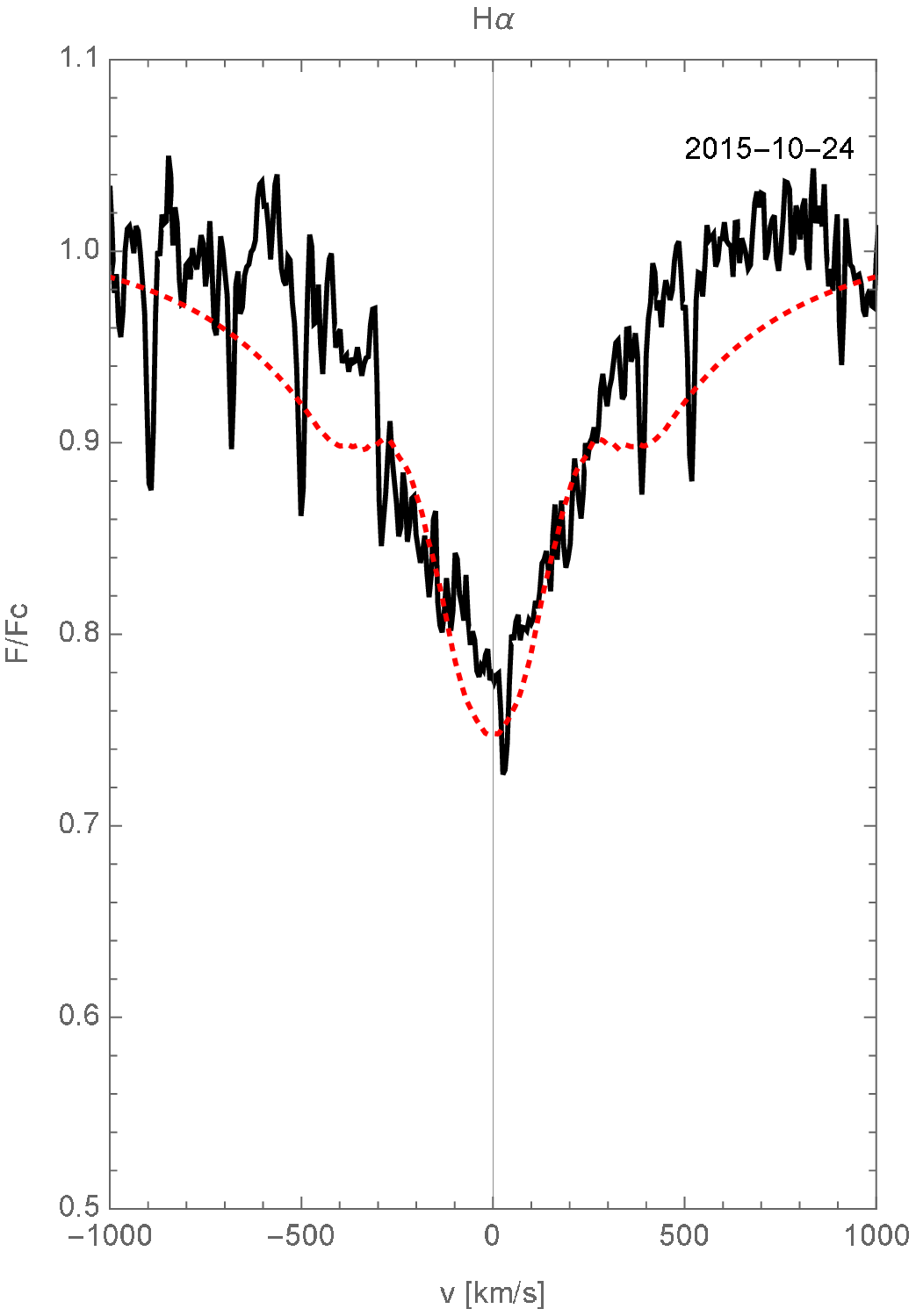}{0.2\textheight}{HD219688}
                    \fig{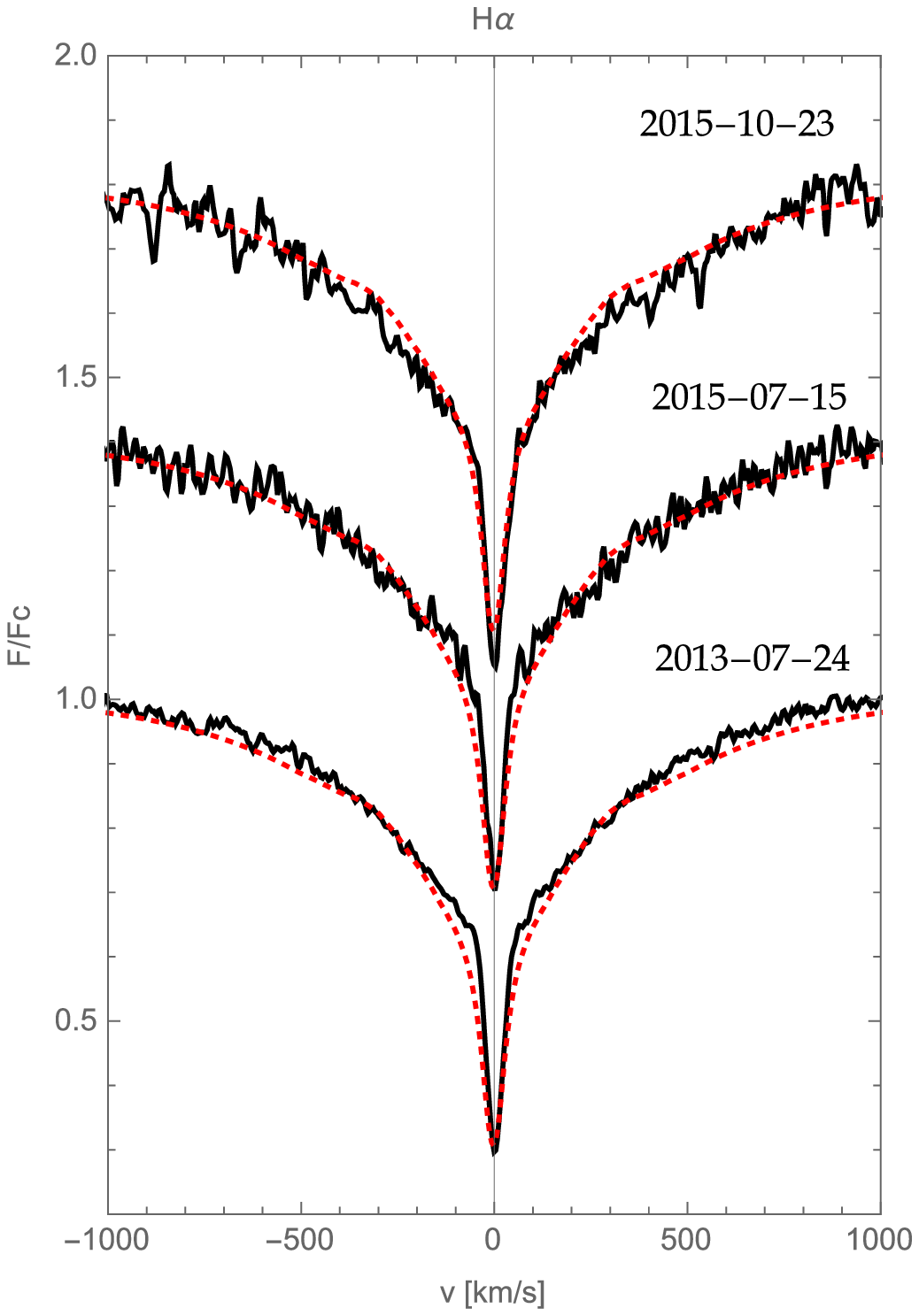}{0.2\textheight}{HD221507}
          }
          
\end{figure}

\newpage
\section{Poor fits} \label{ApC}
The same as Appendix~\ref{ApA} except for program stars with poor fits. See Section~\ref{sec:discussion:e} for details.
\begin{figure}

          \gridline{\fig{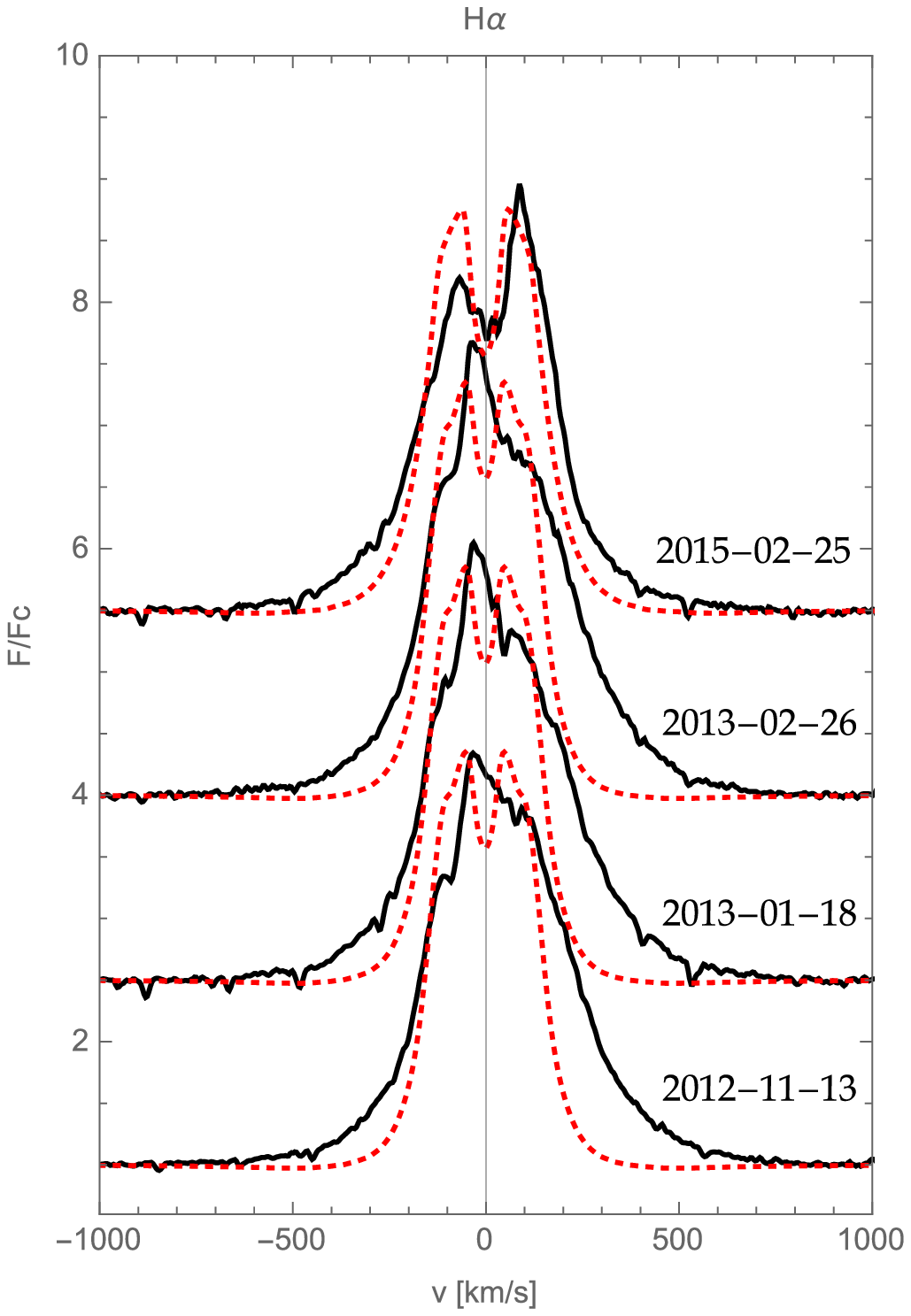}{0.2\textheight}{HD35439}
                    \fig{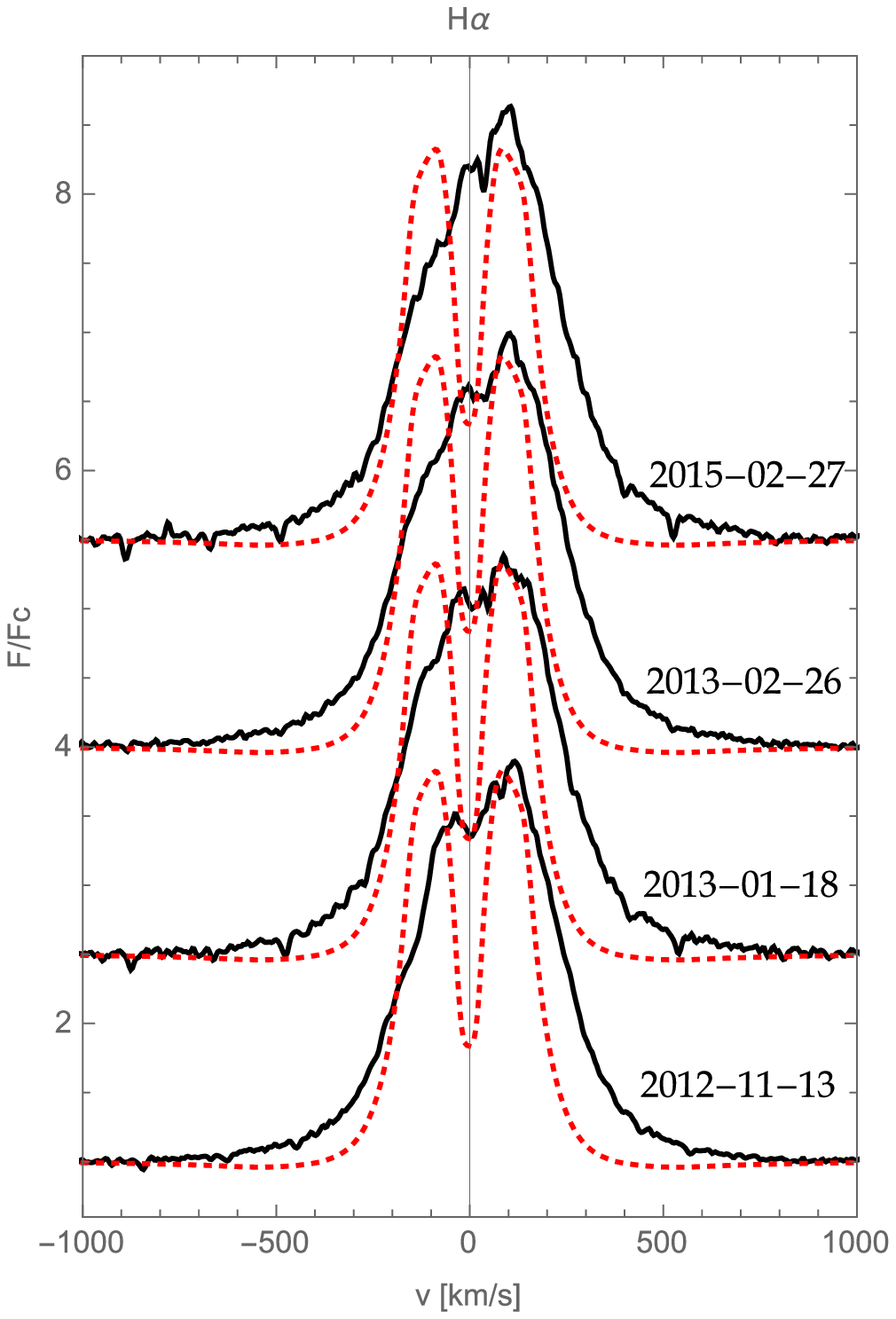}{0.2\textheight}{HD41335}
                    \fig{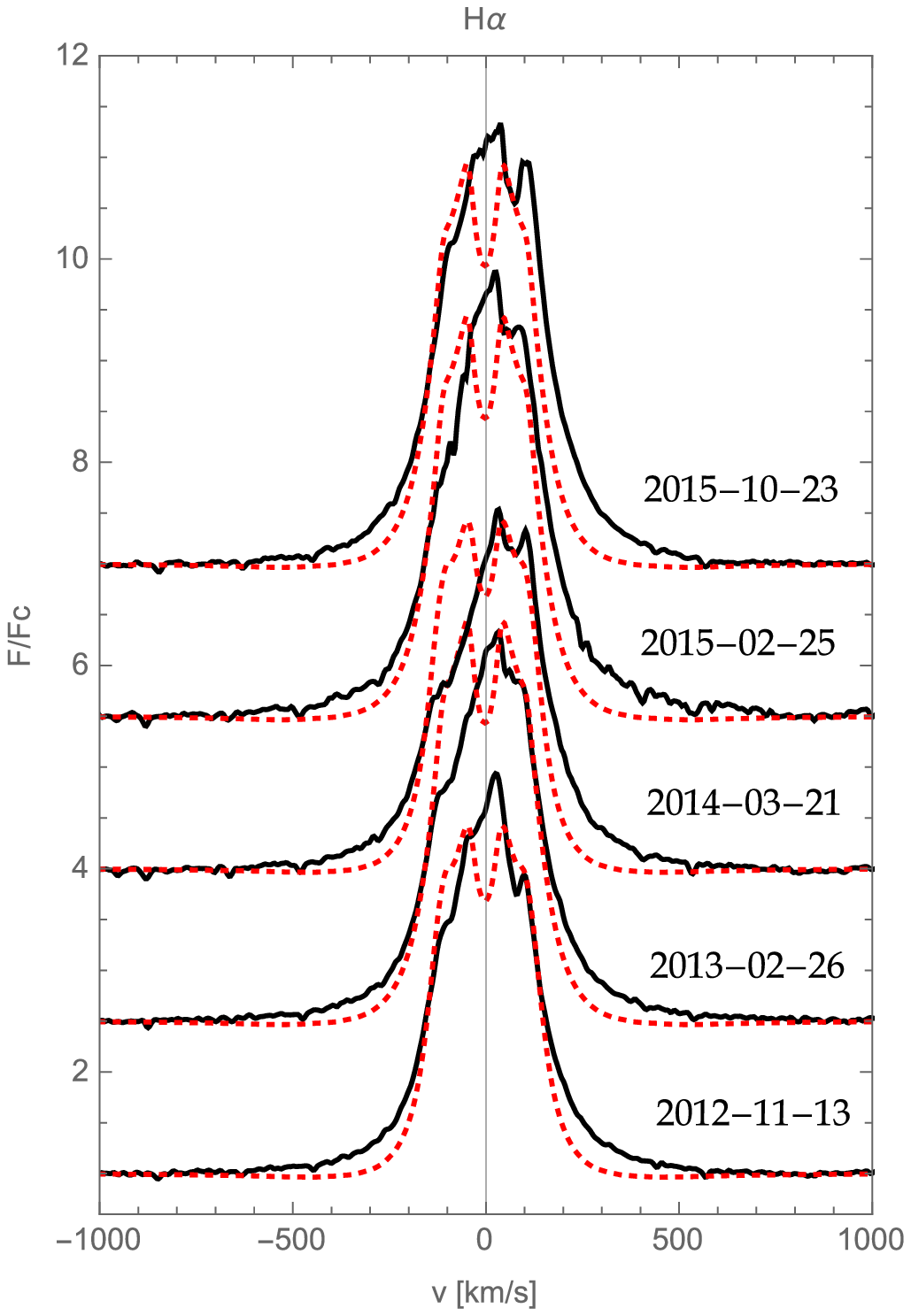}{0.2\textheight}{HD50013}
          }
          
          \gridline{\fig{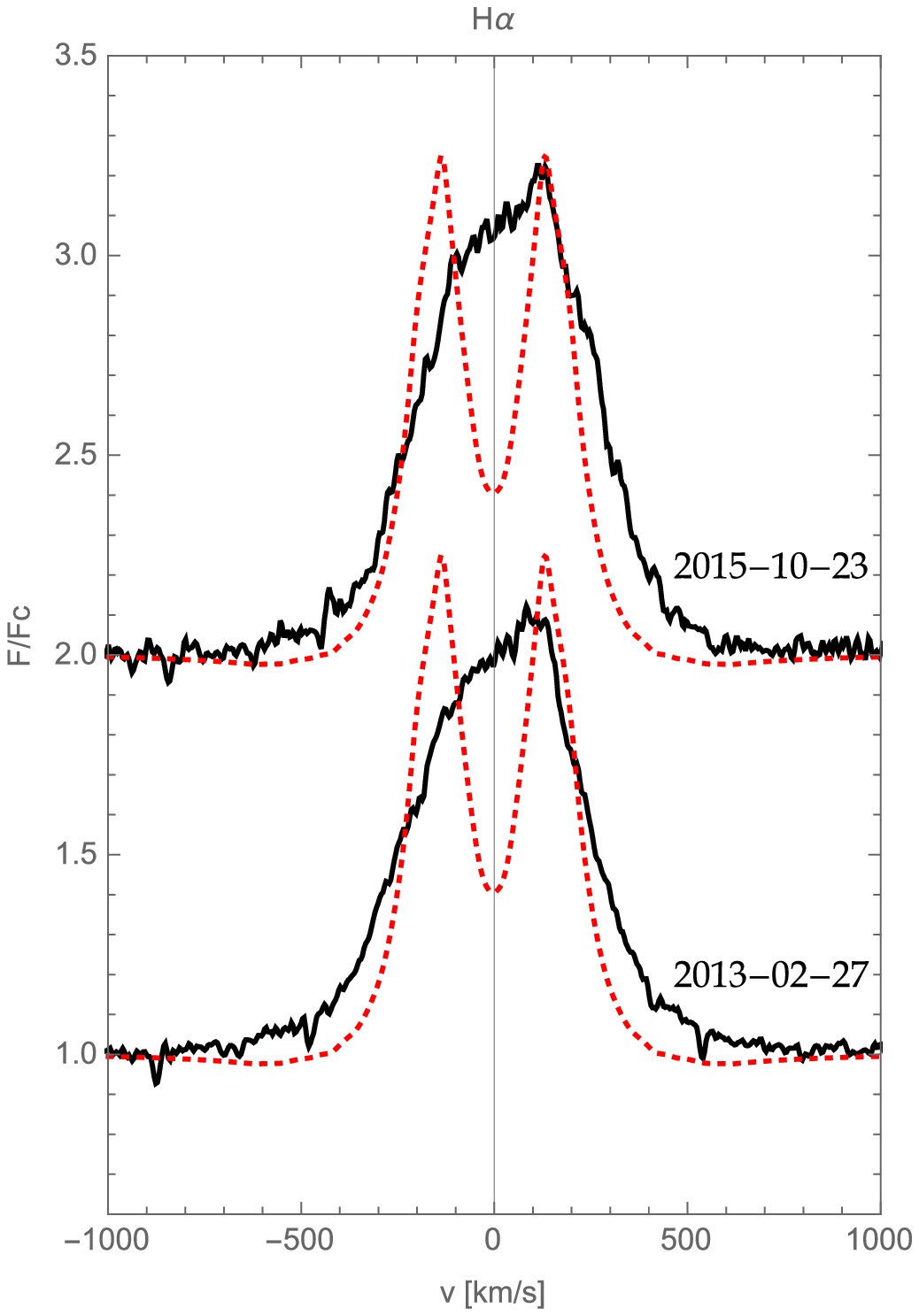}{0.2\textheight}{HD63462}
                    \fig{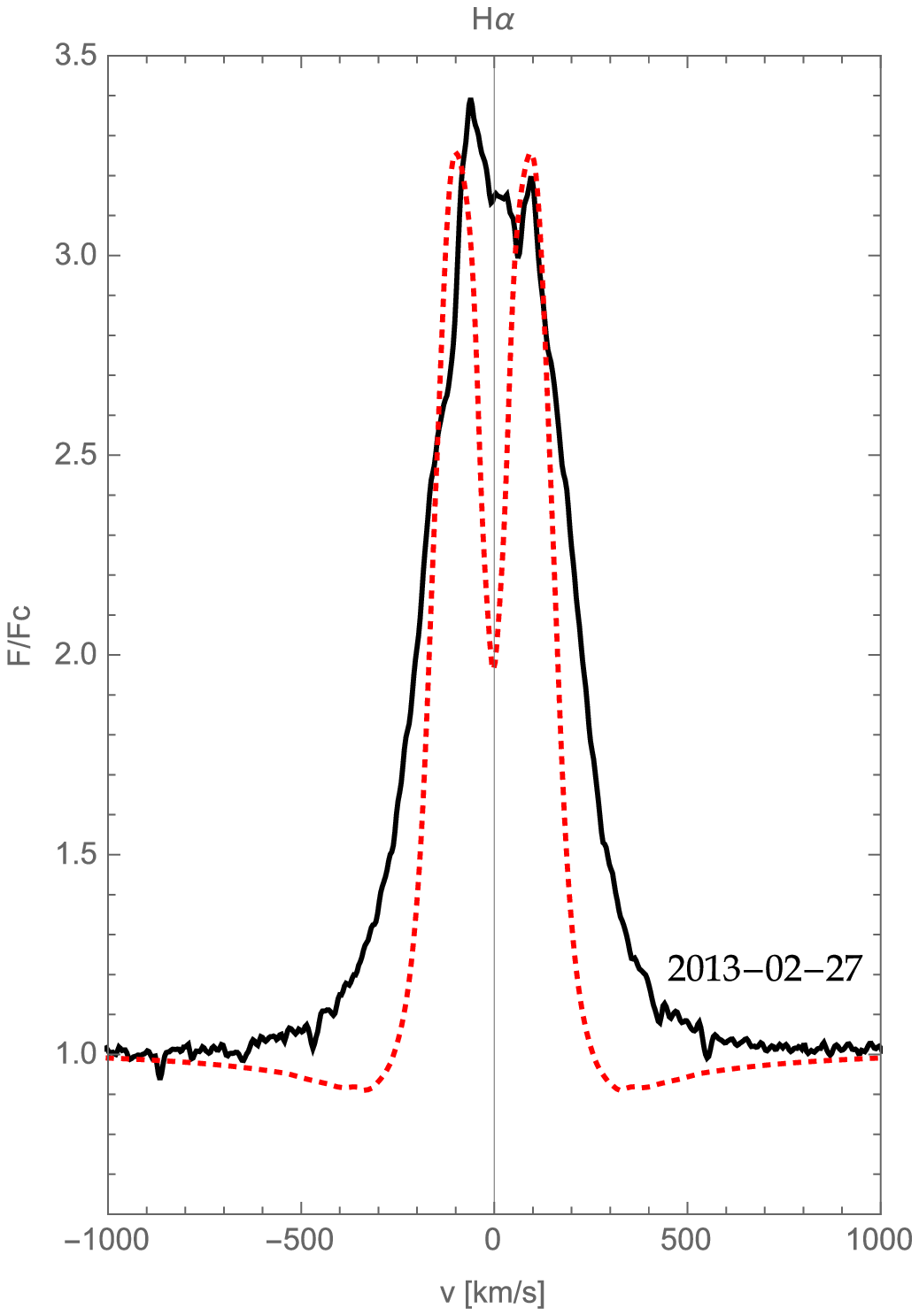}{0.2\textheight}{HD83953}
                    \fig{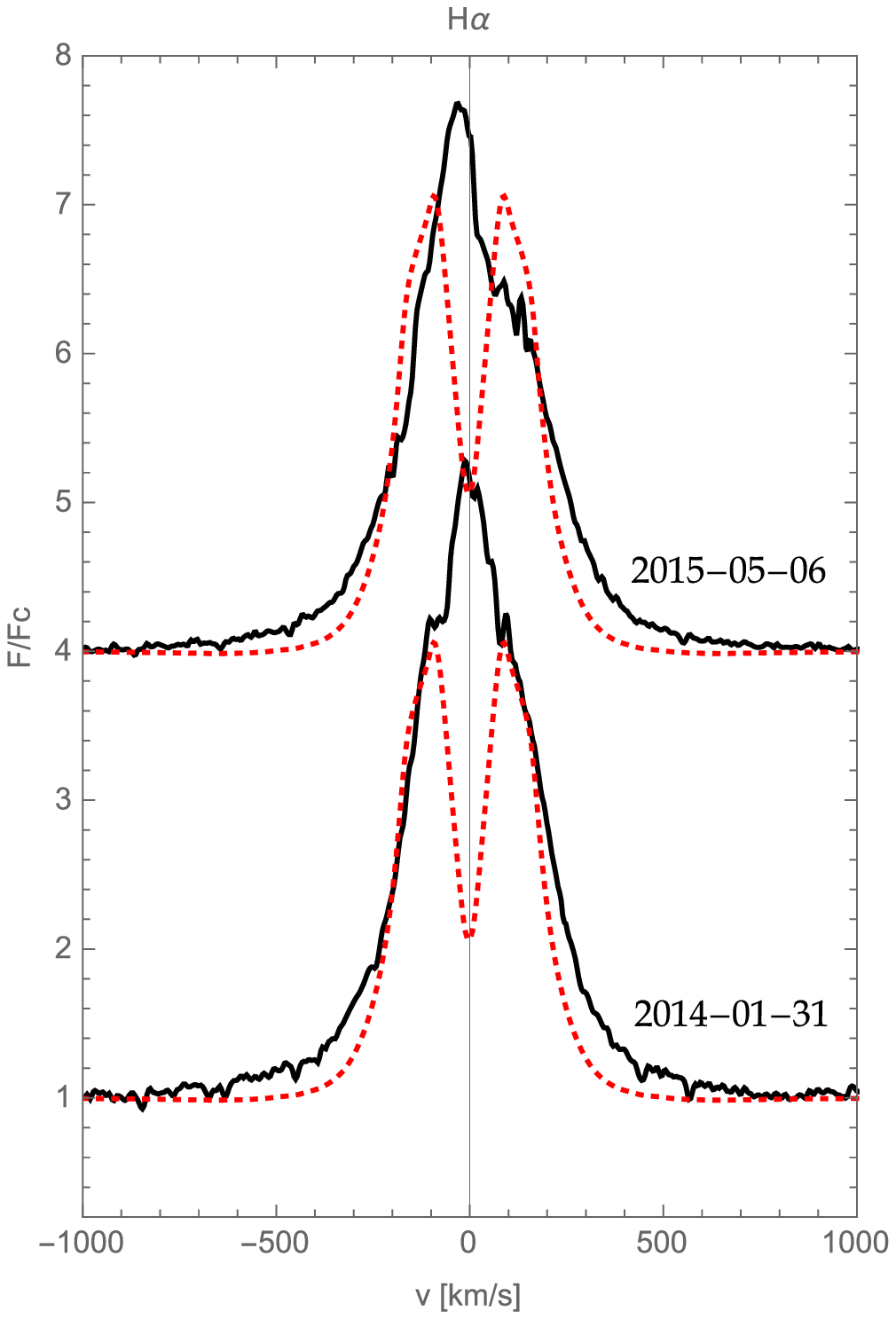}{0.2\textheight}{HD110432}
          }

\end{figure}

\end{document}